\documentclass[12pt]{article}
\usepackage[utf8,ansinew]{inputenc}

\pdfoutput=1

\usepackage{jheppub}
\usepackage{appendix}
\usepackage{color}
\usepackage{epsfig, palatino}
\usepackage{pstricks,pst-node,pst-tree}
\usepackage{epic}
\usepackage{mathrsfs}
\usepackage{ae} 
\usepackage[T1]{fontenc}
\usepackage{amsmath}
\usepackage{physics}
\usepackage{bbm}
\usepackage{amssymb}
\usepackage{bm}
\usepackage{graphicx}
\usepackage{tensor}
\usepackage{ulem}
\usepackage{mathtools}

\usepackage{empheq}
\usepackage[most]{tcolorbox}

\newtcbox{\mymath}[1][]{
    nobeforeafter, math upper, tcbox raise base,
    enhanced, colframe=blue!30!black,
    boxrule=1pt,
    #1}

\def\be#1\ee{\begin{align}#1\end{align}}

\definecolor{darkblue}{cmyk}{0.9,0.9,0,0}
\definecolor{darkgreen}{cmyk}{0.9,0,0.9,0}
\definecolor{blueblue}{cmyk}{0.73,0.28,0,0.5}
\definecolor{lightblue}{RGB}{55,171,200}
\definecolor{grey}{gray}{0.55}
\definecolor{pink}{cmyk}{0., 0.9859943977591037, 0.3571428571428571, 0.16000000000000003}
\definecolor{lightpink}{cmyk}{0., 0.5, 0.5, 0.}
\definecolor{lightgreen}{cmyk}{0.24175824175824182, 0., 0.9615384615384616, 0.28627450980392155}

\usepackage{hyperref}
\usepackage{wasysym}
\usepackage{varioref}
\usepackage{makeidx}
\usepackage[english]{babel}
\usepackage{simplewick}
\usepackage{array}
\usepackage{multirow}
\usepackage{tabularx}

\usepackage{caption}
\usepackage{subcaption}

\def\({\left(}
\def\){\right)}
\def\[{\left[}
\def\]{\right]}
\def\<{\langle}
\def\>{\rangle}

\newcommand{\beq}{\begin{equation}}
	\newcommand{\eeq}{\end{equation}}
\newcommand{\beqq}{\begin{equation*}}
	\newcommand{\eeqq}{\end{equation*}}
\newcommand\beqa{\begin{eqnarray}}
	\newcommand\eeqa{\end{eqnarray}}

\newcommand{\la}[1]{\label{#1}}

\usepackage{tikz}
\usetikzlibrary { decorations.pathmorphing, decorations.pathreplacing, decorations.shapes, }
\usetikzlibrary{shapes.misc}

\tikzset{cross/.style={cross out, draw=black, minimum size=2*(#1-\pgflinewidth), inner sep=0pt, outer sep=0pt},
cross/.default={1pt}}

\preprint{NORDITA 2023-069 \\ \vspace{1pt} \hfill QMUL-PH-23-23}
\title{The O(N) Monolith reloaded:\\
Sum rules and Form Factor Bootstrap}
\author[a]{Luc\'ia C\'ordova,}
\author[a,b,c]{Miguel Correia,}
\author[d,e]{Alessandro Georgoudis}
\author[b]{and Antoine Vuignier}
\affiliation[a]{CERN, Theoretical Physics Department,
CH-1211 Geneva 23, Switzerland}
\affiliation[b]{Fields and Strings Laboratory, Institute of Physics, Ecole Polytechnique Federale de Lausanne (EPFL),
CH-1015 Lausanne, Switzerland}
\affiliation[c]{Department of Physics, McGill University, 3600 Rue University, Montr\'{e}al, H3A 2T8, QC Canada}
\affiliation[d]{NORDITA, Stockholm University and KTH Royal Institute of Technology, Hannes Alfv\'ens v{\"a}g 12, SE-106 91 Stockholm, Sweden}
\affiliation[e]{Centre for Theoretical Physics, Department of Physics and Astronomy, Queen Mary University of London, Mile End Road, London E1 4NS, United Kingdom}

\emailAdd{lucia.gomez.cordova@cern.ch, miguel.correia@epfl.ch, a.georgoudis@qmul.ac.uk, antoine.vuignier@epfl.ch}

 \abstract{
We revisit the space of gapped quantum field theories with a global O(N) symmetry in two spacetime dimensions. Previous works using S-matrix bootstrap revealed a rich space in which integrable theories such as the non-linear sigma model appear at special points on the boundary, along with an abundance of unknown models hinting at a non conventional UV behaviour.
We extend the S-matrix set-up by including into the bootstrap form factors and spectral functions for the stress-energy tensor and conserved O(N) currents. Sum rules allow us to put bounds on the central charges of the conformal field theory (CFT) in the UV. We find that a big portion of the boundary can only flow from CFTs with infinite central charges. We track this result down to a particular behaviour of the amplitudes in physical kinematics and discuss its physical implications.

}

\begin{document}

\maketitle

\section{Introduction}

The S-matrix bootstrap allows us to explore the space of consistent Quantum Field Theories (QFTs). In practice, the imposed consistency conditions are unitarity, analyticity and crossing symmetry of the $2 \to 2$ scattering amplitude. When establishing bounds in parameters, like couplings or Wilson coefficients, one often finds a much larger allowed space than the one covered by known theories \cite{Paulos:2016but,Doroud:2018szp,Paulos:2017fhb,He:2018uxa,Cordova:2018uop,Guerrieri:2018uew,Homrich:2019cbt,EliasMiro:2019kyf,Paulos:2018fym,Karateev:2019ymz,Bercini:2019vme,Cordova:2019lot,Kruczenski:2020ujw,Guerrieri:2020bto,Guerrieri:2020kcs,Correia:2020xtr,Hebbar:2020ukp,Sinha:2020win,Guerrieri:2021ivu,Correia:2021etg,Tourkine:2021fqh,Karateev:2022jdb,EliasMiro:2021nul,He:2021eqn,Guerrieri:2021tak,Chowdhury:2021ynh,Chen:2022nym,Miro:2022cbk, Guerrieri:2022sod,Haring:2022sdp,Correia:2022dyp, Marucha:2023vrn, He:2023lyy,Acanfora:2023axz}.  This might mean there are other fully consistent theories we did not know before or that we must go beyond the $2 \to 2$ set of constraints in order to discard unphysical theories. 

In this work we focus on gapped QFTs with O(N) internal symmetry in $d=2$ spacetime dimensions. The space of allowed S-matrices carved out by the $2 \to 2$ set of constraints was first studied in \cite{Cordova:2018uop,He:2018uxa,Cordova:2019lot} and shows a number of interesting features. In the absence of O(N) symmetry the boundary of the allowed space can be spanned by continuous families of integrable S-matrices \cite{Paulos:2016but,Paulos:2017fhb,Chen:2021pgx,Chen:2022nym}. On the other hand, in the O(N) case the Yang-Baxter (YB) relations become non-trivial and for $N>2$ integrability is only achieved at isolated boundary points. For example, the O(N) non-linear sigma model (NLSM) is found at a kink. 

\begin{figure}[t!]
\centering
\includegraphics[width=.88\textwidth]{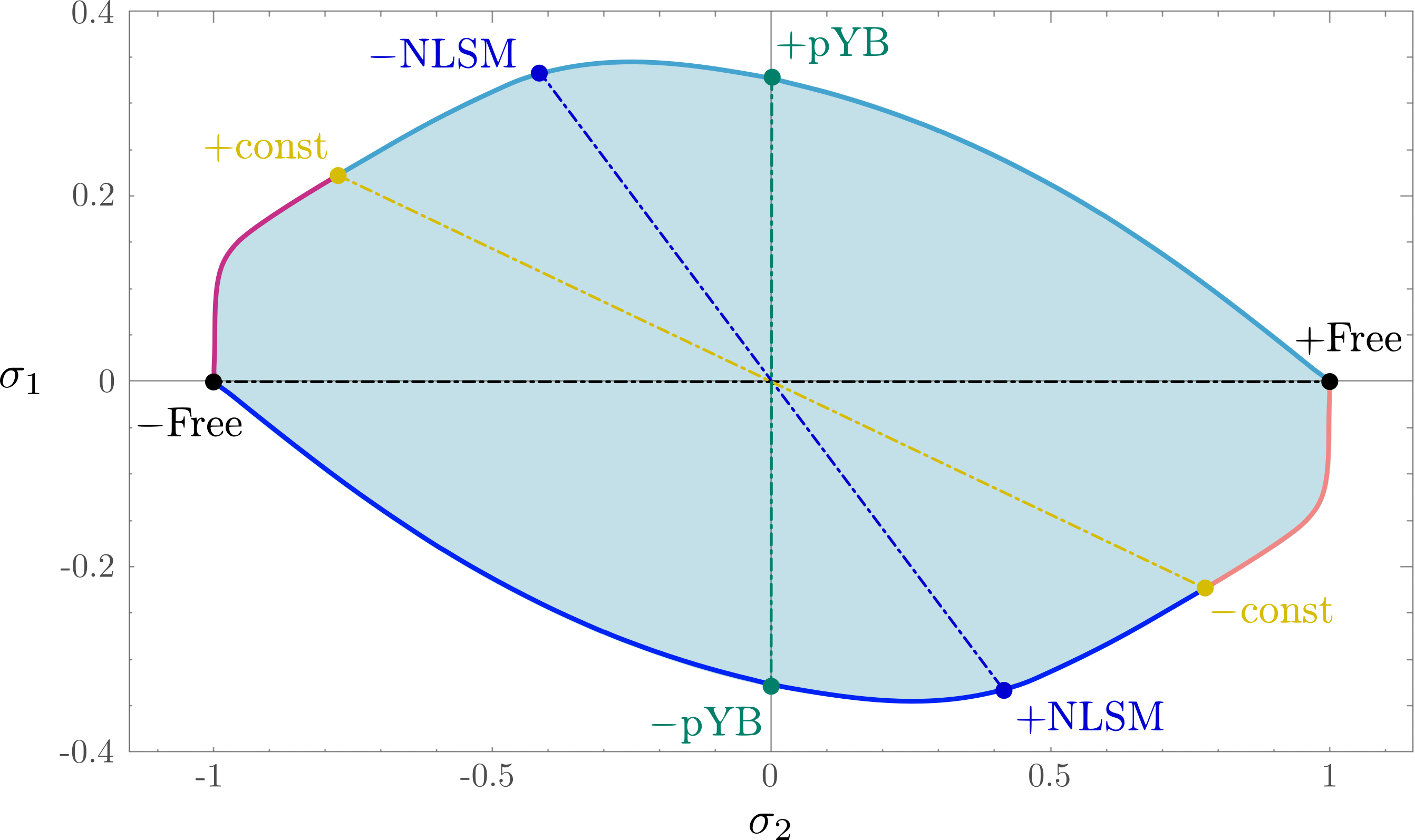}
\caption{Two-dimensional slice of the O(N) monolith for $N=7$  taken from \cite{Cordova:2019lot}. The shaded blue region are the allowed values for the amplitudes at the crossing symmetric point, the axes given by $\sigma_{1,2}=\sigma_{1,2}(s=2m^2)$. The highlighted points are different exact S-matrices as explained in the main text. The lines connecting them are the sections used in \ref{sec:minckmono}.
}
\label{fig:2Dmonolith}
\end{figure}

The space of O(N) S-matrices -- the O(N) monolith -- exhibits a variety of other interesting and still mysterious features, as highlighted in figure \ref{fig:2Dmonolith}. Namely, the `constant' 
amplitudes on which the extremal S-matrices are inelastic (a rare observation of non-unitarity saturation at the boundary), or the `pre-vertices' where the periodic-Yang-Baxter (pYB) S-matrices are found. The latter satisfy the Yang-Baxter relations but 
a corresponding physical model or Lagrangian formulation is still unknown. 
The pYB amplitudes -- and generically the extremal amplitudes at the boundary -- show a rich analytic structure with infinitely many resonances arranging themselves in a periodic manner, which raises questions about the UV nature of these putative QFTs.\footnote{Similar periodicities are also observed in the S-matrices coming from $T\bar{T}$ deformations, which are known to be incompatible with conventional local UV-complete QFTs \cite{Smirnov:2016lqw,Camilo:2021gro}. } Are these structures unphysical? Is the NLSM the only non-trivial physical theory at the boundary? 

These questions call for an enlargement of the subset of constraints and observables. One option would be to include multi-particle constraints. However, these are plagued with fundamental and technical difficulties such as the presence of anomalous thresholds \cite{Hannesdottir:2022bmo,Correia:2022dcu,Caron-Huot:2023ikn}. A recent viable alternative, which allows to probe the UV nature of these theories, is to include local operators such as the stress-energy tensor \cite{Karateev:2019ymz}. In this extended framework, form factors and spectral densities join the S-matrix in a more powerful bootstrap setup \cite{Karateev:2019ymz,Karateev:2020axc,Correia:2022dyp,He:2023lyy}. 

In particular, if the local operators are chosen to be the stress tensor and the conserved O(N) currents, information about the conformal field theory (CFT) at the UV fixed point can be included via the so-called \emph{sum rules}. These relate CFT properties -- the central charges $c$ and $k$,\footnote{We define $c$ and $k$ as coefficients in the two point function of the stress-energy tensor and the O(N) conserved currents in the CFT, respectively. 
If the currents define an affine Lie algebra, the current central charge $k$ defines the `level' of the algebra.} respectively -- to the spectral density of these operators in the QFT \cite{Karateev:2020axc}. These CFT quantities can now be targeted by the QFT bootstrap, and be used to probe the O(N) monolith. By minimizing $c$ and $k$ across the O(N) monolith we can constrain the possible CFTs from which these S-matrices in the infrared (IR) could flow from. 

In simple terms, this framework let us see explicitly how conditions in the IR trickle up along the flow onto the UV, allowing us to constrain the UV CFTs via the S-matrix principles. So, we may also ask concretely, what is the allowed space of central charges $c$ and $k$ that is compatible with the unitarity, crossing symmetry and analyticity of the S-matrix? Where do Wess-Zumino-Witten (WZW) models -- for which $c$ and $k$ are related by the Sugawara construction -- sit inside this space? What differentiates the amplitudes leading to finite/infinite central charges? 

The goal of this work is to implement this framework and address these questions. The organization of this paper is as follows. In section \ref{sec:setup} we briefly overview the O(N) monolith space of S-matrices and review how form factors and spectral densities can be included into the O(N) bootstrap setup. In section \ref{sec:results} we show our numerical results for the bounds on the central charges $c$ and $k$. In section \ref{sec:analytic_structure_S} we use analytical form factor bootstrap methods to argue for a relation between the behaviour of the amplitudes in physical kinematics and the finiteness of the central charges. We then discuss our results in section \ref{sec:discussion}. Our conventions, derivations of dual optimization problems, numerical implementation and various analytical results are collected in appendices \ref{app:exactamps}, \ref{sec:conventions}, \ref{app:dualproblems}, \ref{sec:numerics} and \ref{sec:analytics}.

\section{Review of O(N) monolith and Form Factor Bootstrap setup}
\label{sec:setup}
In this work we will be dealing with three different types of observables: $2\to 2$ scattering amplitudes of identical massive particles $S(s)$, two particles form factors of the trace of the stress-energy tensor $F_\Theta(s)$ and of the O(N) conserved currents $F_J(s)$ and their spectral densities, $\rho_\Theta(s)$ and $\rho_J(s)$. 

In the first part of this section we review some properties and known bootstrap results for amplitudes of particles transforming in the vector representation of O(N). In the second part we explain how to include the stress tensor and O(N) currents and set up the bootstrap problem of minimizing the associated central charges.

\subsection{Pure S-matrix bootstrap and O(N) monolith}
The particles we are scattering are states transforming in the vector representation of O(N) and carry a label $i=1,2,...,N$. The scattering amplitude is defined as 
\be 
{}_{kl}^{out} \langle p_1, p_2 | k_1, k_2 \rangle^{in}_{ij} \equiv (2\pi)^2 \delta^{(2)}(p_1+p_2-k_1-k_2) \mathcal{N}_2(s) S_{ij}^{kl}(s)\,, \label{eq:norminout}
\ee
where $\mathcal{N}_2(s) \equiv 2 \sqrt{s} \sqrt{s-4m^2}$ and $s$ is the center of mass energy squared $s=-(p_1+p_2)^2$. Following group theoretic considerations, the scattering amplitude is decomposed into the singlet ($\bullet$), antisymmetric ($A$) and symmetric ($S$) irreducible representations as follows\footnote{The projectors are given by 
\be
T^\bullet_{ij,kl} \equiv \frac{1}{N} \delta_{ij} \delta_{kl}, \qquad T^A_{ij,kl} \equiv \frac{\delta_{ik} \delta_{jl}- \delta_{il} \delta_{jk}}{2}, \qquad T^S_{ij,kl} \equiv \frac{\delta_{ik} \delta_{jl}+\delta_{il} \delta_{jk}}{2} - \frac{1}{N} \delta_{ij} \delta_{kl}. \nonumber
\ee}
\be 
S_{ij}^{kl}(s) =  S_\bullet(s)\,T^\bullet_{ij,kl} + S_A(s)\,T^A_{ij,kl}\ +  S_S(s)\,T^S_{ij,kl}\,,
\ee 

The three different channel amplitudes $S_{a}(s)$  ($a=\bullet,A,S$) obey the usual unitarity constraint $|S_a(s)|\leq 1$, for $s\geq4m^2$, but get mixed under crossing:
\beq
S_a(4m^2-s)=C_{ab} S_b(s)\,, \qquad C_{ab}=\left(
\begin{array}{ccc}
\frac{1}{N}&-\frac{N}{2}+\frac{1}{2}\,\,&\frac{N}{2}+\frac{1}{2}-\frac{1}{N}\\
-\frac{1}{N}&\frac{1}{2}&\frac{1}{2}+\frac{1}{N}\\
\frac{1}{N}&\frac{1}{2}&\frac{1}{2}-\frac{1}{N}
\end{array}
\right) \,,
\la{eq:SmatrixCrossing}
\eeq
where $C_{ab}$ is the crossing matrix. There is an alternative basis $ S_{ij}^{kl}(s) = \sigma_1(s)\, \delta_{ij}\delta_{kl} +  \sigma_2(s)\, \delta_{ik}\delta_{jl} + \sigma_3(s)\, \delta_{il}\delta_{jk}\, , $ in which crossing symmetry is more straightforward: $\sigma_1(s) = \sigma_3(4m^2-s)~  \text{and}~\sigma_2(s) = \sigma_2(4m^2-s). $ The map between the bases is
\be S_\bullet(s) = \sigma_2(s) + \sigma_3(s) + N \sigma_1(s)\,, \quad S_A(s) = \sigma_2(s) - \sigma_3(s)\,, \quad S_S(s) = \sigma_2(s) + \sigma_3(s)\,.  \ee

As for the analytic properties, we will focus on theories without bound states, so that each amplitude $S_a(s)$ is an analytic function of $s$ away from the multi-particle thresholds starting at $s=4m^2$ and $s=0$ (from crossing).

The space of amplitudes satisfying the above requirements was studied in \cite{Cordova:2019lot}. We review next some of the findings there. Using functionals of the form $\mathcal F=\sum_a n_a S_a(s_*)$
one can get slices of this infinite dimensional space of allowed amplitudes $S_a(s)$. Figure~\ref{fig:2Dmonolith} shows a two-dimensional slice of this space we call the monolith, constructed by putting bounds on the value of the amplitude at the crossing symmetric points $\sigma_{1,2}(s_*=2m^2)$, which one can see as effective quartic couplings. 
A simple feature of these bounds is that they are symmetric under the simultaneous change of sign for all channels $S_a\rightarrow -S_a$.\footnote{This goes back to the fact that we are considering theories without bound states poles, whose residue should be positive in unitary theories.}
As usual in these optimization problems, one has a unique solution when the bounds are saturated, so that one can read off the amplitudes at the boundary of the monolith.  

The picture for $N>2$ is the following, there are several special amplitudes sitting at special points of the boundary (the explicit expressions can be found in appendix~\ref{app:exactamps}):
\begin{itemize}
    \item The free boson theory where $S_a(s) = 1$ (+Free in figure \ref{fig:2Dmonolith});
    \item The free fermion theory where $S_a(s) = -1$ ($-$Free);
    \item The integrable O(N) non-linear sigma model (+NLSM);
    \item The `negative' NLSM S-matrix, in which one changes sign on all NLSM amplitudes (-NLSM);
    \item The periodic Yang-Baxter S-matrices ($\pm$pYB) which are periodic in rapidity $S_a^\text{pYB}(\theta)=S_a^\text{pYB}(\theta+\tau^\text{pYB})$;\footnote{The relation between the center of mass energy $s$ and the rapidity is $s=4m^2\cosh^2\(\theta/2\)$.}
    \item The constant S-matrices  ($\pm$const), inelastic in the symmetric channel given by $S_{\bullet,A,S}=\pm\,\(1,\,-1,\,-\frac{N-2}{N+2}\)$.
\end{itemize}

For the S-matrices of the last three points it is currently not known if a UV completion exists or if they correspond to any physical model. Nonetheless, except for the constant inelastic S-matrices, all of the listed S-matrices are integrable, i.e. they satisfy the Yang-Baxter relations.\footnote{For explicit solutions of Yang-Baxter's equations in theories with O(N) symmetry see \cite{Zamolodchikov:1978xm}.}

Generically, the S-matrices on the boundary (even if not at a kink) enjoy several common features. The first one is that they saturate unitarity $|S(s)|=1$, even if they are not integrable (except for the constant S-matrices on the last point). We expect however that these amplitudes are good approximations to the physical ones --at least at low energies-- in which some particle production is present (see e.g. discussion section in \cite{Cordova:2018uop}). The second one is that they exhibit a very rich structure of infinite resonances (seen as zeros in the physical sheet) arranging themselves in periodic fashion. That is, a generic point at the boundary in figure~\ref{fig:2Dmonolith} will obey  $S^\text{bdy}(\theta)=S^\text{bdy}(\theta+\tau)$ for a given period $\tau$ that changes along the boundary (see figure~8 in \cite{Cordova:2019lot}). 
The last point is quite puzzling since it gives us an observable which is periodic in a given parametrization of energy $\theta\sim\ln s$. In fact, one of the motivations for this work is to understand if such S-matrices can be compatible with a unitary UV completion.

\subsection{Adding currents: sum rules and central charge minimization}

In order to extract more information about the UV behaviour of the theories found with the S-matrix bootstrap, we will now include into the setup the existence of operators such as the stress tensor and O(N) currents. As we explain in the following, the inclusion of these operators will give us access to the central charges of the ultraviolet conformal theories. For clarity we discuss first the setup for theories without global symmetry and a generic operator $\mathcal O(x)$ and later generalize to the O(N) case.

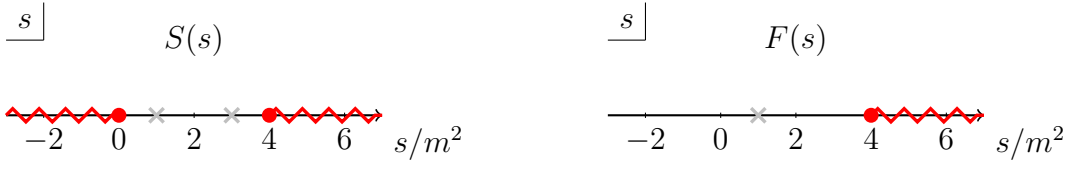
\begin{figure}[ht]
\centering
\begin{tikzpicture}
\draw[thick,->]  (-6.5,0) -- (-1.5, 0) node[anchor=north west] {$s/m^2$};
\draw[thick,->]  (1.5,0) -- (6.5, 0) node[anchor=north west] {$s/m^2$};
\foreach \x in {-2,0,2,4,6}
   \draw ( \x/2-5 ,1pt) -- (\x/2-5 ,-1pt) node[anchor=north] {$\x$};
\foreach \x in {-2,0,2,4,6}
   \draw ( \x/2+3 ,1pt) -- (\x/2+3 ,-1pt) node[anchor=north] {$\x$};
\draw[decorate,decoration=zigzag][line width=0.45mm, red ]  (-6.5,0) -- (-5, 0);
\draw[decorate,decoration=zigzag][line width=0.45mm, red ]  (-3,0) -- (-1.5, 0);
\draw[decorate,decoration=zigzag][line width=0.45mm, red ]  (5,0) -- (6.5, 0);
\node at (-5,0)[circle,fill,inner sep=2pt,red]{};
\node at (-3,0)[circle,fill,inner sep=2pt,red]{};
\node at (5,0)[circle,fill,inner sep=2pt,red]{};
\draw (-3.5,0) node[cross=4pt,lightgray,line width=0.45mm] {};
\draw (-4.5,0) node[cross=4pt,lightgray,line width=0.45mm] {};
\draw (3.5,0) node[cross=4pt,lightgray,line width=0.45mm] {};
\node at (-4,1) {$S(s)$};
\node at (4,1) {$F(s)$};
\draw (-6.5,1)--(-6,1);
\draw (-6,1)--(-6,1.5);
\node at (-6.25,1.25) {$s$};
\draw (1.5,1)--(2,1);
\draw (2,1)--(2,1.5);
\node at (1.75,1.25) {$s$};
\end{tikzpicture}
\caption{Analytic structure of the scattering amplitude $S(s)$ and two-particle form factor $F(s)$. The light gray crosses are possible simple poles due to bound states, set to zero in the rest of this paper, and red dots are branch points with the cuts along the real axis, coming from multi-particle thresholds.}
\label{fig:SFanalyt}
\end{figure}

As proposed in the pioneering work \cite{Karateev:2019ymz}, the main ingredient is the matrix of inner products of three different type of states: the first two are the two-particle in and out states (as in the previous section) and the third one is the state given by the operator $\mathcal O(x)$ acting on the vacuum
\begin{equation}
    \ket{\psi_1} = \omega\ket{p_1,p_2}^{in}, \qquad 
 \ket{\psi_2} = \omega\ket{p_1,p_2}^{out}, \qquad
 \ket{\psi_3} = \int d^2 x\, e^{ip\cdot x}\, \mathcal O(x)\ket{0}\,.
\end{equation}
 where $p=p_1+p_2$, $s=-p^2$ and $\omega=\left(2\sqrt s \sqrt{s-4}\right)^{-1/2}$.\footnote{The latter kinematic factor $\omega(s)$ is included in order to absorb the $\mathcal N_2(s)=\omega(s)^{-2}$ appearing in the normalization \eqref{eq:norminout}.} Unitarity translates to the positive semi-definiteness of the matrix of inner products $B$
 \begin{equation}
     \bra{\psi_i}\ket{\psi_j}=B_{ij} (2\pi)^2 \delta^2(p-p')\,, \qquad  B = \begin{pmatrix} 1 & S^*(s) & \omega F^*(s) \\ S & 1 & \omega F(s) \\ \omega F(s) & \omega F^*(s) & 2\pi \rho(s) \end{pmatrix}\succeq 0 \,,\quad s\geq4m^2
     \,.
 \end{equation}
The overlap between the \textit{in} and \textit{out} states gives the usual two-particle amplitude $S(s)$. If we take the inner product of two-particle  \textit{in} or \textit{out} state with the state created by the operator we get the two-particle form factor 
\begin{equation}
    F(s)= \langle 0 | \mathcal O(0) | p_1,p_2\rangle^{in}\,. 
\end{equation}
As a function of the complex variable $s$, the two particle form factor $F(s)$ is an analytic function except for the right hand cut at $s\geq 4m^2$ and possible bound state poles as depicted in figure~\ref{fig:SFanalyt}. 

Lastly the overlap $\langle\psi_3|\psi_3\rangle$ gives the Fourier transform of Wightman two point function of the operator, known as the spectral density of the operator $\mathcal O$
\begin{equation}\label{eq:spectraldef}
    2\pi\,\rho(s)=\int d^2x\, e^{-ip \cdot x} \bra{0} \mathcal O(x) \mathcal O(0)\ket{0}\,,
\end{equation}
which is a real positive function.

The generalization to our O(N) setup is straightforward. Now, because unitarity constrains each channel separately, we need to consider three different  matrices $B_{\bullet, A, S}$, one for each representation. The size of these matrices depend on which operator(s) we want to include. One of the operators we study is the trace of the stress tensor $\Theta(x)=T^\mu_\mu(x)$ which is a singlet of O(N) and hence appears in $B_\bullet$. The second operator we consider is the O(N) global current $J^\mu (x)$ which transforms in the adjoint of the group and therefore appears in the antisymmetric channel $B_A$. The explicit states we use are (for more details see \ref{app:conventions}):
\begin{gather}  
\ket{\psi_1}_a = \omega\ket{p_1,p_2}_a^{in}, \qquad 
 \ket{\psi_2}_a = \omega\ket{p_1,p_2}_a^{out}, \label{eq:statesON}\\
 \ket{\psi_3}_\bullet = \int d^2 x\, e^{ip\cdot x} \,\Theta(x)\ket{0}, \qquad \ket{\psi_3}_A = \int d^2 x\, e^{ip\cdot x} \,\frac{q_\mu J^\mu(x)}{q^2}\ket{0}, \qquad \ket{\psi_3}_S = 0, \nonumber
\end{gather}
where again $a$ labels the representation and $q \equiv p_1-p_2$. As before, unitarity becomes the positive semi-definiteness of the following matrices 

\begin{equation}
\label{eq:uniB}
\resizebox{\hsize}{!}{  $  
B_\bullet = \begin{pmatrix} 1 & S_\bullet^* & \omega F_\Theta^* \\ S_\bullet & 1 & \omega F_\Theta \\ \omega F_\Theta & \omega F_\Theta^* & 2\pi \rho_\Theta \end{pmatrix}\succeq0, \quad B_A = \begin{pmatrix} 1 & S_A^* & i\omega F_J^* \\ S_A & 1 & i\omega F_J \\ -i\omega F_J &-i \omega F_J^* & \frac{2\pi s}{s-4m^2} \rho_J \end{pmatrix}\succeq0, \quad B_\textbf{S} = \begin{pmatrix} 1 & S_S^*\\ S_S & 1  \end{pmatrix}\succeq0,
$}
\end{equation}
where all functions are evaluated at physical values of the energy $s\geq4m^2$. Details on the derivation of the different elements in the matrices are given in appendix~\ref{app:conventions}.

The stress tensor and O(N) currents are natural operators to consider since we are studying QFTs with global O(N) symmetry. Moreover, the fact that they are conserved currents gives rise to sum rules that  allow us to probe the central charges $c$ and $k$ of the conformal theory in the ultraviolet, defined as the coefficients appearing in the CFT two point functions\footnote{We have used the complex coordinates in Euclidean signature $z=x+i y$ and $T\equiv T_{zz}$, and $J\equiv J_z$. If the holomorphic $J$ and antiholomorphic $\bar J\equiv J_{\bar z}$ currents are separately conserved, the current central charge $k$ is the level of the associated affine Lie algebra.}
\be 
\label{ckdef}
\langle T(z) T(0)\rangle_\text{CFT}= \frac{c/2}{z^4}, \qquad \langle J(z)J(0) \rangle_\text{CFT}= \frac{k}{z^2}.
\ee

Conservation of the stress energy tensor gives Zamolodchikov's famous c-theorem \cite{Zamolodchikov:1986gt} which we can express as a sum rule integral for the spectral density of $\Theta(x)=T^\mu_\mu(x)$
\begin{equation}
   c = 12\pi \int_0^\infty ds\, \frac{\rho_\Theta(s)}{s^2}\,.
   \label{eq:csumrule}  
\end{equation}
In the left hand side of the equation above we have only the UV central charge $c = c_{UV}$  and not the difference between UV and IR since we are dealing with gapped theories for which $c_{IR}=0$. 

Similarly, there exists a $k$-theorem for the central charge of the global symmetry currents $J^{\mu}(x)$ \cite{Vilasis-Cardona:1994oke}. As reviewed in appendix~\ref{app:sumrules}, conservation of the currents implies the following sum rule
\begin{equation}
    k = \frac{\pi}{2} \int_0^\infty ds\, \rho_J(s)\,.
    \label{eq:ksumrule}
\end{equation}

The above sum rules allow us to extract the central charges $c$ and $k$ from the spectral densities in the QFT. Since these central charges characterize the conformal theory in the ultraviolet, they are a natural target for bootstrap bounds. Moreover, for a fixed finite number of flavors $N$, we expect that typical physical theories have finite central charges. With these considerations in mind, the bootstrap problem we will solve is the following: 
\begin{empheq}[box=\mymath]{equation}
   \textsf{Minimize $c$, $k$ subject to } B_a\succeq 0\,, \label{eq:optimization}
\end{empheq}
where the amplitude, form factor and spectral density entering $B_a$ have analyticity and crossing built in (see appendix~\ref{sec:numerics} for explicit ans\"atze). The bounds we find when performing the above minimizations are meaningful since the normalization of the operators we are considering is not arbitrary. Indeed, the normalization of the stress tensor and O(N) currents is fixed by the fact that there are conserved charges with canonical normalization associated to them. We show how to fix their normalization in appendix~\ref{app:norm}.

\paragraph{Numerics}
To solve the optimization problem \eqref{eq:optimization} we resort to numerical methods. There are two different complimentary approaches we use. One is the \textit{primal} method in which one explores the space of functions consistent with all our assumptions and was put forward in \cite{Paulos:2016but,Paulos:2017fhb}. In practice one writes ans\"atze for $S_a(s)$, $F_{\mathcal O}(s)$ and $\rho_{\mathcal O}(s)$ that are compatible with the required analyticity and crossing, and then one solves unitarity \eqref{eq:uniB} numerically using SDPB \cite{Simmons-Duffin:2015qma, Landry:2019qug}. We explain the numerical implementation in appendix~\ref{app:primal}. In the \textit{dual} approach we introduce dual variables which play the role of Lagrange multipliers for the constraints and solve an optimization problem for these \cite{Cordova:2019lot,Guerrieri:2020kcs,He:2021eqn,Guerrieri:2021tak}. In appendix~\ref{app:dualproblems} we show how to construct the dual optimization problems and in \ref{app:dual} how to implement them numerically. The dual method has the important advantage of giving rigorous bounds, whereas we have found easier to understand analytic properties of the optimal functions using the primal approach. 

\paragraph{Saturation of unitarity constraints}
Before we present our results in the next section, let us mention a generic feature of the optimal solutions. The inequalities coming from the unitarity constraints $B_a\succeq0$  in \eqref{eq:uniB} are usually saturated. The implication is that the spectral densities and central charges we find are given exactly by their two-particle contributions. To see this, recall that the positive semi-definiteness of $B_a$ is equivalent through Sylvester's criterion to the positivity of all principal minors. In particular the 2x2 upper left minor yields
\begin{equation}
    |S_a| \leq 1\,. \label{eq:unitarity_ineq_1} 
\end{equation}
 When above inequality is saturated we have the usual unitarity saturation for the two-particle amplitude $S_a$ found in pure S-matrix bootstrap calculations. The full 3x3 determinant gives 
\begin{equation}
\begin{gathered}
\label{eq:unitarity_ineq_3}
 \omega^2 {S_\bullet^*} {F_\Theta}^2 + \omega^2 {S_\bullet} {F_\Theta^*}^{2} -2 \omega^2 |F_\Theta|^2 + 2\pi \rho_\Theta(1-|S_\bullet|^2) \geq 0\,, \\
\omega^2 {S_A^*} {F_J}^2 + \omega^2 {S_A} {F_J^*}^{2} -2 \omega^2 |F_J|^2 + \frac{s}{s-4m^2}2\pi \rho_J(1-|S_A|^2) \geq 0\,,
\end{gathered}
\end{equation}
which reduces to Watson's equation $F(s)=S(s) F^*(s)$ \cite{Watson:1954uc} when unitarity is saturated. The latter equation allows one to construct the form factor from the S-matrix as explained in section \ref{sec:analytic_structure_S}. The positivity of the 2x2 lower right minor implies
\begin{equation}
    2 \pi \rho_\Theta \geq \omega^2 |F_\Theta|^2, \qquad \frac{s}{s-4m^2}2\pi \rho_J \geq \omega^2 |F_J|^2\,,\label{eq:unitarity_ineq_2}
\end{equation}
which follows from inserting a complete set of states in \eqref{eq:spectraldef}, so that the spectral density is given by the sum of all $n$-particle form factors. Schematically we have $\rho_{\mathcal O}``= "|F_{\mathcal O}|^2+\sum_{n>2}|F^{(n)}_{\mathcal O}|^2=\rho^{(2)}_{\mathcal O}+\sum_{n>2}\rho^{(n)}_{\mathcal O}$ where $F^{(n)}_{\mathcal O}$ is the $n$-particle form factor and the inequality \eqref{eq:unitarity_ineq_2} arises when truncating the right hand side to the two-particle form factors. Therefore when we are solving our optimization problem and \eqref{eq:unitarity_ineq_2} is saturated we are really finding the two particle contribution to the spectral density only $\rho^\text{opt}_{\mathcal O}=\rho^{(2)}_{\mathcal O}$.\footnote{Whereas the equation $|S^{(2)}(s)|=1$ for all physical energies can be consistent in integrable theories which have factorized scattering, the condition $\rho_{\mathcal O}(s)=\rho^{(2)}_{\mathcal O}(s)$ is only true in free theories. That is, the $n$-particle form factors are in general non trivial in interacting theories, even in integrable ones.} 
Similarly the central charges can be decomposed into their n-particle contributions $c=c^{(2)}+\sum_{n>2}c^{(n)}$. The bounds we find are again given by the two-particle contribution only $c^\text{opt}=c^{(2)}$, which is of course a valid bound since $c^{(n)}\geq0$. All this traces back to the fact that we are only considering two-particle in and out states in our setup. \\

The bounds found by solving the optimization problem \eqref{eq:optimization} shed light on the conformal theories that can flow to the gapped O(N) QFTs described in the previous section. It should be noted that the $c$-minimization problem was already considered in \cite{Karateev:2019ymz} for two lines in the monolith of figure~\ref{fig:2Dmonolith}. As shown in the next section, in this work we optimize both $c$ and $k$ for each point inside the monolith, which allow us to have a more fine-grained picture of the QFTs and associated UV fixed points. 

\section{Bounds on stress tensor and current central charges}\label{sec:results}
In this section we discuss the different minimization problems we solve. First, in section \ref{sec:minckmono}, we focus on the $N=7$ monolith shown on figure \ref{fig:2Dmonolith} and minimize the central charges $c$ and $k$ at each point. We find that for some of the points at the boundary the minimum central charges diverge. Then in \ref{sec:mincNkN} we find the global minima for $c$ and $k$ as a function of the number of flavors $N$. Finally in \ref{sec:cvsk} we do a mixed bootstrap in which we fix one of the central charges, $c$, and minimize the other one, $k$. 

\subsection{\texorpdfstring{Minimum $c$ and $k$ on the O(N) monolith}{Minimum c and k on the monolith}}\label{sec:minckmono}

\begin{figure}[ht!]
\centering
\includegraphics[width=.95\textwidth]{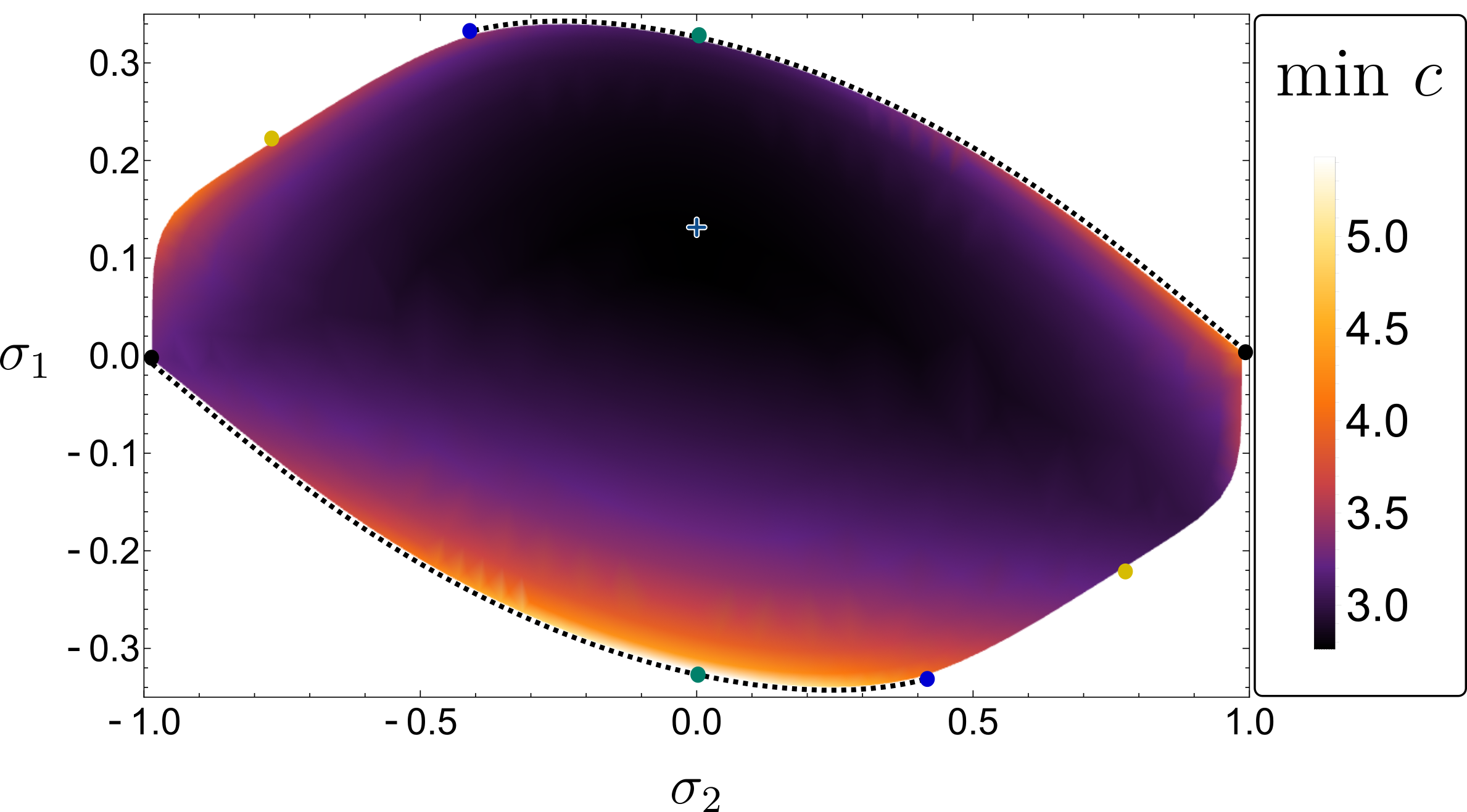}\vspace{0.2cm}
\caption{Temperature plot for minimum central charge $c$ on the $N=7$ monolith. The dashed lines at the boundary are the amplitudes for which $c$ diverges, they include the periodic Yang-Baxter solutions. The blue cross indicates the position of the global minimum. Results obtained from dual numerics with parameters $N_\text{max}=50$ and $N_\text{grid}=100$. 
}
\label{fig:ctemperature}
\end{figure}

\begin{figure}[th!]
\centering
\includegraphics[width=.95\textwidth]{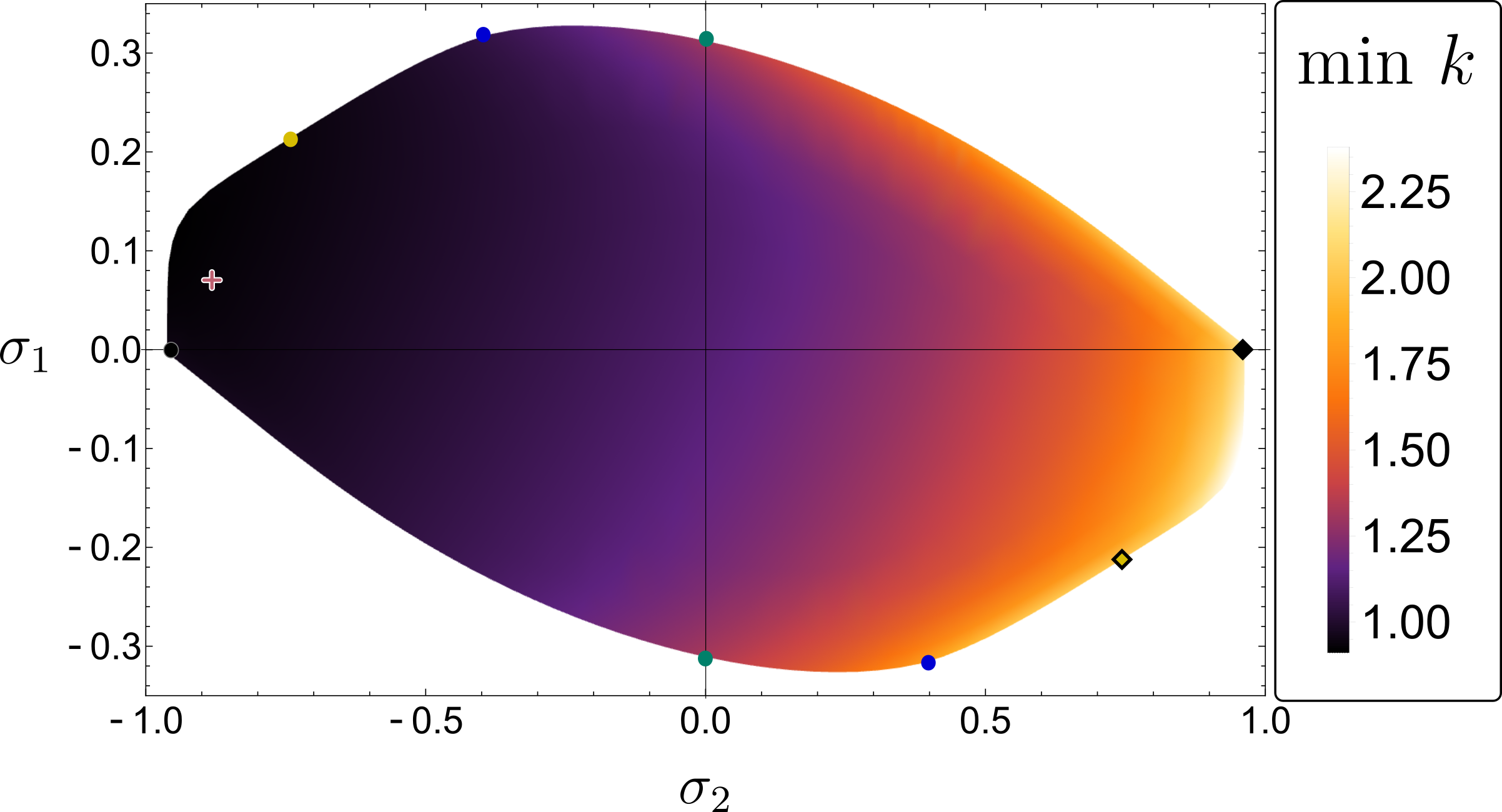}\vspace{0.2cm}
\caption{Temperature plot for minimum current central charge $k$ on the $N=7$ monolith. The two diamonds indicate the points where $k$ diverges (free boson and -constant solution). The pink cross marks the position of the global minimum. Results obtained from dual numerics with parameters $N_\text{max}=50$ and $N_\text{grid}=100$. 
}
\label{fig:ktemperature}
\end{figure}

Our first application is to study the lower bound on the central charges $c$ and $k$ for the space of gapped, O(N) symmetric QFTs with no bound states described in the previous section. When $N=7$, the numerical primal and dual minimization procedures yield the global minima:
\be
c_{\text{min}}^{primal} &= 2.916110...\,, \qquad c_{\text{min}}^{dual} = 2.756025...\,, \\ k_{\text{min}}^{primal} &= 0.971155...\,, \qquad k_{\text{min}}^{dual} = 0.904049...\,.
\ee
We then generalize this minimization procedure 
to every point covering the monolith of figure~\ref{fig:2Dmonolith}, parametrized by the value of the amplitudes at the crossing symmetric point $\sigma_1(2m^2)$ and $\sigma_2(2m^2)$. The results are shown on the temperature plots of figures \ref{fig:ctemperature} and \ref{fig:ktemperature}. In each plot we mention the values of $N_\text{max}$ and $N_\text{grid}$, giving respectively the size of the numerical ansatz and the number of points where we evaluate the unitarity constraints.

A notable feature of the obtained bounds for $c$ is the rapid growth of the minimum central charges near the boundary. Most importantly, we find no lower bound for the dashed regions on figure \ref{fig:ctemperature}, lying on the boundary of the monolith between NLSM and Free theories. This suggests that no unitary CFT with finite central charge $c$ can flow to a gapped phase described by those S-matrices, which in particular include pYB. As we argue in section~\ref{sec:analytic_structure_S}, amplitudes which give rise to infinite central charges exhibit a particular decrease of the phase shift at large energies.

As for $k$, the only places where there is no lower bound are free boson and one of the constant solutions ($-$const). The fact that both models give the same result is obvious from the equality of the amplitudes in the antisymmetric channel $S_A=1$. One can understand the divergence in $k$ near these points in different ways. One is with a direct computation from the free boson Lagrangian as in \ref{app:freeboson}. Another way is to recall that the UV fixed point we are dealing with describes $N$ free non-compact bosons $\phi$ which are not proper primary fields since their two-point functions are logarithmic. The O(N) current is $J\propto\phi\partial\phi$ so its two-point function has also logarithmic divergences. On the other hand, the finiteness of $c$ can be traced to the fact that the stress tensor is proportional to $T\propto \partial \phi \partial \phi$ and $\partial \phi$ is a well defined primary.

\begin{figure}[ht!]

\begin{subfigure}{0.49\textwidth}
\includegraphics[width=1\textwidth]{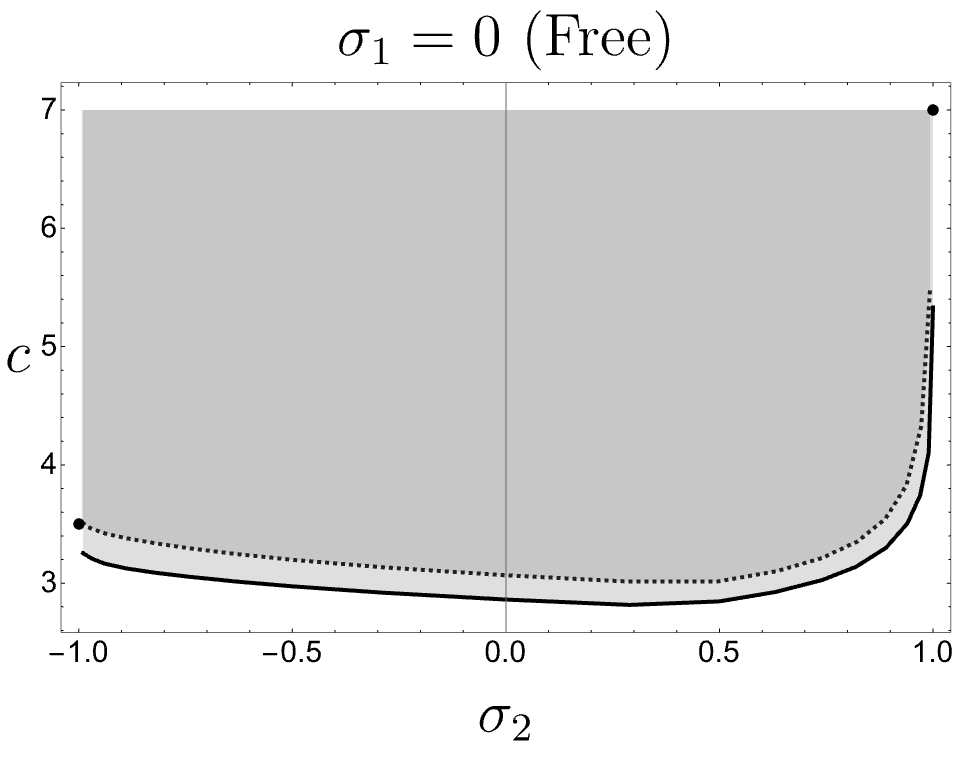}
\caption{}
\end{subfigure}
\begin{subfigure}{0.49\textwidth}
\includegraphics[width=1\textwidth]{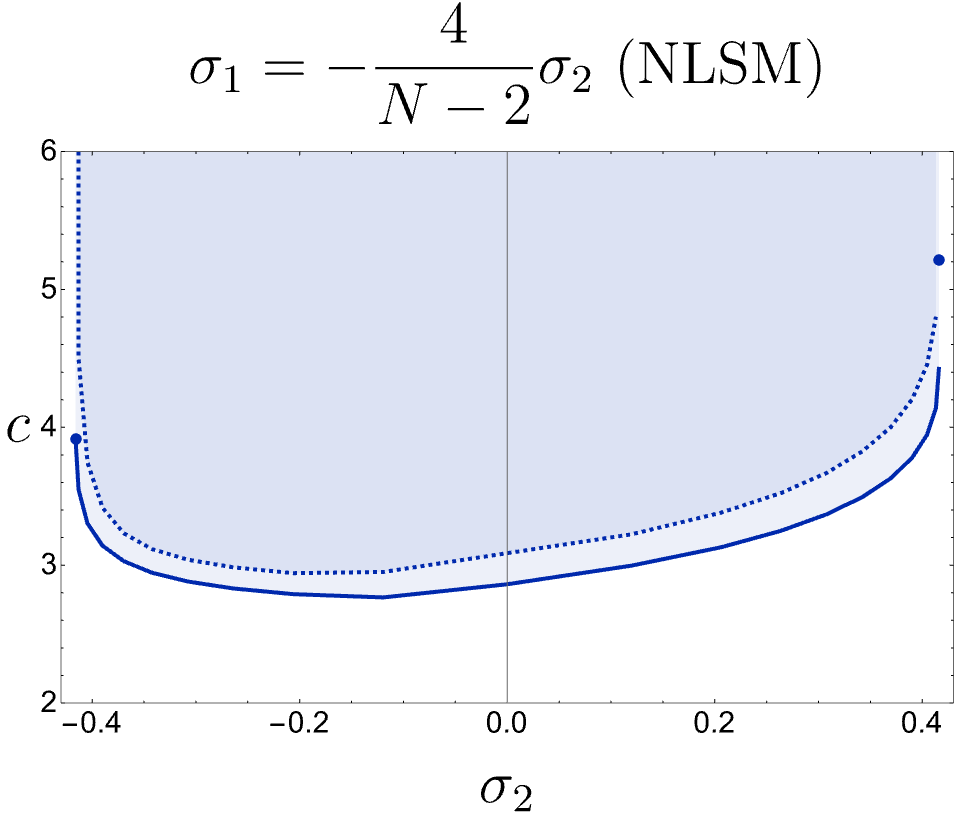}
\caption{}
\end{subfigure}\vspace{0.6cm}

\begin{subfigure}{0.49\textwidth}
\includegraphics[width=1\textwidth]{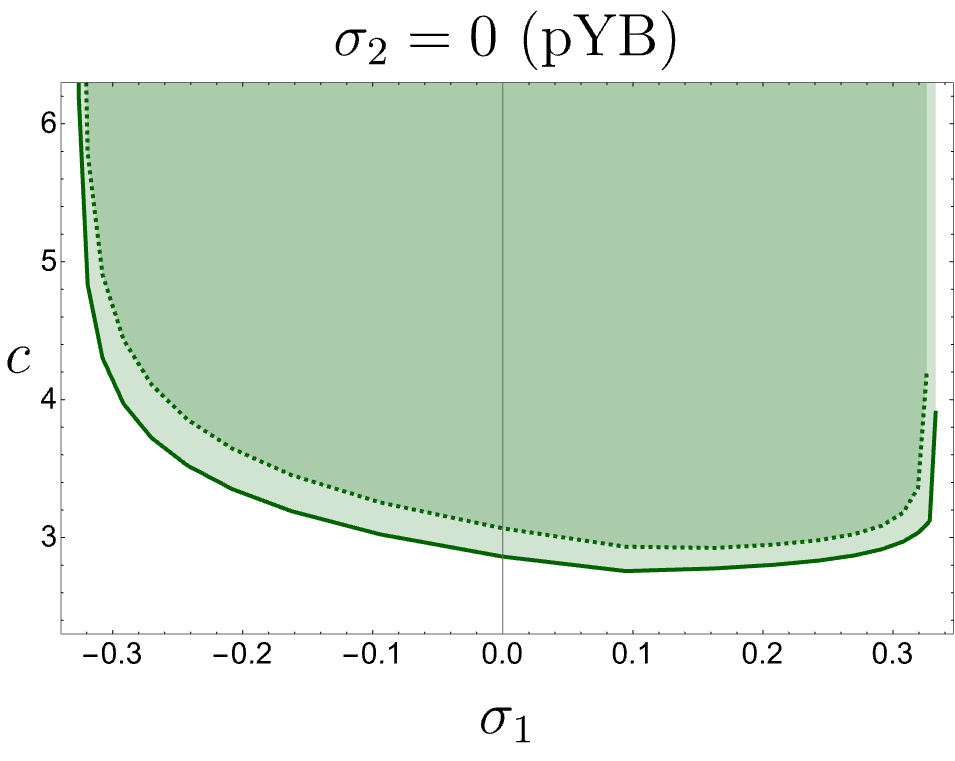}
\caption{}
\end{subfigure}
\vspace{0.1cm}
\begin{subfigure}{0.49\textwidth}
\includegraphics[width=1\textwidth]{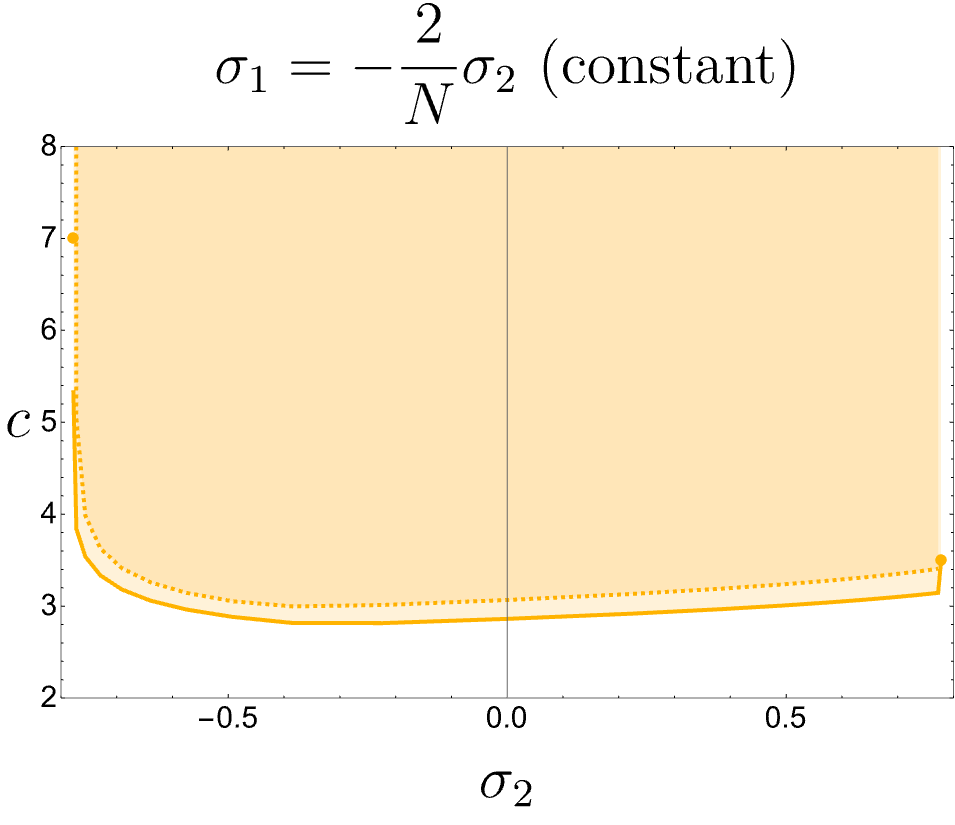}
\caption{}
\end{subfigure}
\caption{Minimum UV central charge $c$ for different sections of the monolith connecting known amplitudes at the boundary, namely: (a) $\pm$ Free, (b) $\pm$ NLSM, (c) $\pm$ pYB and (d) $\pm$ constant (c.f. figure~\ref{fig:2Dmonolith}). The solid (dashed) lines are the lower bounds obtained with the dual (primal) methods, both with $N_\text{max}=50$ and $N_\text{grid}=300$. The shaded regions are the allowed values for $c$. The endpoints are the two particle contribution to the central charge computed with the analytic form factor bootstrap from the known S-matrices as explained in \ref{app:exactamps}. 
}
\label{fig:csections}
\end{figure}

\begin{figure}[ht!]

\begin{subfigure}{0.49\textwidth}
\includegraphics[width=1\textwidth]{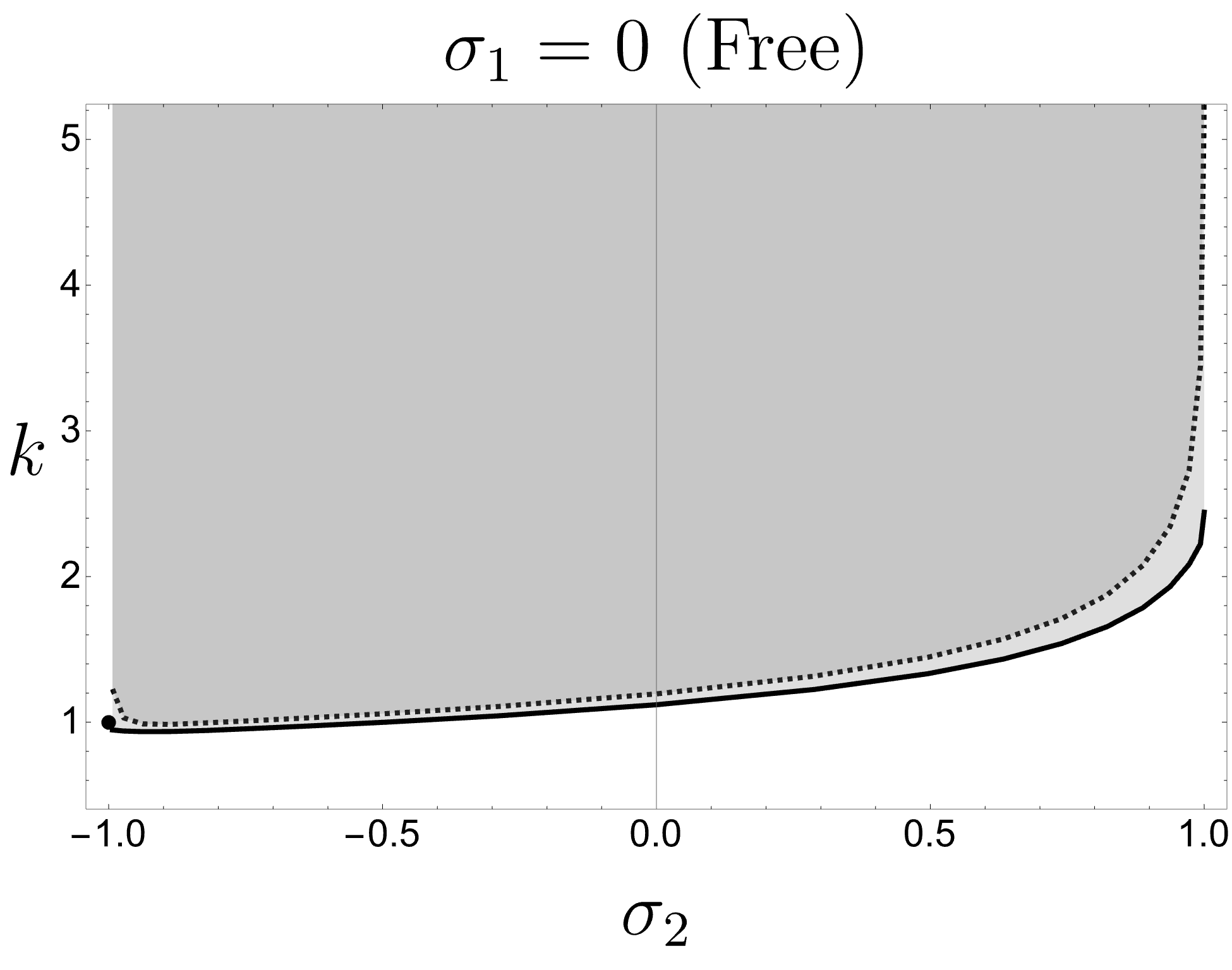}
\caption{}
\end{subfigure}
\begin{subfigure}{0.49\textwidth}
\includegraphics[width=1\textwidth]{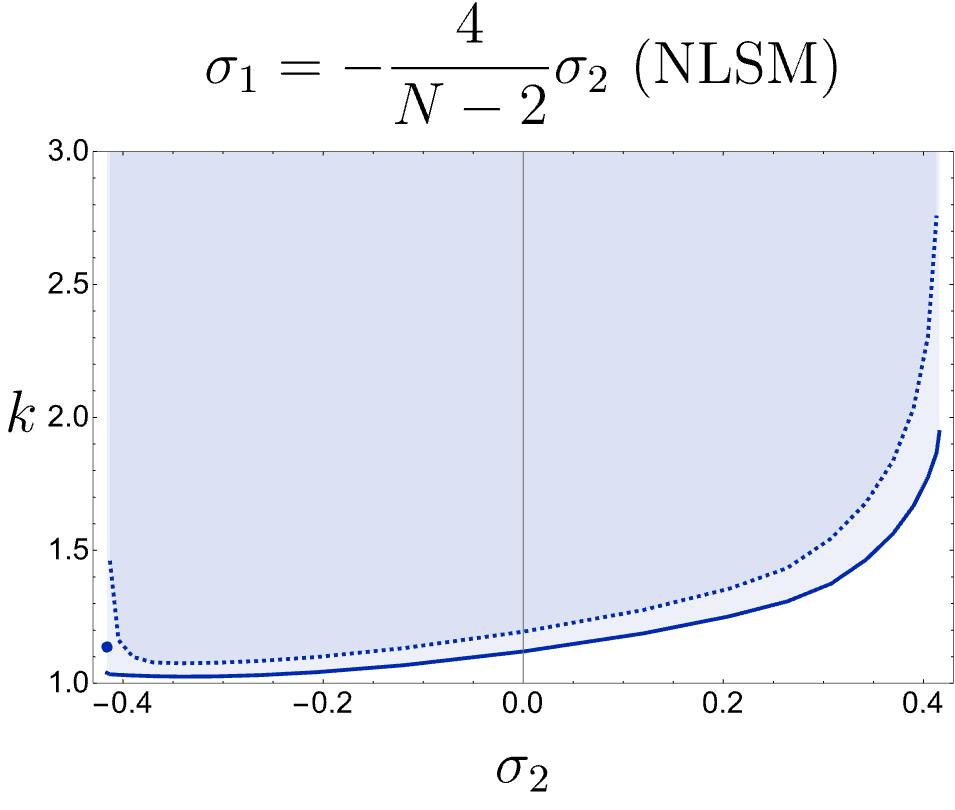}
\caption{}
\end{subfigure}\vspace{0.6cm}
\begin{subfigure}{0.49\textwidth}
\includegraphics[width=1\textwidth]{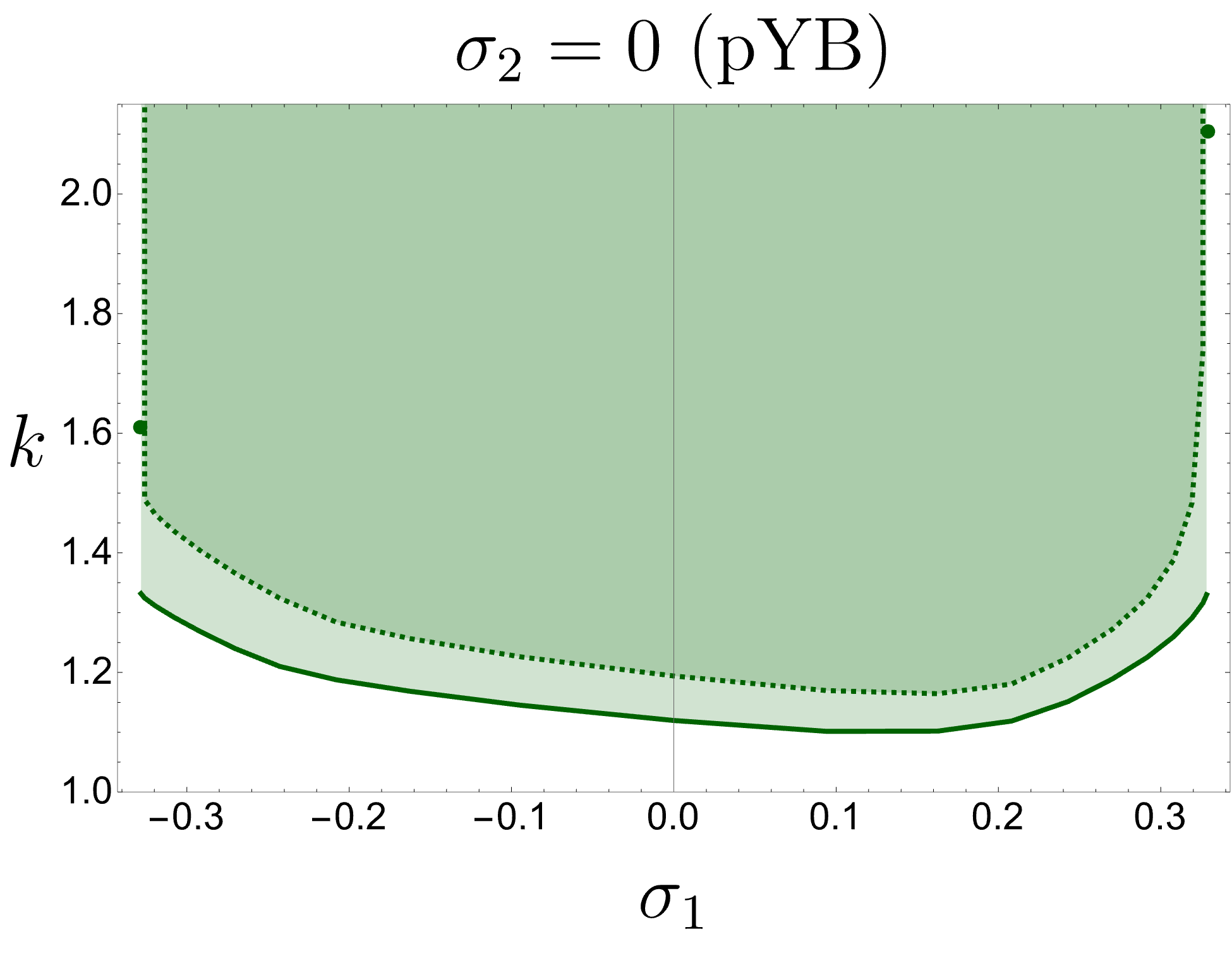}
\caption{}
\end{subfigure}
\vspace{0.1cm}
\begin{subfigure}{0.49\textwidth}
\includegraphics[width=1\textwidth]{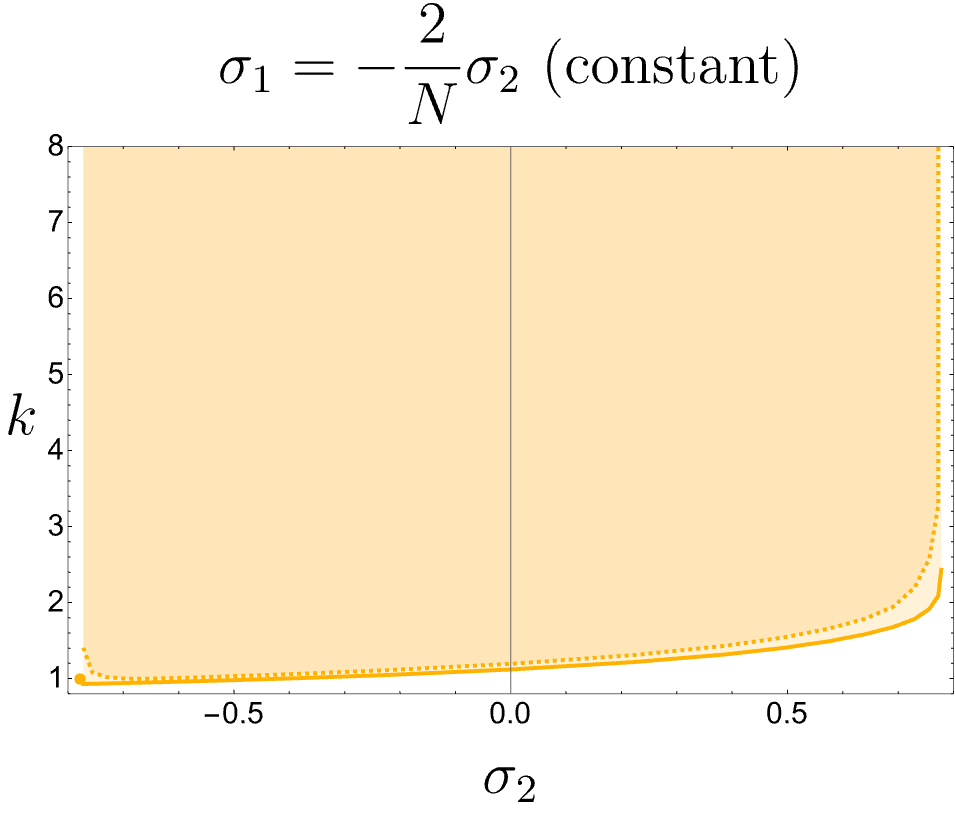}
\caption{}
\end{subfigure}
\caption{Minimum UV current central charge $k$ for different sections of the monolith: (a) $\pm$ Free, (b) $\pm$ NLSM, (c) $\pm$ pYB and (d) $\pm$ constant (see figure~\ref{fig:2Dmonolith}). The solid (dashed) lines are the lower bounds obtained with the dual (primal) methods, both with $N_\text{max}=50$ and $N_\text{grid}=300$. The shaded regions are the allowed values for $k$. The endpoints are the analytic values computed in \ref{app:exactamps}.
}
\label{fig:ksections}
\end{figure}

In figures~\ref{fig:csections} and \ref{fig:ksections} we show various sections of the temperature plots connecting special points on the boundary of the monolith, namely: Free, NLSM, pYB and constant solutions. They are the same sections depicted by dashed lines in figure~\ref{fig:2Dmonolith}, with the same color coding. 

In the plots we show the bounds obtained by primal (dashed lines) and dual numerics (solid). By construction the dual (rigorous) bounds are below the primal ones, so that the optimal one lies between the two.\footnote{As we increase the numerical precision by considering larger ans\"atze (parametrized by $N_\text{max}$) the gap between dual and primal bounds closes, as exemplified later in figure~\ref{fig:cvsk}.}
For most sections we find that the convergence of the bounds close to the boundary is challenging. This is particularly evident for the primal bounds, which might even stop converging before reaching the actual boundary.\footnote{This is what happens in the constant section of figure~\ref{fig:csections} (d), where the dual endpoint seems to be above the primal result, but the primal bound is actually infinity at that point.} In many cases one can trace back this difficulty to the rapid growth of the minimum central charges as one gets close to the boundary from a radial direction. 
However, a clear understanding of which parameters in the numerical ans\"azte are most important for good numerical convergence eludes us at this point. In this sense we see the clear advantage of using both primal and dual numerics to bracket the optimal bound. 

The endpoints of these sections can be computed from the analytic S-matrices as explained in appendix~\ref{app:exactamps} (see table~\ref{table:c2k2values} for explicit values). Note for instance that the endpoints of the free and constant sections are related since $S_\bullet^{\pm\text{Free}} = S_\bullet^{\pm\text{const}}$ and $S_A^{\pm\text{Free}} = -S_A^{\pm\text{const}}$ (see appendix~\ref{app:exactamps0} for all the endpoint amplitudes).

\subsection{\texorpdfstring{Minimum $c$ and $k$ as a function of $N$}{Minimum c and k as a function of N}}\label{sec:mincNkN}

\begin{figure}[ht!]
\begin{subfigure}{0.49\textwidth}
\includegraphics[width=1\textwidth]{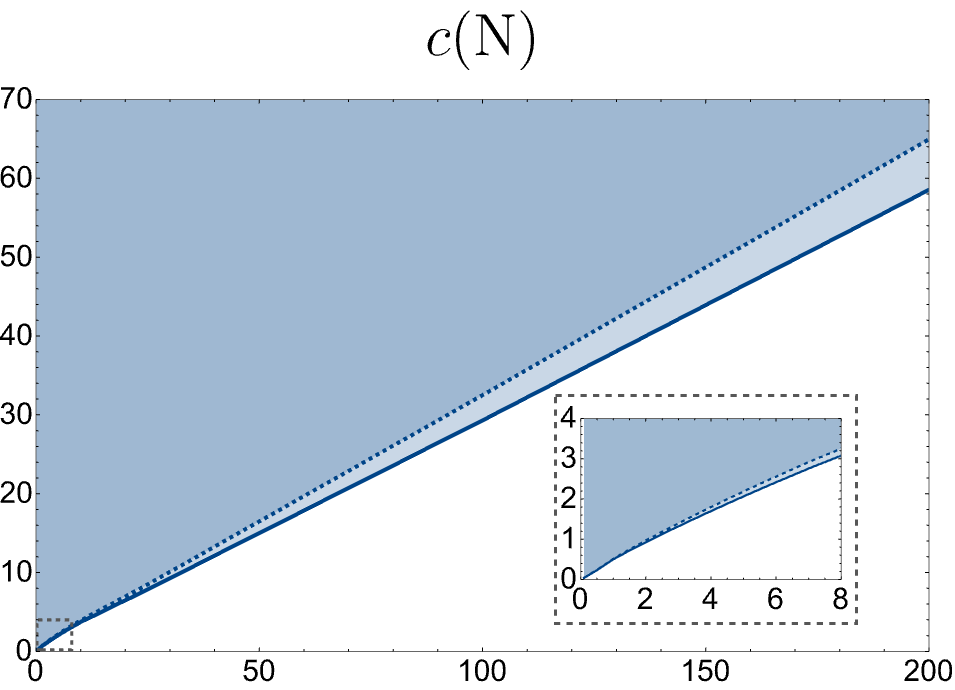}
\caption{}
\end{subfigure}\hspace{0.3cm}
\begin{subfigure}{0.49\textwidth}
\includegraphics[width=1\textwidth]{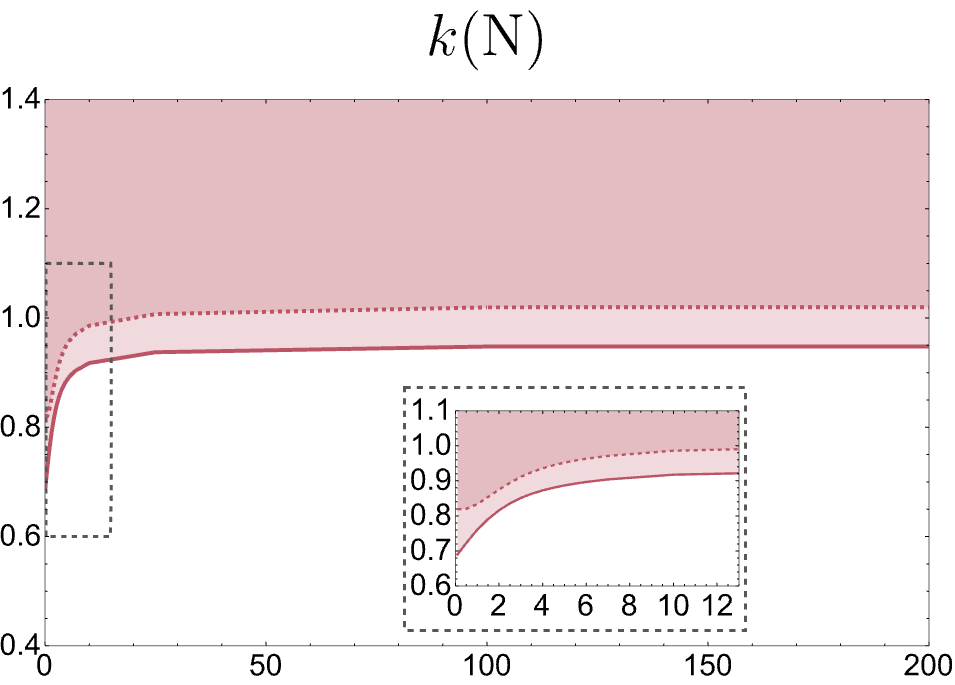}
\caption{}
\end{subfigure}
\caption{Global minima for central charges $c$ and $k$ as a function of the number of flavors $N$. The dual (primal) bounds are given by the solid (dashed) lines. The allowed values for the central charges appear in the shaded regions. In (a) we see the linear dependence with $N$ as in \eqref{eq:asymptcN}. In (b) the current central charge $k$ remains constant at large $N$ and close to the free fermion value $k=1$. Both primal and dual numerics were done with $N_\text{max}=50$ and $N_\text{grid}=300$.
}
\label{fig:ckN}
\end{figure}

 Now we consider the problem of finding the global minimal central charges $c$ and $k$ as we vary the number of flavors $N$ (i.e. without fixing $S_a(s=2m^2)$). The results of our optimization procedures are shown on figure \ref{fig:ckN}. At large $N$, the bound for the central charge $c$ becomes linear.\footnote{It is expected in general that the central charge $c$ scales with $N$ since we have states in an $N$-dimensional representation, with each component contributing to the stress tensor. In $d=2$ it can also be argued from Cardy's formula \cite{Cardy:1986ie}.} The slope is 

\be 
c_{\text{min}}^{primal}(N) \sim 0.3221 N\,, \qquad  c_{\text{min}}^{dual}(N) \sim 0.2925 N\,, \label{eq:asymptcN}
\ee 
which is compatible from the previous primal results of \cite{Karateev:2019ymz}. Instead for $k$ the bounds are saturated by a constant 
\be 
k_{\text{min}}^{primal}(N) \sim 1.0250\,, \qquad  k_{\text{min}}^{dual}(N) \sim 0.9513\,. \label{eq:asymptkN}
\ee

The optimal bound for $k$ at large $N$ is compatible with 1, which is the value taken by the free fermion. The rigorous dual bound sits slightly below 1, but since the numerics only consider 2 particles contributions, we expect that this bound will increase when including $n>3$ contributions (see the discussion below \eqref{eq:unitarity_ineq_2}). Therefore we believe that the true lower bound should be saturated by the free fermion.

\subsection{\texorpdfstring{Bounds on the $(c,k)$ plane}{Bounds on the (c,k) plane} }\label{sec:cvsk}
Our last application is to consider the lower bound on the current central charge $k$ for fixed central charge $c$. Analyzing both the stress tensor and O(N) currents at the same time gives us a more detailed idea of which CFTs can flow to the gapped theories described by the monolith. 

\begin{figure}[ht!]
    \centering
    \begin{subfigure}{0.88\textwidth}
    \includegraphics[width=\textwidth]{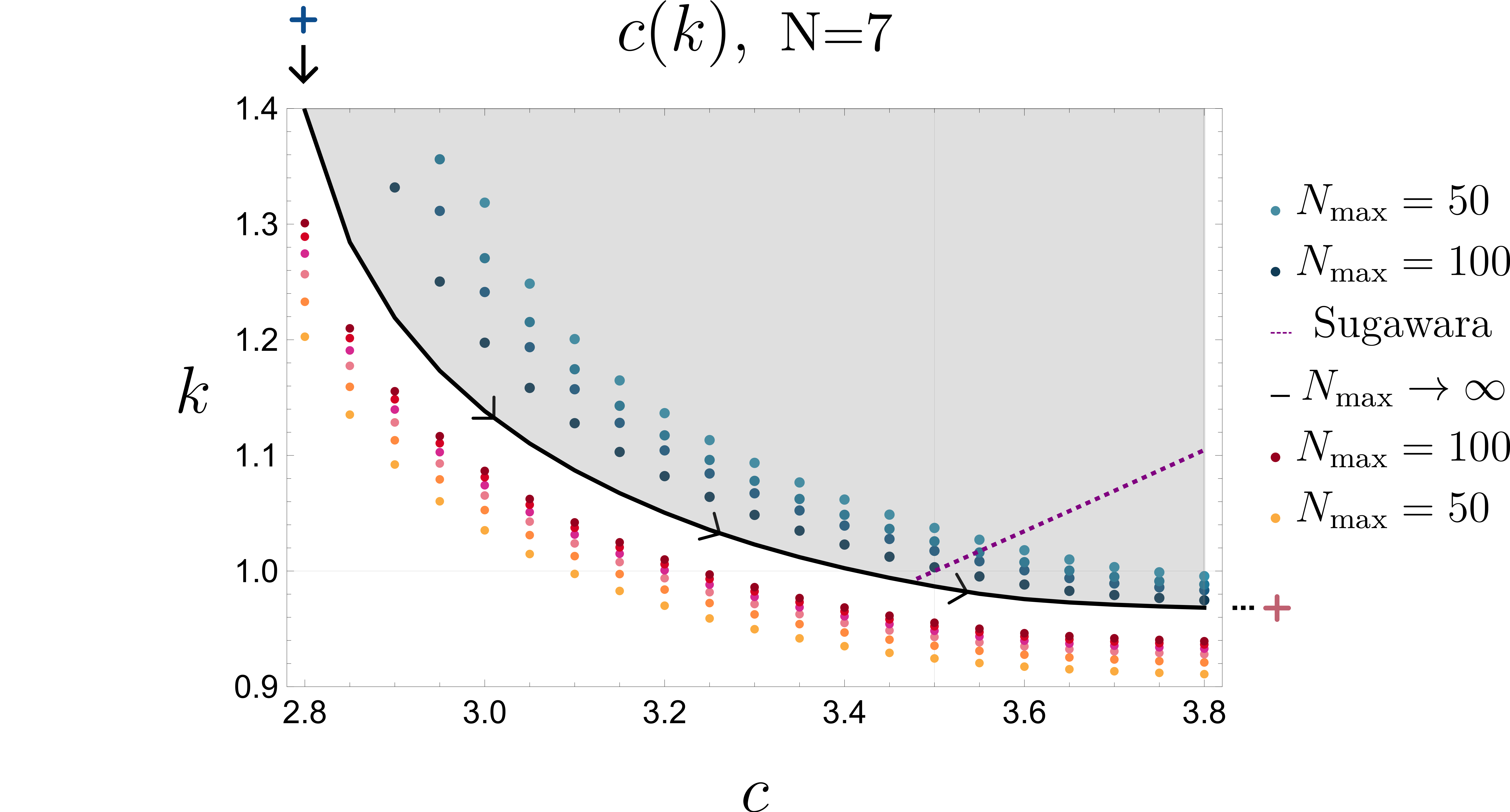}
    \caption{}
    \end{subfigure}
    \begin{subfigure}{0.77\textwidth}
    \vspace{0.3cm}
    \includegraphics[width=\textwidth]{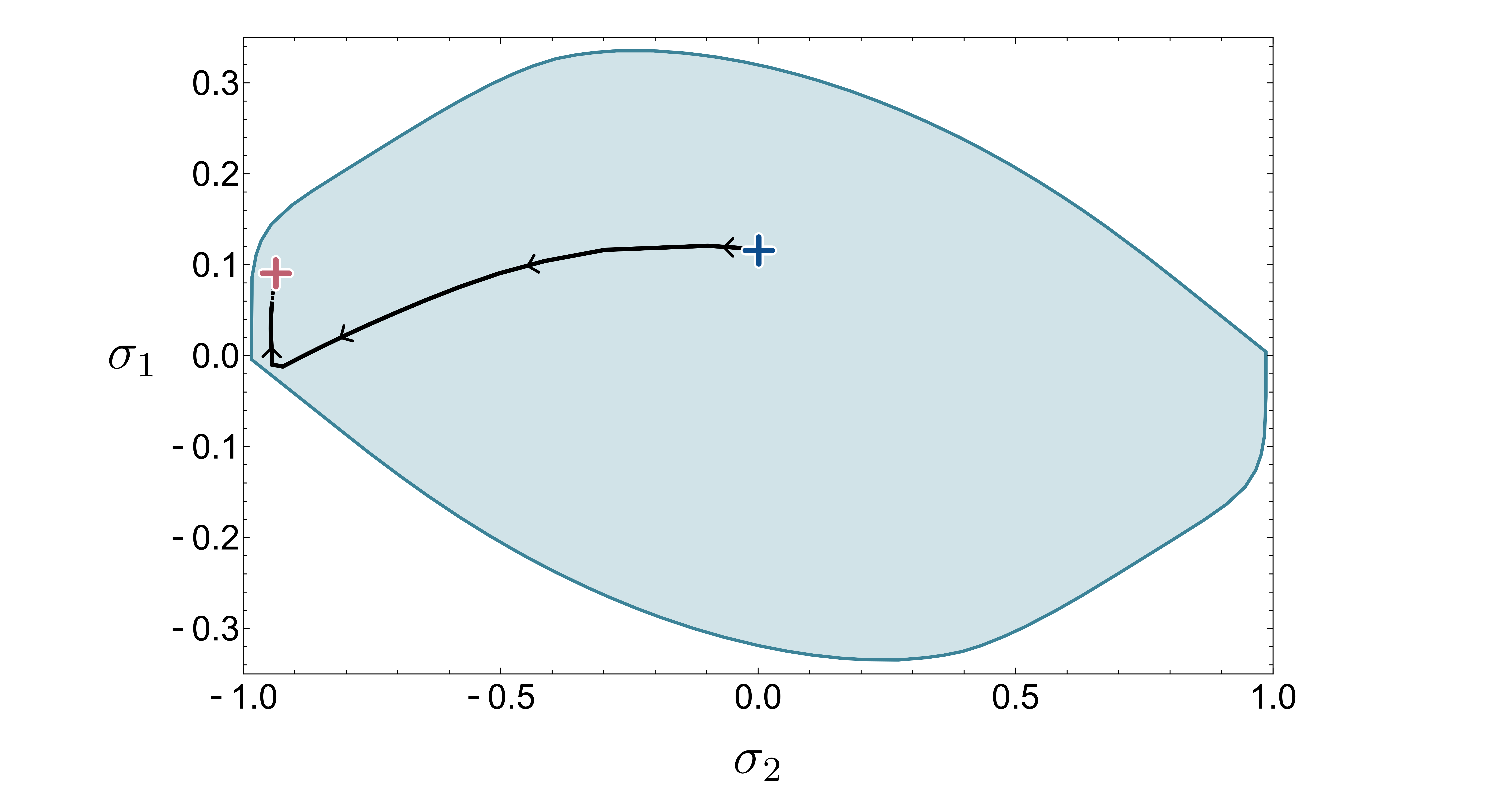}
    \caption{}
    \end{subfigure}
    \caption{(a) Allowed values for $c$ and $k$ central charges for $N=7$. The  red (blue) points are obtained using dual (primal) numerics for different $N_\text{max}$. The black line is a simple linear extrapolation in $1/N_\text{max}$ and the purple dashed line are the values obtained with the Sugawara construction. The curve starts at the global $c$ minimum (blue cross outside plot) and ends at the global $k$ minimum (pink cross). We expect that compact unitary CFTs lie to the right of this Sugawara line. (b) The path followed by the curve in (a) inside the monolith.}
    \label{fig:cvsk}
\end{figure}

In figure \ref{fig:cvsk} we show these bounds for $N=7$. In \ref{fig:cvsk}(a) we see the numerical primal and dual bounds for different size of the ans\"atze $N_\text{max}$ along with a $N_\text{max}\rightarrow\infty$ extrapolation.  In \ref{fig:cvsk}(b) we show the path followed by the lower bound curve inside the monolith. 

Let us discuss known unitary conformal theories which might live in the allowed region in figure~\ref{fig:cvsk}. The purple dashed line is given by the following equation
\begin{equation}
c_\text{Suga}(k)
=\frac{k\, N(N-1)/2}{k+N-2}\,.
\la{eq:cSugaON}
\end{equation}
This relation between $c$ and $k$ follows from the Sugawara construction which we briefly review next. Recall that the conservation equation for the current of a global symmetry group $g$ reads $\partial_z \bar J+\partial_{\bar z} J=0$. If the conformal theory is compact (i.e. with discrete spectrum) it is expected that the holomorphic and antiholomorphic currents are separately conserved $\partial_z \bar J=\partial_{\bar z} J=0$. In this case the OPE of the currents defines an affine Kac-Moody algebra whose level is given by $k$
\begin{equation}
J^a(z) J^b(w)\sim \frac{k \delta^{ab}}{(z-w)^2}+\frac{if_{abc}}{z-w}J^c(w)\,,
\end{equation}
with the group indices $a$ taking values $a=1,\ldots,\text{dim}g$ and $f_{abc}$ are the structure constants of the algebra. Importantly, one can construct the stress-energy tensor of the theory from the normal ordered product of these currents
\beq
T(z)=\frac{1}{k+h^\vee} \sum_a:J^a(z)J^a(z):\,,
\eeq
where the prefactor is fixed by symmetry and $h^\vee$ is the dual Coxeter number of the group. The above construction is known as Sugawara construction and it gives a relation between the central charge $c$ and the level $k$ of the algebra
\beq
c_\text{Suga}(\hat g_k)=\frac{k\, \text{dim}\,g}{k+h^\vee}\,, \label{eq:Sugawara}
\eeq
which reduces to \eqref{eq:cSugaON} for $g=so(N)$. Notably this affine algebra structure can be realized by the Wess-Zumino-Witten (WZW) models, whose Lagrangian is defined by a nonlinear sigma model and a topological term. These models have also free field representations in terms of free fermions or free bosons and ghosts. For instance, for $so(N)$ 
at level one we have $N$ real free fermions giving $c=N/2$. This is indeed the point $(c,k)=(7/2,1)$ being approached from different directions by our primal and dual bounds in figure~\ref{fig:cvsk}.\footnote{Of course we see as well the amplitude, form factor and spectral density approaching those of the free fermion.}

So far we have looked at possible models lying exactly at the Sugawara line in figure~\ref{fig:cvsk}. Starting from this line one can populate the right hand side of the plot by taking the direct sum of WZW models and CFTs whose content is given by O(N) singlets. In this way one increases the value of the central charge $c$ without modifying $k$. Note that a priori there is no reason to assume our gapped theory should arise from a compact CFT. As such, the allowed region of figure~\ref{fig:cvsk} could include non compact CFTs, in particular the region to the left of Sugawara line. In fact, the simplest amplitude we can write $S_a=1$ for free boson leads to a CFT lying in this left region for $N>3$ since $(c,k)=(N,\infty)$. 

Finally there is also a small region below the Sugawara line with $k<1$. As explained before, we believe that once we include multi-particle contributions into the bootstrap this region will no longer be allowed, so that the true minimum for $k$ coincides with the free fermion value.\footnote{We expect this to be true for $N\geq3$ since for $N=2$ the minimum $k$ sits at the boundary of the monolith and corresponds to one of the $\gamma\rightarrow\infty$ sine-Gordon amplitudes.}

\section{\texorpdfstring{Infinite central charge S-matrices}{Infinite central charge S-matrices}}
\label{sec:analytic_structure_S}
Given that some regions on the monolith have infinite central charges, a natural question to ask is what characterizes the amplitudes which lead to divergent sum rules? In this section we use analytic bootstrap methods to derive conditions on the $2\to 2$ amplitudes giving rise to infinite central charges. 

\begin{figure}[ht!]
\begin{subfigure}{0.49\textwidth}
\includegraphics[width=1\textwidth]{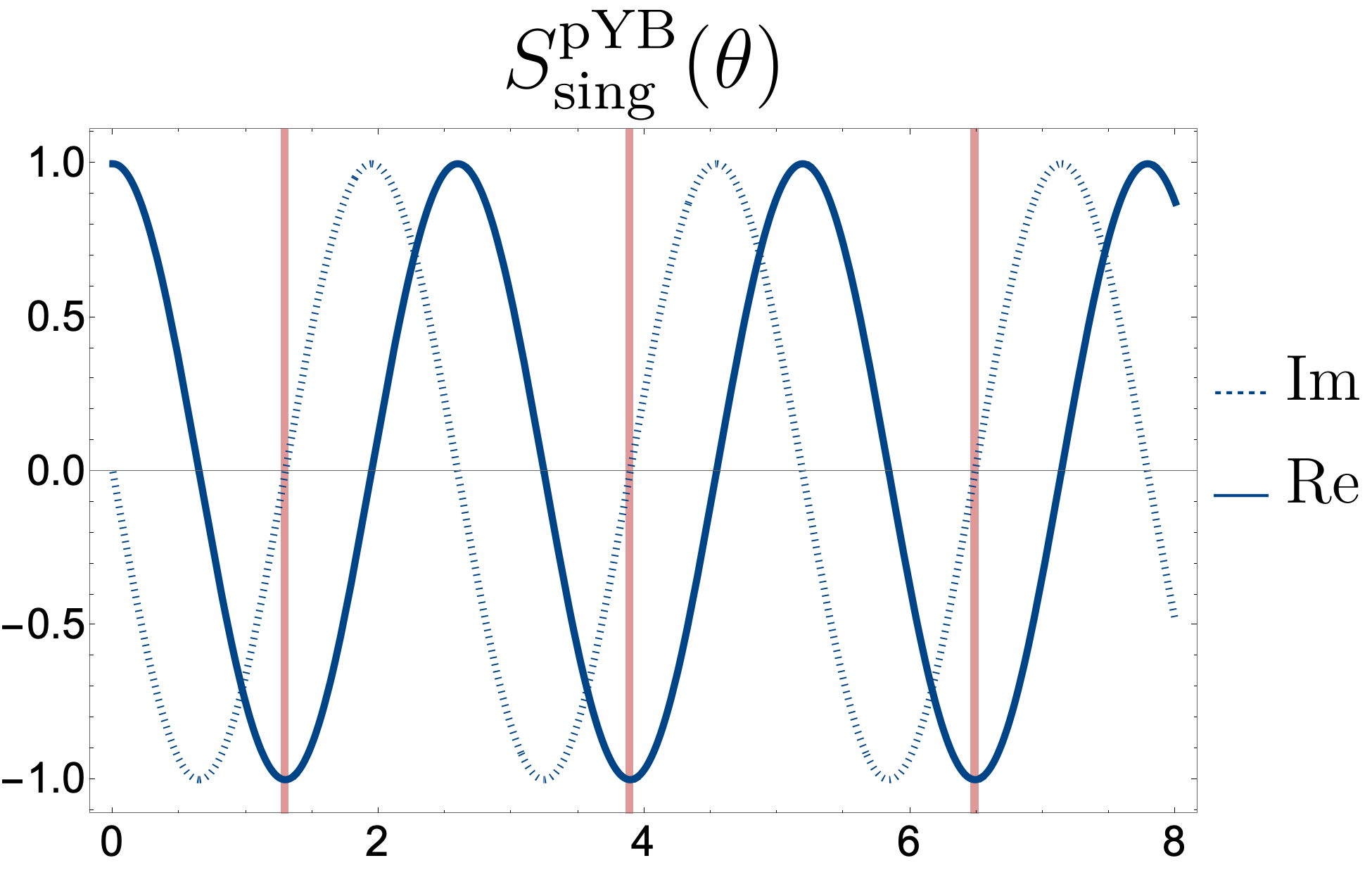}\vspace{0.2cm}
\caption{}
\end{subfigure}
\begin{subfigure}{0.49\textwidth}
\includegraphics[width=1\textwidth]{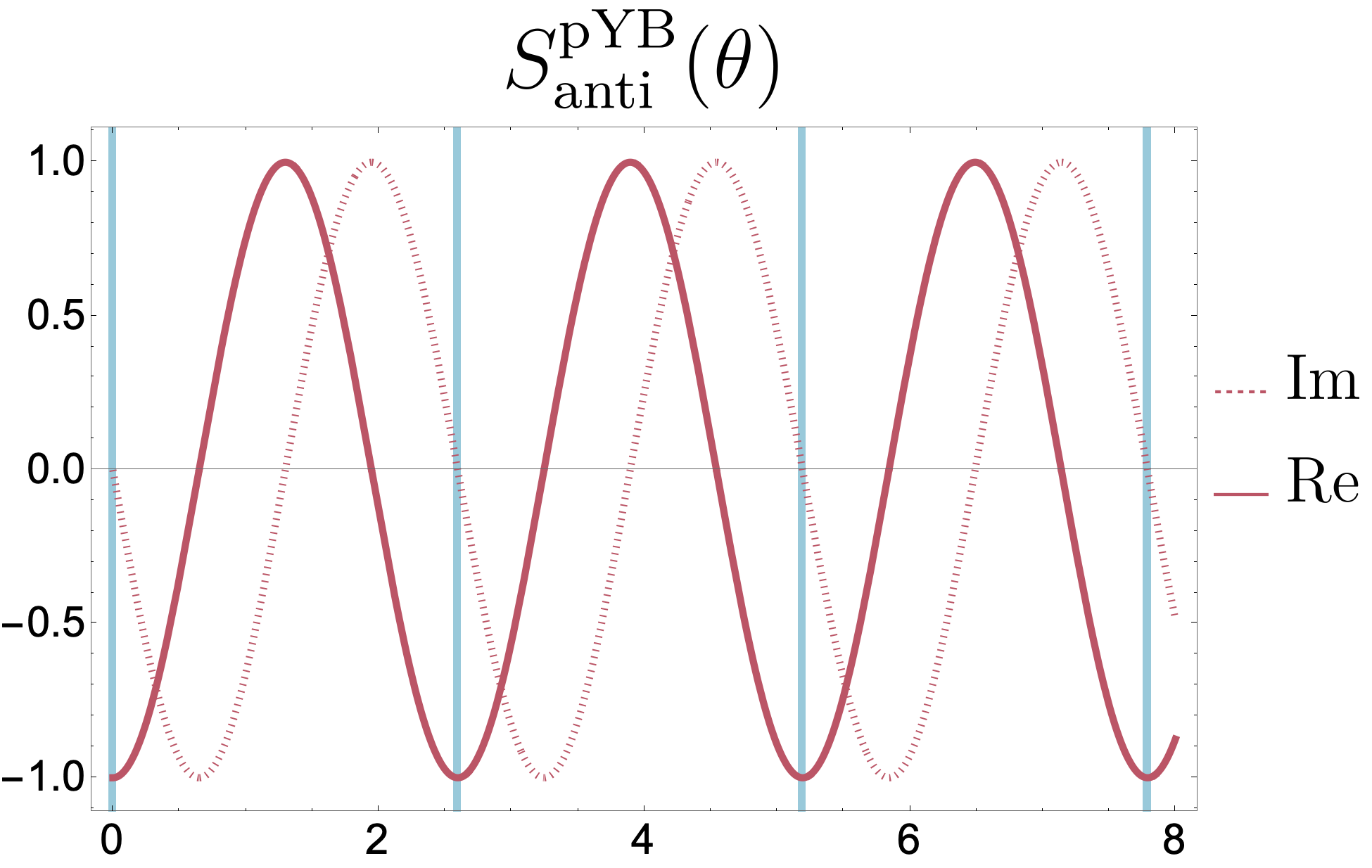}\vspace{0.2cm}
\caption{}
\end{subfigure}
\caption{Periodic Yang-Baxter amplitude in the (a) singlet and (b) antisymmetric channels for physical kinematics in the rapidity variable $\theta$ (related to $s$ as $s=4m^2\cosh(\theta/2)$). We have set $N=2000$ which gives a period in $\theta$ of $\tau^\text{pYB}\approx2.6$. In (a) we see the singlet channel having infinite dangerous points highlighted by the vertical red lines leading to $c\rightarrow\infty$. Instead, for the antisymmetric amplitude in (b) we have only safe points marked with blue vertical lines and $k$ finite.
}
\label{fig:SpYB}
\end{figure}

Let us define the \textit{dangerous} and \textit{safe} points as the positions $s_0$ in the physical region $s\geq 4m^2$ where the amplitude obeys
\begin{eqnarray}
    \textit{dangerous:}& s_0 \text{ such that } S(s_0)=-1 \text{ and } \text{Im } S'(s_0)>0 \,,\\ 
    \textit{safe:}& s_0 \text{ such that } S(s_0)=-1 \text{ and } \text{Im } S'(s_0)<0 \,. \label{eq:s0dangsafe}
\end{eqnarray}
Denoting by $\textbf{d}_a$ ($\textbf{s}_a$) the number of \textit{dangerous} (\textit{safe}) points in the representation $a$, the conditions leading to infinite $c$ or $k$ central charges read
\begin{empheq}[box=\mymath]{align}
&\textbf{d}_\bullet\,-\textbf{s}_\bullet\,>1 \implies c\rightarrow\infty\,,
\label{eq:cconditionsINFTY}
\\
&\textbf{d}_A-\textbf{s}_A>0 \implies k\rightarrow\infty \,.\label{eq:kconditionsINFTY}
\end{empheq}
These conditions offer a very simple explanation for why the dashed regions at boundary of the monolith in figure~\ref{fig:ctemperature} have infinite $c$ but finite $k$. As illustrated in figure~\ref{fig:SpYB} for periodic Yang-Baxter, the singlet channel has infinite dangerous points whereas the antisymmetric amplitude has infinite safe points.

In the rest of this section we first review in \ref{sec:analyticFFBootstrap} how to construct the form factors and spectral densities starting from a known S-matrix, and then in \ref{sec:proof_lemma} we present an argument for the conditions in \eqref{eq:cconditionsINFTY} and \eqref{eq:kconditionsINFTY}. In \ref{sec:numerics_safe_and_dangerous} we show how dual and primal numerics behave close to a point on the monolith with infinite central charge and how these extremal S-matrices with large but finite central charges smartly evade these criteria.

\subsection{Review of analytic Form Factor Bootstrap}
\label{sec:analyticFFBootstrap}

There is a well-known procedure to compute n-particles form factors in integrable theories known as \textit{Form Factor Bootstrap}  (see for instance \cite{Smirnov:1992vz} or \cite{Omnes:1958hv}). 
Here we will restrict to the two-particle form factor and its contribution to the spectral density and follow this procedure also for non-integrable amplitudes.  The main ingredient is Watson's equation $F(s)=S(s)F^*(s)$ which gives a direct relation between the two particle form factor and amplitude for energies where there is no particle production (for pedagogical purposes we ignore the representation indices).  As reviewed earlier, the optimization problem we are considering leads to functions saturating the inequality constraints, so that we have unitarity saturation and Watson's equation for all energies. Therefore, even for non integrable theories, we can perform the optimization \eqref{eq:optimization} analytically if we know the amplitude.

We start by writing the two particle form factor as
\begin{equation}
    F(s)=B(s)\, e^{\alpha(s)-\alpha(0)}\,, \label{eq:FFBalpha}
\end{equation}
where $B(s)$ is a real analytic function and $\alpha(s)$ takes care of the discontinuity at $s\in[4m^2,\infty)$. We have chosen the difference $\alpha(s)-\alpha(0)$ in the exponent so that the normalization of the form factor as in  \eqref{eq:normFA} and \eqref{eq:normF0} is fixed by $B(0)$. Given a unitarity saturating amplitude, we can write it as a pure phase $S(s)=e^{i \varphi(s)}$ for physical energies $s\in[4m^2,\infty)$. Using Watson's equation, we see that the imaginary part of $\alpha(s)$ is proportional to the argument of $S(s)$:
\begin{equation}
    F(s)=S(s)F^*(s) \implies \text{Im}\, \alpha(s)=\varphi(s)/2=  - i \log S(s)/2\,, \qquad s\in[4m^2,\infty)\,.
    \label{eq:alphi}
\end{equation}
Now $\alpha(s)$ should not grow as $s \to \infty$ otherwise $F(s)$ in equation \eqref{eq:FFBalpha} would have an essential singularity. We therefore assume $\alpha(s)$ to obey a once-subtracted dispersion relation, 
\begin{equation}
    \alpha(s)-\alpha(0)=\frac{s}{\pi} \int\limits_{4m^2}^\infty ds'\, \frac{\text{Im}\,\alpha(s')}{s'(s'-s)}\,. \label{eq:alphadisp}
\end{equation}
which via equation \eqref{eq:alphi} `solves' the form factor in terms of the amplitude up to the freedom in the function $B(s)$ which drops out of Watson's equation \eqref{eq:alphi}.\footnote{This is a similar trick to the one often used in integrable models, where one finds an integral representation in the rapidity $\theta$ for the S-matrix first $ S(\theta)=\int\limits_0^\infty \frac{dt}{t} f(t) \sinh\(\frac{t \theta}{i\pi}\)$ and then writes the "minimum" two-particle form factor as $F_\text{min}(\theta)=\mathcal N \int\limits_0^\infty \frac{dt}{t} f(t) \, \csch t\,\sin^2\(\frac{t \theta}{2\pi}\)$ (see also appendix~\ref{app:exactamps}). In our notation the latter is the exponential factor in \eqref{eq:FFBalpha} $F_\text{min}(\theta)=e^{\alpha(\theta)-\alpha(0)}$.}
\par
We may then insert into the sum rules for $c$ and $k$ equations \eqref{eq:csumrule} and \eqref{eq:ksumrule}, making use of the relation \eqref{eq:unitarity_ineq_2} between $\rho(s)$ and $F(s)$ when unitarity is saturated in
\be
\label{eq:sumrulesF}
c = 3\int_{4m^2}^\infty ds\, \frac{|F_\Theta(s)|^2 }{s^2 \sqrt{s} \sqrt{s- 4m^2} } , \qquad k =  \frac{1}{8}\int_{4m^2}^\infty ds\,\frac{\sqrt{s - 4m^2}}{ s \sqrt{s}  } |F_J(s)|^2 .
\ee

\paragraph{Free theory examples}

To illustrate how to fix $F(s)$, let us show how to compute the free fermion and boson form factors for both $c$ and $k$ (for direct computation from the free Lagrangians see appendices~\ref{app:freefermion},\ref{app:freeboson}). Starting from free boson $(+)$ we have (in all representations) $S^+(s)=1\implies \varphi=0$, which sets $\alpha^+=0$, so all we have left to fix is $B^+(s)$. In principle this could be any polynomial with fixed normalization $B(0)=F(0)$. However, as we will now see, the degree of this polynomial is bounded from above in order for the central charge sum rules to converge. Take for instance $c$ in \eqref{eq:csumrule}
\begin{equation}
    c^+=12\pi \int\limits_{4m^2}^\infty ds\, \frac{\rho^+_\Theta(s)}{s^2} =3 \int\limits_{4m^2}^\infty ds\, \frac{{B^+_\bullet}^2(s)}{\sqrt{s}\sqrt{s-4m^2}\, s^2}\,. 
\end{equation}
In order for the integral to converge at infinity the degree of such polynomial should be zero. Hence $B_\bullet^+(s)=B_\bullet^+(0)=-2\sqrt{N}m^2$ which leads to the expected central charge $c^+=N$ and matches the result from \ref{app:freeboson}. 

From \eqref{eq:ksumrule} for $k$ we have instead
\begin{equation}
    k^+= \frac{\pi}{2} \int\limits_{4m^2}^\infty ds\, \rho^+_J(s) = \frac{1}{8} \int\limits_{4m^2}^\infty ds\, \frac{\sqrt{s-4m^2}}{ s^{3/2}} \,{B^+_A}^2(s)\,.
\end{equation}
Note that if $B_A^+(s)$ is a constant ($B_A^+(s)=B_A^+(0)=2$) we get a divergence since the integral does not decay fast enough at infinity. Moreover, considering $B_A(s)$ a higher degree polynomial would only make the divergence worse. Therefore we have $k^+=\infty$ in accordance to direct computation from the Lagrangian in \ref{app:freeboson}

Now let us discuss the free fermion case. For $S^-= e^{i\varphi^-}=-1$ we have $\varphi^-= \pi$  (we choose the usual principal branch $\varphi \in (-\pi,\pi]$). This leads to $e^{\alpha^-}=(1 - s/4m^2)^{-1/2}$.
For the central charge $c$ to be finite we must have $B_\bullet^-(s)=  \sqrt{N} (4 m^2 - s)/2$ so that ${|F_\bullet^-}(s)|^2=m^2 N (s-4 m^2)$ and we find the expected result,
\begin{equation}
    c^-=3 \int\limits_{4m^2}^\infty ds\, \frac{m^2 N (s-4 m^2)}{\sqrt{s}\sqrt{s-4m^2}\, s^2}=\frac{N}{2}\,.\label{eq:cfreefermion}
\end{equation}
For the current central charge $k$ instead we have $B_A^-(s) = -2$ and we find 
\begin{equation}
    k^-=\frac{1}{8} \int\limits_{4m^2}^\infty ds \frac{16 m^2}{s^{3/2} \sqrt{s - 4m^2}} =1\,,
\end{equation}
which matches the computation in appendix~\ref{app:freefermion}. 

\subsection{\texorpdfstring{Derivation of conditions on $S(s)$ }{Derivation of conditions on S(s)}}
\label{sec:proof_lemma}
Now that we have understood how to fix the two-particle form factor starting from the amplitude in some simple examples, let us try to derive the conditions \eqref{eq:cconditionsINFTY} and \eqref{eq:kconditionsINFTY}.

Recall that in order to solve Watson's equation we have $\mathrm{Im}\, \alpha(s) =  - \frac{i}{2} \log S(s)$ as in \eqref{eq:FFBalpha}. We take the usual convention in which the logarithm has the branch cut along the negative real axis so that 
whenever $S(s_0) = - 1$ we go on top of the branch cut. Close to this point we can expand\footnote{Smoothness of $\mathrm{Im}\, S(s)$ for $s > 4m^2$ is guaranteed due to elastic unitarity.} 
\be
S(s \sim s_0) = - 1 + i (s - s_0) \, \mathrm{Im}\, S'(s_0) + \dots
\ee
Depending on the sign of $\mathrm{Im}\, S'(s_0)$, we either cross the branch cut from below or above. If $\mathrm{Im}\, S'(s_0) > 0$ the branch cut is crossed from below, which we denote as a \textit{dangerous} point
\be
s_0 \text{ \textit{dangerous} point:} \qquad\mathrm{Log}\,  S(s_0^+) = \mathrm{Log}\, S(s_0^-) - 2\pi i\,.
\ee
if instead $\mathrm{Im}\, S'(s_0) < 0$ the branch cut is crossed from above and we have
\be
s_0 \text{ \textit{safe} point:} \qquad \mathrm{Log}\, S(s_0^+) = \mathrm{Log}\, S(s_0^-) + 2\pi i,
\ee
where the '$\mathrm{Log}\,$' is the principal branch logarithm whose imaginary part is restricted to $(-\pi, \pi]$. Hence, as we take $s \to \infty$, and we go over $\mathbf{s}$ safe points and $\mathbf{d}$ dangerous points,  we must have $\mathrm{Log}\, S(s) \to \mathrm{Log}\, S(s)  - 2 \pi i (\mathbf{d} - \mathbf{s})$ or
\be
\label{eq:imalpha}
\mathrm{Im} \,\alpha(s) = \mathrm{Im}\, \alpha_0(s) -  \pi (\mathbf{d} - \mathbf{s}), \qquad s \to \infty\,,
\ee
where $\mathrm{Im} \,\alpha_0(s) \in [-\pi/2,\pi/2)$ is on the principal branch of the logarithm. 

Plugging into the dispersion relation \eqref{eq:alphadisp} we then have
\be
\alpha(s) =  \alpha_0(s) + (\mathbf{d} - \mathbf{s}) \log(-s) + \text{constant}, \qquad s \to \infty.
\ee

Similarly for $\alpha_0(s)$ given the upper bound $\Im \alpha_0(s) \leq \frac{\pi}{2}$ we observe the lower bound\footnote{This bound can be shown to apply in the case without flavour symmetry where $\Im \alpha_0(s)$ is smooth at infinity and admits a Taylor series (see appendix A of \cite{Correia:2022dyp}). In this case we have
\be
\label{eq:alphasmooth}
\Im \alpha_0(s) =  \sum_{n = 0}^\infty \frac{a_n}{s^n}, \qquad s \to \infty,
\ee
and also 
\be
\label{eq:boundsa0}
\Im \alpha_0(s) \leq \frac{\pi}{2} \implies a_0 \leq \frac{\pi}{2}\,,
\ee
Plugging into the dispersion relation \eqref{eq:alphadisp} we find
\be
\alpha_0(s) = - \frac{a_0}{\pi} \log(-s), \qquad s \to \infty
\ee
Taking the real part and making use of \eqref{eq:boundsa0} yields the bound on the real part in  equation \eqref{eq:Realpha}. See e.g. \cite{PhysRev.137.B720, Bronzan:1974jh,1987CzJPh..37..297F} for similar arguments. With O(N) symmetry the behavior \eqref{eq:alphasmooth} is not necessarily true. For example for pYB we have $\Im \alpha_0(s) \to - \frac{\mathrm{arccosh}(N/2)}{ 2 \pi} \log(s)$, but we still observe that the bound \eqref{eq:Realpha} applies. It would be interesting to extend this proof for generic O(N) amplitudes. \label{footnote:smooth} 
}
\be
\label{eq:Realpha}
\Re \alpha_0(s) \gtrsim - \frac{1}{2} \log(s), \qquad s \to \infty.
\ee
Finally, since $B(s)$ is a polynomial, we find the following lower bound on the growth of the form factor $F(s) = B(s) \,e^{\alpha(s) - \alpha(0)}$,
\be
|F(s)| \gtrsim s^{\mathbf{d} - \mathbf{s} - \frac{1}{2}}, \qquad s \to \infty.
\ee
Plugging the above behaviour into the sum rules for $c$ and $k$ in equation \eqref{eq:sumrulesF} and demanding that the integrand falls off faster than $1/s$ as $s \to \infty$ leads to the conditions \eqref{eq:cconditionsINFTY} and \eqref{eq:kconditionsINFTY}.

\begin{figure}[th!]
\begin{subfigure}{0.95\textwidth}
\centering
\includegraphics[width=1\textwidth]{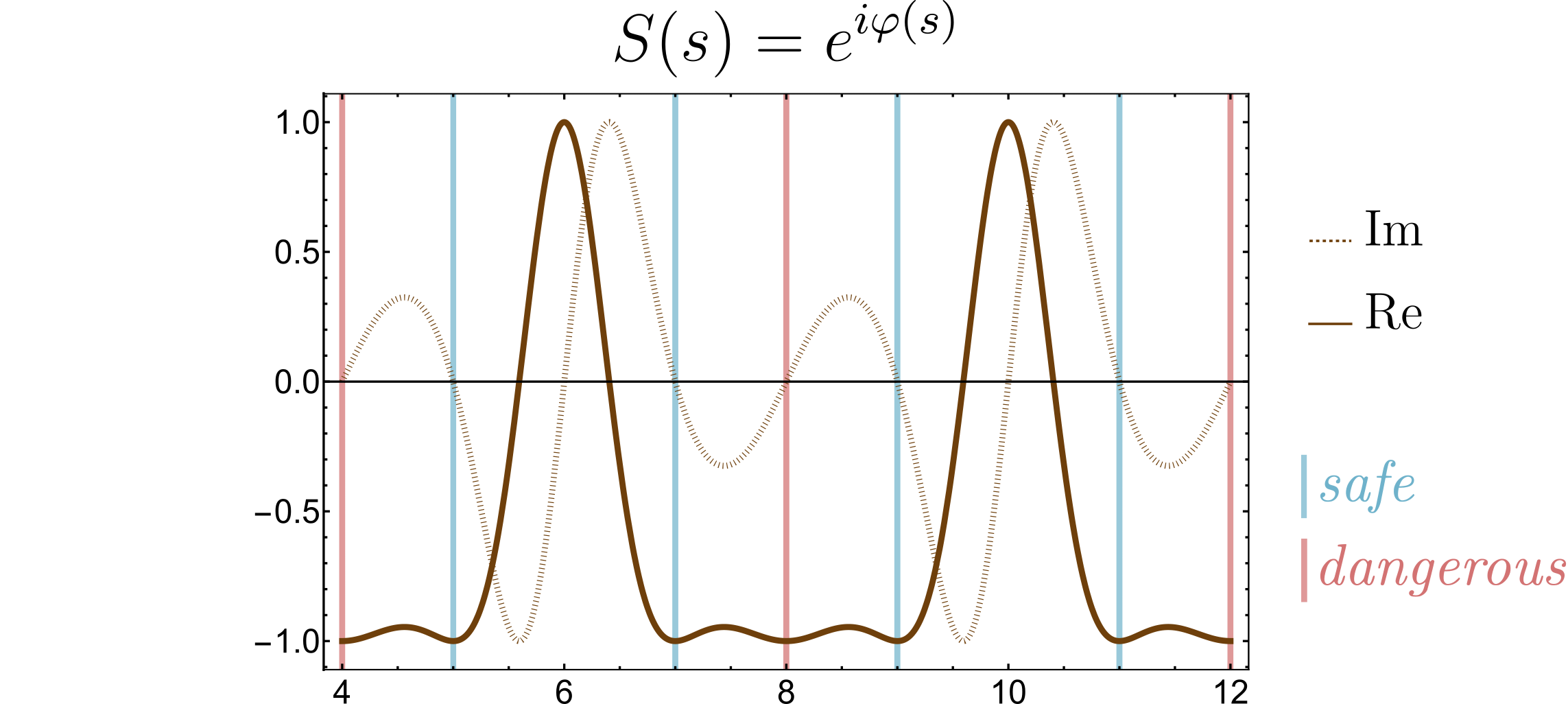}
\caption{}
\vspace{0.2cm}
\end{subfigure}
\begin{subfigure}{0.95\textwidth}
\centering
\includegraphics[width=1\textwidth]{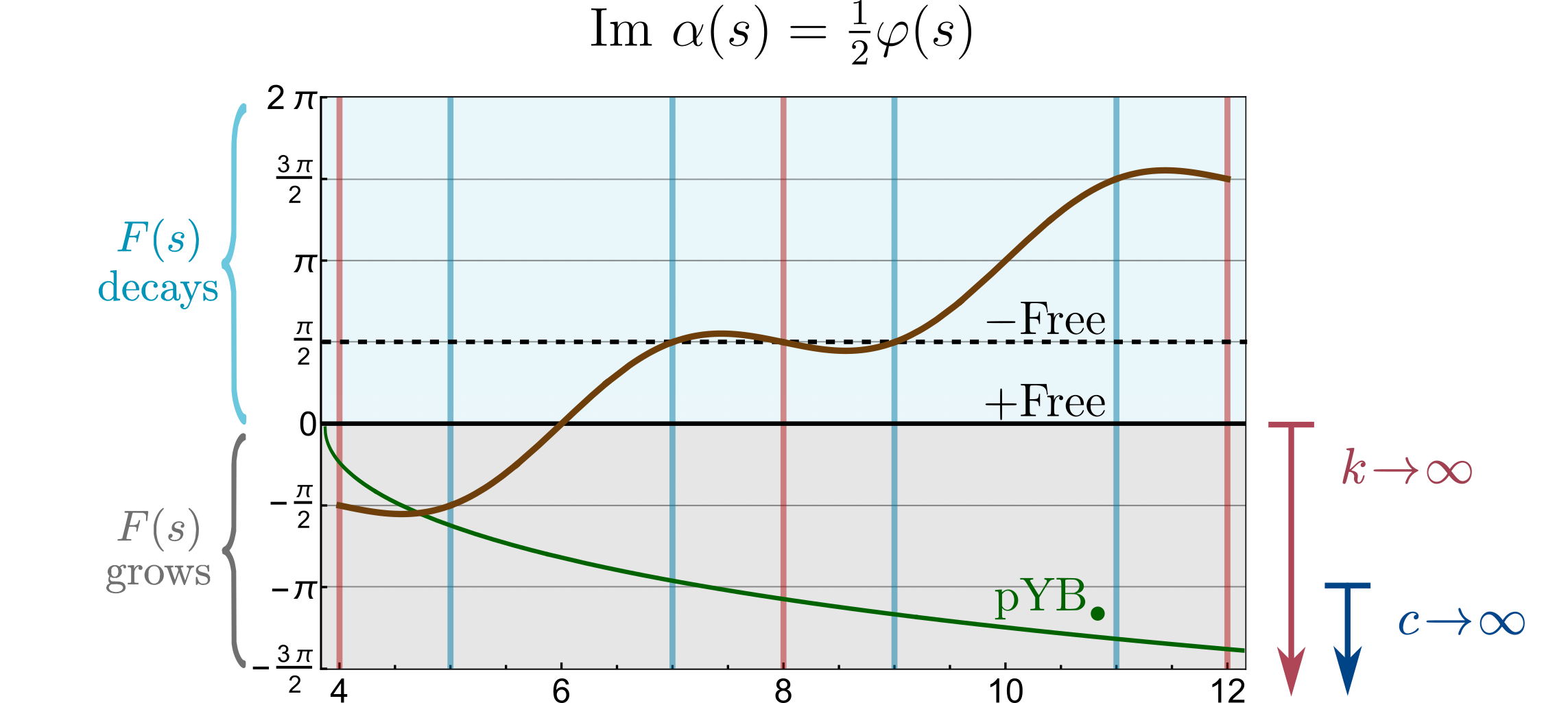}
\caption{}
\end{subfigure}
\caption{(a) Example of \textit{dangerous} and \textit{safe} points for an elastic amplitude in physical kinematics. Whenever $S(s_0)=-1$ we have a dangerous (red) or safe (blue) point according to \eqref{eq:s0dangsafe}.
(b) At each of these points, the imaginary part of $\alpha(s)$ crosses $\pi/2$ (mod $\pi$). As explained in the main text, depending on the limiting value $\Im\alpha_\infty$, the form factor either grows or decays with $s$. As a consequence, the sum rules for $k$ and $c$ diverge if $\Im\alpha_\infty\leq0$ and $\Im\alpha_\infty\leq\pi$, respectively. The solid brown line is $\Im\alpha(s)$ corresponding to the amplitude in (a). We also show curves for free boson (black solid) with $k\to\infty$ and $c$ finite, free fermion (black dashed) with both $k$ and $c$ finite, and periodic Yang-Baxter in the singlet channel with $c\to\infty$.
}
\label{fig:dangsafe}
\end{figure}

In figure~\ref{fig:dangsafe} we illustrate these dangerous/safe points in an example. The conditions \eqref{eq:cconditionsINFTY} and \eqref{eq:kconditionsINFTY} make up for a quick test as to why the dashed regions on the monolith of figure~\ref{fig:ctemperature} have infinite $c$ and suggest why they have finite $k$. Counting the number of dangerous versus safe points provides a practical bookkeeping tool to understand which amplitudes lead to infinite central charges. However, it is important to note that the conditions \eqref{eq:cconditionsINFTY} and \eqref{eq:kconditionsINFTY} are only sufficient and that some limiting cases may have infinite central charges without having enough dangerous points. One clear example is the free boson for which $k = \infty$ and has $\mathbf{d}_A - \mathbf{s}_A = 0 $. 

Let us derive a more general criteria for the divergence of the central charges which also includes the free boson.
Suppose the amplitude is such that $\Im \alpha(s) \to \Im \alpha_\infty$ as $s\to\infty$. Repeating the same steps as before we now have
\be
\alpha(s) \to - \frac{\Im \alpha_\infty}{\pi} \log(-s)\,, \qquad s \to \infty\,,
\ee
and therefore $|F(s)| \to B(s) \, s^{- \Im \alpha_\infty / \pi}$. 

Requiring that the integrand in the sum rules \eqref{eq:sumrulesF} falls off faster than $1/s$ puts again bounds on $\Im \alpha_\infty$. These read
\be
&\Im \alpha_\infty \leq - \pi \implies c \to \infty\,,  \\
&\Im \alpha_\infty \leq 0 \implies k \to \infty\,.
\ee
The above rules now explain the free boson case where $\Im \alpha_\infty = 0\implies k \to \infty$. They are of course compatible with the previous conditions \eqref{eq:cconditionsINFTY} and \eqref{eq:kconditionsINFTY}, which imply $\Im \alpha_\infty \leq - 3 \pi /2$ for the first and $\Im \alpha_\infty \leq - \pi/2$ for the latter.
In figure \ref{fig:dangsafe}(b) we show the different regions in $\Im \alpha$ leading to divergent central charges along with various examples.  

To finish this section, note that if there are bound states in the theory they would show up as poles in the form factor below $s < 4m^2$. These are therefore included into $B(s)$ that can now decay as $B(s) \gtrsim s^{- n_B}$, where $n_B$ is the number of bound states. So the condition for diverging central charge \eqref{eq:cconditionsINFTY} changes to 
\begin{align}
&\mathbf{d}_\bullet - \mathbf{s}_\bullet - n_B > 1 \implies c \to \infty, \\
&\mathbf{d}_A - \mathbf{s}_A - n_B> 0 \implies k \to \infty
\end{align}
where for $c$ it is understood that $n_B$ is the number of bound states in the singlet channel while for $k$, $n_B$ is the number of bound states in the antisymmetric channel.

Interestingly, we appear to observe that (see also discussion at the end) that in the absence of O(N) symmetry where the CDD solution for the S-matrix holds exactly, the above conditions are never satisfied, as CDD zeros only give rise to safe points, whereas CDD poles only give rise to dangerous points whose effect is canceled out by the fact that a bound state pole 
behaves as a safe point in the conditions above. This seems to explain why diverging central charges were not observed e.g. in the work \cite{Correia:2022dyp} which only dealt with flavourless scattering.

\subsection{An example from numerics}
\label{sec:numerics_safe_and_dangerous}
It is interesting to understand the mechanisms used by the dual and primal numerics to produce large but finite central charges without contradicting the conditions \eqref{eq:cconditionsINFTY} and \eqref{eq:kconditionsINFTY} derived above. For instance, in the vicinity of the regions on the monolith in which $c$ diverges (dashed lines in figure~\ref{fig:ctemperature}) there is a tension between having finite central charge, i.e. not too many dangerous points, and S-matrices close to those at the boundary with an infinite number of them. As we will see explicitly in an example, this tension gets resolved by the emergence of new safe points in physical kinematics, related to resonances in the complex $s$ or $\theta$ plane.

\begin{figure}[t!]
\includegraphics[width=.99\textwidth]{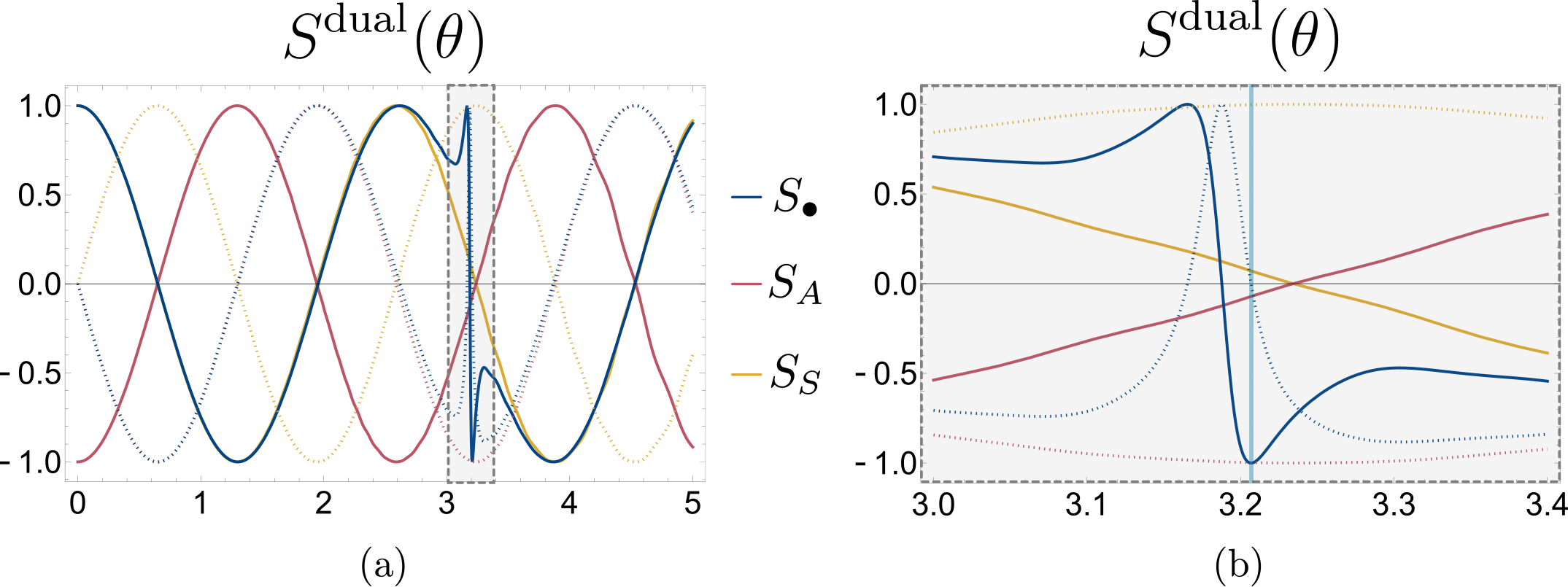}
\caption{
Dual numerical amplitudes close to periodic Yang-Baxter point for physical values of the rapidity $\theta$ and $N=2000$. In (a) we see an amplitude similar to pYB which however has a finite central charge $c \approx 5.38 N$ due to new --as compared to pYB-- safe points. In (b) we zoom into the first safe point in the singlet channel around $\theta=3.18$, marked with a vertical blue line. Numerics done with $N_\text{grid}=400$ and $N_\text{max}=240$.
}
\label{fig:S2000}
\end{figure}

\begin{figure}[th!]
\includegraphics[width=.99\textwidth]{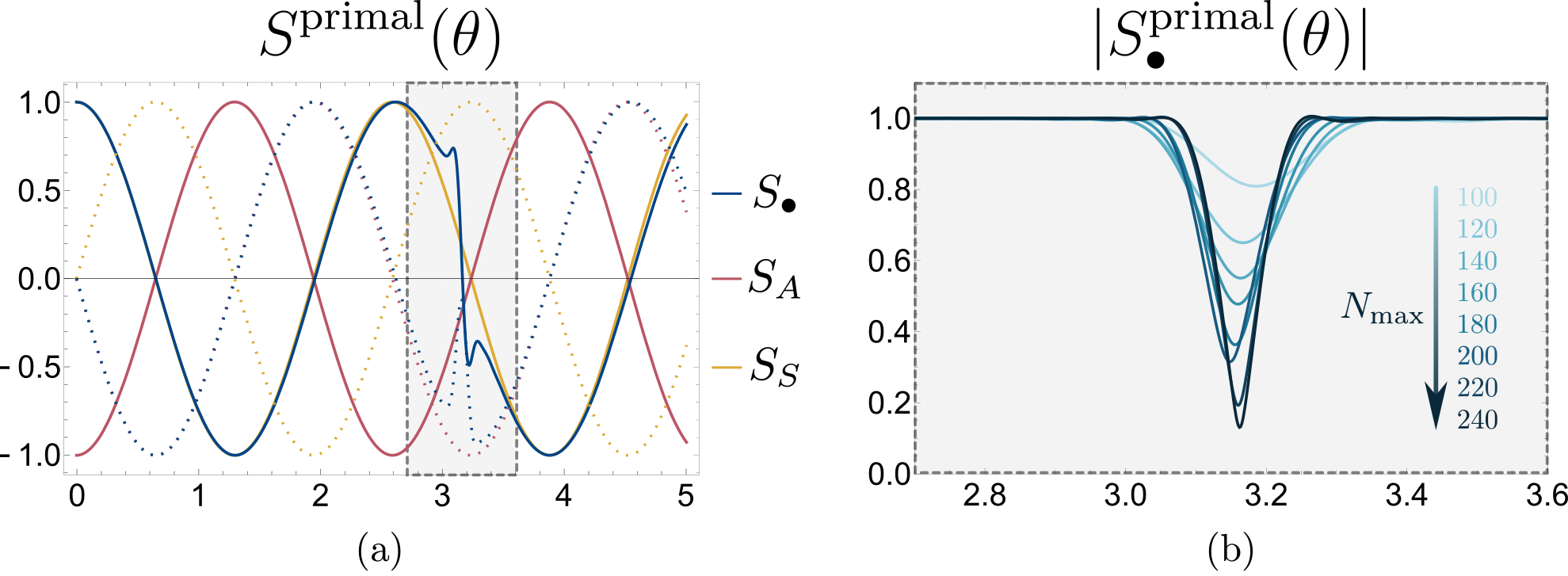}
\caption{
Primal numerical amplitudes close to periodic Yang-Baxter point for physical values of the rapidity $\theta$ and $N=2000$. The central charge minimization gives a large but finite value $c\approx7.06N$. 
In (a) we see an amplitude similar to pYB with an extra 'jump' in the singlet channel around $\theta=3.18$. In (b) we zoom in around this point for which the amplitude exhibits some inelasticity that gets localized as we increase $N_\text{max}$. Numerics done with $N_\text{grid}=400$ and $N_\text{max}=240$ for (a) and the different specified values in (b).
}
\label{fig:S2000UNIT}
\end{figure}

First, a word of caution about numerics. As it is very often the case in S-matrix bootstrap problems, the optimization objectives ($c$ and $k$ in our case) are not very sensitive to changes in the large energy behaviour of the observables. In particular, the integral sum rule for $c$ \eqref{eq:csumrule} has a $s^{-2}$ damping factor which makes it very difficult to extract reliable information about the optimal functions at large energies. This is usually not too worrisome since the objective rapidly converges when increasing $N_\text{max}$. Moreover, when working with both primal and dual approaches we can get a good idea to how far we are from the true optimal bound.  As for the observables like amplitudes, at small enough energies we see saturation of unitarity for the primal observables, so that primal and dual functions agree and describe the true optimal function in this range of energies. 

For concreteness we focus on $c$-minimization for a point close to close to pYB in the $\sigma_2=0$ section. We choose a large number of flavors $N=2000$ so that we have various dangerous points at low enough energies and fix $\sigma_1=0.02227$.\footnote{Recall that the pYB period in rapidity is $\tau_\text{pYB}=2\pi^2/\text{arccosh}(N/2)$.} In figures~\ref{fig:S2000} and \ref{fig:S2000UNIT} we show the amplitudes obtained from the dual and primal numerics, respectively. The left panels show an amplitude very similar to pYB  (c.f. figure~\ref{fig:SpYB}), except for some new structure in the singlet channel close to $\theta=3.18$. The minimization gives a finite central charge of $c\approx 10,753\approx 5.38 N$ for the dual and $c\approx13,043\approx6.52N$ for the primal.

As evident in figures~\ref{fig:S2000} and \ref{fig:S2000UNIT}, the dual and primal numerics evade conditions \eqref{eq:cconditionsINFTY} with different mechanisms. Namely, the former clearly shows the existence of new safe points whereas the latter does not saturate unitarity at all energies and therefore is in no contradiction with \eqref{eq:cconditionsINFTY}. Nonetheless, both mechanisms point at the same physics.  

In the dual approach, the amplitudes are constructed from the dual variables assuming unitarity saturation (see appendix~\ref{app:dualproblems}), so that the jump in the singlet amplitude close to $\theta=3.18$ is a true safe point (see figure~\ref{fig:S2000}(b)). In the primal numerics what we see instead is that at finite $N_{max}$ unitarity is not saturated at high energies so that the assumptions made to derive \eqref{eq:cconditionsINFTY} do not hold. 
In figure~\ref{fig:S2000UNIT}(a) we see a similar jump of the singlet amplitude around  $\theta=3.18$. However, the jump is not a clear safe point
since unitarity is not saturated around this value, as figure~\ref{fig:S2000UNIT}(b) shows. This inelasticity is a finite $N_\text{max}$ artifact. As we increase the size of our ansatz the inelasticity is sharper and more localized around a given point, which often signals the presence of a resonance close to the real line.

\begin{figure}[t!]
\centering
\includegraphics[width=.5\textwidth]{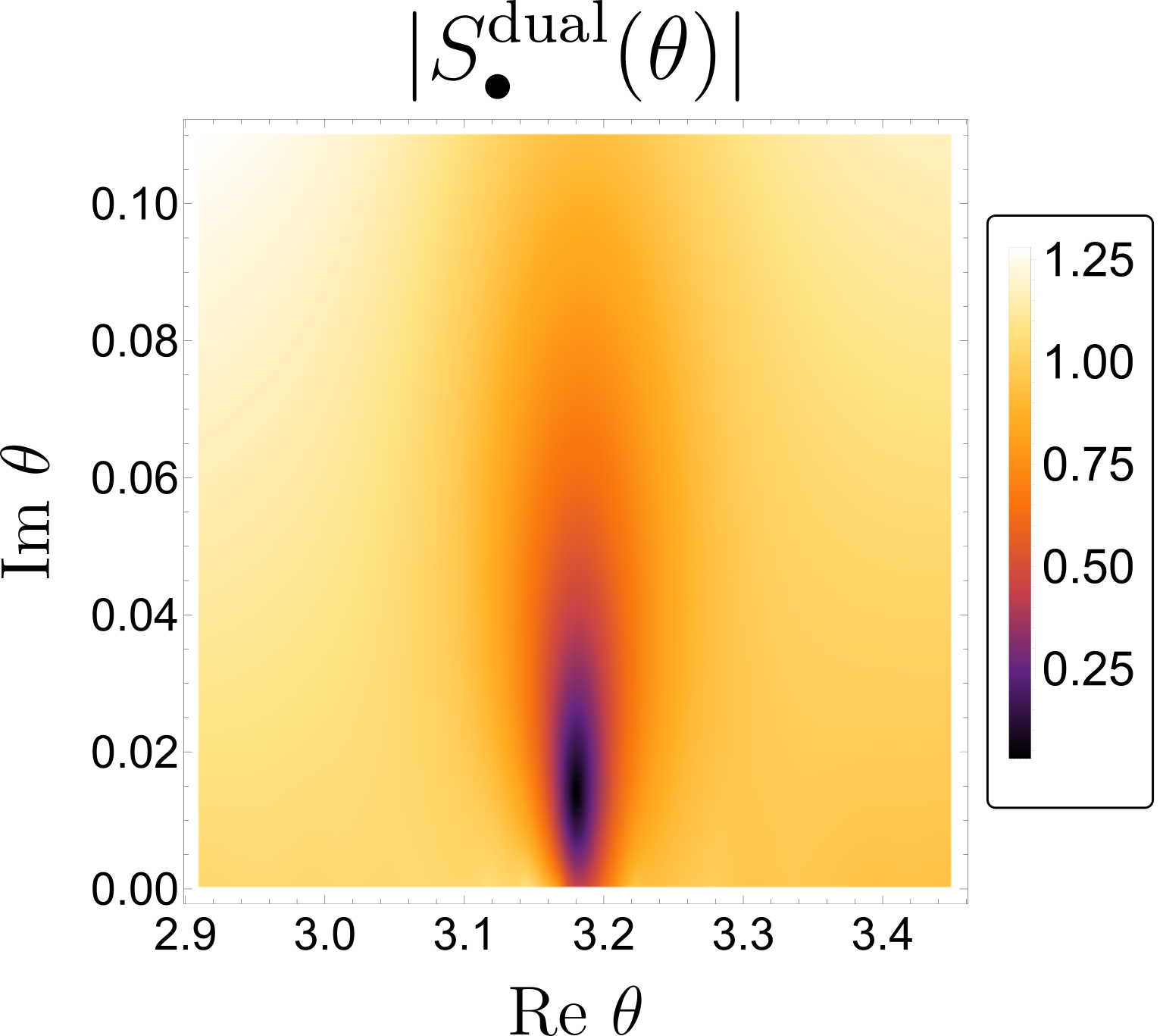}
\caption{
Dual numerical amplitude in the complex rapidity plane featuring a resonance close to the real line. This resonance is responsible for the safe point of figure~\ref{fig:S2000}.
}
\label{fig:densityplot}
\end{figure}

The resonance is apparent as we plot the dual amplitude in the complex rapidity $\theta$ plane as in figure~\ref{fig:densityplot}.\footnote{In principle the dual amplitudes are only defined on the real line, but one can write a dispersion relation from the imaginary part and the known subtraction constants.} There we see a definite zero sitting very close to the real line, which explains the trouble of the primal numerics to correctly reproduce the resonance. At higher energies where the numerics are less reliable we expect the periodic nature of the amplitude to continue, with many new safe points and resonances countering the infinite dangerous points.

In this example we see how the existence of new safe points, necessary to achieve a finite central charge, have a direct physical implication of including more resonances in the theory.

\section{Discussion}
\label{sec:discussion}

In this work we studied gapped, O(N) symmetric QFTs using numerical and analytical bootstrap approaches focused on $2\rightarrow2$ scattering amplitudes and two-particle form factors of the trace of the stress tensor $\Theta(x)$ and of the O(N) conserved currents $J_A^\mu(x)$. 
We now summarize our results and discuss some implications and future directions.

\paragraph{Minimum central charges and WZW models.}
In section \ref{sec:minckmono} we found the minimum of the central charges $c$ and $k$ of the UV CFT across the O(N) monolith space of S-matrices. The results are shown in figures \ref{fig:ctemperature} and \ref{fig:ktemperature} for $N = 7$ (we expect similar plots for any integer $N > 2$). In figures \ref{fig:csections} and \ref{fig:ksections} we plot the minimum $c$ and $k$ across several sections connecting special points of the O(N) monolith. We observe rapid growth of the central charges in several directions towards the boundary of the O(N) monolith, with $c$ becoming infinite over the majority of the boundary. This severely constrains the CFTs from which these putative QFTs can flow from (note that $N$ is \emph{finite}). See further discussion below.

In section \ref{sec:mincNkN} we studied how the global minimum for $c$ and $k$ change with $N$. As shown in figure \ref{fig:ckN} we find that $c$ grows linearly while $k$ goes to a constant as $N \to \infty$. Another observation is that the global minimum central charges $c$ and $k$ for $N>2$ are not on the boundary but inside the bulk of the O(N) monolith. This means that the S-matrix which minimizes the central charge is \emph{not} the same S-matrix that extremizes some coupling in S-matrix space.\footnote{At least none of the `quartic' couplings that span the O(N) monolith.} This is in sharp contrast with e.g. the single scalar exchange bootstrap of \cite{Correia:2022dyp} where the extremal S-matrices coincided (and allowed for an analytical derivation of the minimum $c$). 

In section \ref{sec:cvsk} we studied the allowed space of central charges $c$ and $k$ given the bootstrap constraints on the S-matrix, form factors and spectral densities. The result is shown in figure \ref{fig:cvsk}. We find that the Sugawara construction realized by WZW models lies insides the bounds, but cannot cover the full allowed space. As always, it is entirely possible that adding more bootstrap constraints would rule out this extra space. However, it is important to note that the Sugawara construction assumes conservation of not only the current but also the individual holomorphic and antiholomorphic components. It would be interesting to see how this constraint could be included in our bootstrap setup or to repeat the same exercise of bounding $c$ and $k$ but via the numerical conformal bootstrap \cite{Chester:2019wfx}, which to our knowledge has not been done yet. It would be interesting to compare these two approaches as it would give us an understanding of how constraining really are the S-matrix principles in the UV, especially with the extra assumptions taken here on the RG flow (in that we are restricting ourselves to CFTs which admit O(N) preserving deformations). 

\paragraph{Infinite central charges, dangerous points and resonances.} As already mentioned, we observe that the minimum central charge $c$ diverges on most of the boundary of the O(N) monolith (see figure \ref{fig:ctemperature}). 
The corresponding extremal S-matrices share the following feature: they exhibit the presence of \textit{dangerous points} on the physical region where $S(s_0) = -1$ and $\mathrm{Im}\, S'(s_0) > 0$. In section \ref{sec:analytic_structure_S} we argued that if the amplitudes have enough dangerous points, then the associated central charges diverge. This criterion is satisfied by the pYB S-matrices, which are located on this portion of the boundary. Curiously, the divergent central charge is not a consequence of the periodicity in the amplitude, as amplitudes can be periodic and lack dangerous points, but to how the phase shift decreases at large energies (see figure \ref{fig:dangsafe}) which is guaranteed by the existence of a sufficient number of dangerous points.

Moving away from pYB slightly into bulk of the O(N) monolith, where the central charge $c$ is large but finite, we observe that the extremal S-matrix deviates from pYB around the second dangerous point (which is located at relatively high energies). The S-matrix coming from primal optimization avoided the creation of dangerous points by generating some inelasticity around that location, see figure \ref{fig:S2000UNIT}. The S-matrix coming from dual optimization, which is elastic by construction, instead creates a safe point (where $\mathrm{Im}\, S'(s_0)$ has the opposite sign), see figure \ref{fig:S2000}. This has repercussions on the analytic structure, as seen in figure \ref{fig:densityplot}, where a close-by resonance is observed. 
There seems to be a relation between the presence of dangerous and safe points on physical kinematics and the existence of bound states and resonances on the complex plane. In the case of  scattering without flavour, where the CDD solution is exact, we see that CDD zeros (i.e. resonances) only lead to safe points, and that bound states are needed in order to have dangerous points.\footnote{This was tested with up to six CDD zeros arbitrarily located in the complex plane.} This suggests that the necessary condition \eqref{eq:cconditionsINFTY} for infinite central charges is never met in the case of scattering without internal symmetry, which is consistent with the findings in \cite{Correia:2022dyp}. In the O(N) case, on the other hand, dangerous points can be present without a corresponding bound state, which leads to the observed diverging central charges. Moreover, regardless of the existence of safe points, for the O(N) S-matrices an infinite amount of resonances across the infinitely many sheets is generically present due to crossing and unitarity \cite{Cordova:2018uop,Cordova:2019lot}. It thus remains to be seen if a sharp connection between resonances/bound states and safe/dangerous points can be made.

\paragraph{Inserting more of the UV and targeting specific deformations.}
In this work we did not target a specific theory whose UV CFT is known. On the contrary, we constrained the class of CFTs from which a generic gapped theory in the IR with O(N) symmetry would flow from. Possible theories which could be targeted are deformation of WZW models.\footnote{For instance, it would be interesting to understand the connection to the integrable deformation known as lambda model and the RG flows studied in \cite{Appadu:2017bnv}.} In this case $c$ and $k$ would be known from the Sugawara construction in equation \eqref{eq:Sugawara} and could be fixed in the bootstrap setup. Then, the space of couplings of the S-matrix and form factors in the IR could be carved out, similarly to how Ising Field Theory was targeted in \cite{Correia:2022dyp}.

A still open problem is of course how to include more information about the UV CFT, for example the dimensions of the deforming operators that give rise to the RG flow. Could we put a bound on these dimensions using the S-matrix bootstrap? The practical aspect of the central charges is that they obey sum rules, which are \emph{integral} relations directly on the spectral density. There exist sum rules for the scaling dimension of operators \cite{DELFINO1996327}, but they involve their vacuum expectation value appearing as a new parameter. A recent development \cite{He:2023lyy} targeting gauge theories,  have used so-called QCD sum rules, which relates the very high energy behavior of the two point function at some large $s_0$ to an integral over the spectral density.\footnote{There is however a slight inconvenience in that the choice of $s_0$ is somewhat arbitrary and it seems hard to estimate the error (see e.g. \cite{Caron-Huot:2023tpw}).} This provides a new way to introduce information about the CFT and the deforming operator(s) using form factor  perturbation theory.

\paragraph{Non-unitarity CFTs, loop models and $N < 2$.}
There are various hints that the space of theories spanned by the S-matrix bootstrap should go outside the usual UV complete paradigm, where we have a unitary CFT in the UV. The first hint we already encountered is the free boson case, where the central charge $k$ diverges due to the logarithmic nature of free massless non-compact bosons. A more interesting example comes from the well studied O(N) loop model, which gives rise to a family of conformal theories in the window $-2\leq N\leq 2$. For $N=1$ we recover the critical Ising model and $N=2$ is the Kosterlitz-Thouless fixed point. For generic non-integer $N$ the conformal theories are logarithmic (see \cite{Gorbenko:2020xya} and references therein).\footnote{It would be interesting to compare our method of computing $k$ to the results recently obtained in \cite{Jacobsen:2023vdf} for these theories.} By perturbing with the energy density operator one triggers an RG flow leading to a gapped integrable QFT. The associated S-matrix given in \cite{Zamolodchikov:1990dg} sits at the boundary of the $|N|\leq2$ monolith, in what becomes (-)pYB for $N>2$. In \cite{Gorbenko:2018dtm,Gorbenko:2018ncu} it was argued that as we continue in $N$ the fixed point above becomes complex. Moreover in  \cite{Gorbenko:2020xya} it was shown that -- at least for $N\gtrsim2$ -- the pYB solution reproduces the expected RG walking behaviour close to the complex fixed point. 

Another indication comes from integrability. The expectation is that most integrable amplitudes one can write down are incompatible with UV completeness, in a similar manner to $T\bar T$ deformed theories \cite{Smirnov:2016lqw,Cavaglia:2016oda}. More precisely, what happens is that the finite size ground state energy computed from the exact S-matrix develops a square root singularity at a size $R^*$ below which the energy becomes complex. This obstacle for UV completeness was shown explicitly for when the S-matrix is a product of two or more CDD zeros \cite{Camilo:2021gro,Cordova:2021fnr}. Intriguingly, this exotic UV behavior goes undetected by the central charge minimization bootstrap in \cite{Correia:2022dyp}, where the space of optimal scalar amplitudes minimizing central charges is spanned with amplitudes of up to three CDD factors. 

We believe the above points highlight the fact that the probes we are using are not fully sensitive to potential non-trivial UV behaviours. After all, we are only imposing positivity of the spectral functions for the stress tensor and O(N) currents, so that we might be restricting to unitary sub-sectors of a more complicated theory. In this light, it would be very interesting to understand which modifications to our bootstrap approach would allow us to test logarithmic/complex CFTs, either to include them as a more general class of theories we should study or to definitely exclude them from our bounds.

\section*{Acknowledgments}
We would like to thank  Nima Afkhami-Jeddi, Victor Gorbenko, Kelian H\"aring, Yifei He, Andrea Manenti, Nafiz Ishtiaque, Denis Karateev, Shota Komatsu, Nat Levine, Jo\~ao Penedones, Aninda Sinha, Emilio Trevisani, Pedro Vieira, Xi Yin and Bernardo Zan for useful discussions. The computations were enabled by resources provided by the National Academic Infrastructure for Supercomputing in Sweden (NAISS) at UPPMAX (Rackham) partially funded by the Swedish Research Council through grant agreement no. 2022-06725. Nordita is partially supported by Nordforsk. AG is also supported by a Royal Society funding, URF\textbackslash R\textbackslash221015. AV is supported by the Simons Foundation grant 488649 (Simons Collaboration on the Nonperturbative Bootstrap) and the Swiss National Science Foundation through the project 200020-197160 and through the National Centre of Competence in Research SwissMAP.  This project has received funding from the European Research Council (ERC) under the European Union's Horizon 2020 research and innovation programme (grant agreement number 949077). 

\newpage
\begin{appendices}

\section{Results for exact amplitudes}\label{app:exactamps}

\subsection{\texorpdfstring{Exact amplitudes on $N>2$ monolith}{Exact amplitudes on N>2 monolith}}\label{app:exactamps0}
We first write the exact amplitudes appearing at the boundary of the $N>2$ monolith. We use the following notation for the different channels $S_a=(S_\bullet,\,S_A,\,S_S)$ and the rapidity variable $\theta$, related to the Mandelstam invariant $s$ through $s=4m^2\cosh(\theta/2)$.
\begin{align}
\textbf{free:  }&\quad S_a=\pm\,(1,1,1)\label{S_free}\,,\\
\textbf{NLSM:  }&\quad  S_a(\theta)= \mp\(1,\;\frac{\theta-i\pi}{\theta+i\pi},\; \frac{\theta-i\pi}{\theta+i\pi}\,\frac{\theta-i\lambda_N}{\theta+i\lambda_N} \) F_{\pi+\lambda_N}(\theta)F_{2\pi}(\theta)\label{S_NLSM}\,,\\
\textbf{pYB:  }&\quad S_a(\theta)=\pm\,\(\frac{\sinh\[\nu\(1-\frac{i\theta}{\pi}\)\]}{\sinh\[\nu\(1+\frac{i\theta}{\pi}\)\]},\;-1,\;1 \)\prod\limits_{n=-\infty}^\infty F_{\pi+\frac{in\pi^2}{\nu}}(-\theta)\,,\label{S_pYB}\\
\textbf{constant:  }&\quad S_a=\pm\,\(1,\,-1,\,-\frac{N-2}{N+2}\)\,,\label{S_const}
\end{align}
where $\lambda_N=\frac{2\pi}{(N-2)}$, $\nu=\text{arccosh}(\tfrac{N}{2})$ and 
\begin{equation}
    F_a(\theta)\equiv \frac{\Gamma \left(\frac{a+i \theta }{2 \pi }\right) \Gamma \left(\frac{a-i \theta +\pi }{2 \pi }\right)}{\Gamma \left(\frac{a-i \theta }{2 \pi }\right) \Gamma \left(\frac{a+i \theta +\pi }{2 \pi }\right)}={}_2F_1\left(-\frac{1}{2},\frac{i \theta }{\pi},\frac{a+i \theta }{2 \pi},1\right) \label{eq:Fablock}\,. 
\end{equation}
Except for the constant solution \eqref{S_const}, all other amplitudes are integrable and saturate unitarity.

\subsection{Integrable Form Factor Bootstrap}
In the following we review some of the conventions used in the integrability literature regarding integral representations of S-matrices and the computation of form factors. Here we restrict to the two-particle form factors, for a detailed review containing higher particle form factors see e.g. \cite{Smirnov:1992vz,Mussardo:2020rxh}. 
Here we deal with amplitudes saturating unitarity. Using the map $s=4m^2\cosh(\theta/2)$, the two particle cuts in the $s$ complex plane get opened so that all different Riemann sheets in $s$ are mapped to strips $n\pi\leq\text{Im}\theta<(n+1)\pi$ of the rapidity plane. Therefore the only non-analyticities of the amplitude are poles (and zeros from their unitarity image). One can exploit this fact and write a compact, often simpler, integral representation for the amplitude
\beq
S(\theta)=\exp{\int\limits_0^\infty \frac{dt}{t}\, f(t)\, \sinh{\frac{t \theta}{i\pi}}}\,. \label{intrep_theta_S}
\eeq
This type of representation can be easily derived when writing the poles and zeros of the amplitude in the following form
\beq
\ln\(\frac{1+i \alpha/\beta}{1-i\alpha/\beta}\)=2i \int\limits_0^\infty \frac{dt}{t}\, e^{\beta t}\, \sin\(\alpha t\) \label{eq:id1}\,.
\eeq

The same function $f(t)$ appearing in \eqref{intrep_theta_S} can then be used to compute the ``minimum'' two-particle form factor
\beq
F_\text{min}(\theta)=\mathcal N \exp{\int\limits_0^\infty \frac{dt}{t}\, \frac{f(t)}{\sinh{t}} \,\sin^2\[\frac{t (i\pi-\theta)}{2\pi}\]} \label{eq:FminINT}\,,
\eeq
which solves Watson's equation and is analogous to the factor $e^{\alpha(s)}\subset F(s)$ appearing in \eqref{eq:FFBalpha}.
The full two-particle form factor will be given by $F_2(\theta)=Q(\theta)/D(\theta) \times F_\text{min}(\theta)$, where $Q$ and $D$ are polynomials in $\cosh{\theta}$. The polynomial in the denominator $D$ is fixed by the possible poles of the S-matrix, whereas $Q$ depends on the operator we are considering. They play the role of our function $B(s)$ in \eqref{eq:FFBalpha}.  \\

Some examples for the functions $f(t)$ are \cite{Karowski:1978vz} \footnote{A useful result for the building block $F_a(\theta)$ in \eqref{eq:Fablock} is its integral representation
\beq
F_a(\theta)=\frac{\Gamma \left(\frac{a+i \theta }{2 \pi }\right) \Gamma \left(\frac{a-i \theta +\pi }{2 \pi }\right)}{\Gamma \left(\frac{a-i \theta }{2 \pi }\right) \Gamma \left(\frac{a+i \theta +\pi }{2 \pi }\right)} \,,\quad f_a(t)=2\frac{e^{-a t/\pi}}{1+e^{-t}}\,.
\eeq}

\begin{equation}
\begin{aligned}[c]
&S^+=1\,,\\ 
&S^-=-1\,,\\
&S^\text{shG}(\theta)=\dfrac{\sinh{\theta}-i\sin{\(\pi \lambda\)}}{\sinh{\theta}+i\sin{\(\pi \lambda\)}}\,,\\
&S^\text{NLSM}(\theta)=\eqref{S_NLSM} \,
\end{aligned}
\quad
\begin{aligned}[c]
&f^+=0\,,\\
 &f^-=2\,,\\
 &f^\text{shG}(t)=2 \left(1-\frac{\cosh \left(\frac{1}{2} \left(1-\frac{2 \lambda }{\pi }\right) t\right)}{\cosh \left(\frac{t}{2}\right)}\right)\,,\\
 &f^\text{NLSM}(t)=2\( \frac{e^{-\frac{2}{N-2}t}+1+e^t+e^{-t}}{1+e^t}\,, \frac{e^{-\frac{2}{N-2}t}-1}{1+e^t}\,, \frac{1-e^{-\frac{2}{N-2}t}}{1+e^{-t}} \)\,.
\end{aligned}
\end{equation}

Knowing these functions $f(t)$ for a given model, the analytic structure of the minimum form factor can be extracted from the identity
\beq
\ln\(1+\frac{\alpha^2}{\beta^2}\)=4\int\limits_0^\infty \frac{dt}{t} e^{-\beta t} \sin^2\frac{\alpha t}{2} \label{eq:id2}\,.
\eeq
In practice, one writes a series expansion for the ratio $f(t)/\sinh t$ appearing in \eqref{eq:FminINT} and uses \eqref{eq:id2} iteratively to find:
\beq
\frac{f(t)}{\sinh t}= 4\sum_n a_n e^{-\beta_n t} \implies F_\text{min}(\theta)=\prod_n \[1+\frac{(i\pi-\theta)^2/\pi^2}{ \beta_n^2}\]^{a_n}\,.
\eeq

\subsection{Examples minimum Form Factors}
Particularly simple examples include free fermion and the non-linear sigma model for $N=3$
\beqa
F_\text{min}^-(\theta)&=&-i\sinh\frac{\theta}{2}\,, \label{eq:Fminfreefermion}\\
F_{a,\text{min}}^\text{NLSM, N=3}(\theta)&=& \( \frac{-\pi^2}{\theta(\theta-2\pi i)}\sinh^2\frac{\theta}{2},\; \frac{\pi^2 (\theta-i\pi)}{2\theta(\theta-2\pi i)}\, \tanh\frac{\theta}{2} ,\; \frac{\theta-i\pi}{2} \,\tanh\frac{\theta}{2}\)\,.
\eeqa
Note that if we have $F_\text{min}$ for a given amplitude $S$ and we want to compute the form factor for $-S$ it is enough to multiply by \eqref{eq:Fminfreefermion}.

\paragraph{Periodic Yang-Baxter}
To evaluate the integral representation of periodic Yang-Baxter's S-matrix \eqref{S_pYB} we will use the following identity
\begin{equation}
    \sum\limits_{k\in\mathbb Z} e^{-i k t}=2\pi \sum\limits_{p=0}^\infty \frac{\delta(t-2\pi p)}{1+\delta_{p,0}}\,, \quad t\in(0,\infty)\,.
\end{equation}
Let us start with the prefactor in the singlet channel of\eqref{S_pYB}
\begin{equation}
    \frac{\sinh{\frac{\nu}{\pi}(\pi-i\theta)}}{\sinh{\frac{\nu}{\pi}(\pi+i\theta)}}\implies f(t)=2 e^{-t} \,\sum\limits_{k\in\mathbb Z} e^{i k t \pi/\nu}=4\nu e^{-t} \sum\limits_{p=0}^\infty \,\frac{\delta(t-2\pi p)}{1+\delta_{p,0}}\,.
\end{equation}
It is a simple exercise to plug in the infinite sum over Dirac deltas in \eqref{intrep_theta_S} and recover the prefactor. \\

For the overall factor we have instead
\begin{equation}
    \prod\limits_{n=-\infty}^\infty F_{\pi+\frac{in\pi^2}{\nu}}(-\theta) \implies f(t)=\frac{-2}{1+e^t}\,\sum\limits_{k\in\mathbb Z} e^{i k t \pi/\nu}=\frac{-4\nu }{1+e^t} \,\sum\limits_{p=0}^\infty \frac{\delta(t-2\pi p)}{1+\delta_{p,0}}\,.
\end{equation}

The delta functions make it easy to perform the integration \eqref{intrep_theta_S} and recover the amplitudes in an alternative representation. For the singlet channel amplitude we have
\begin{equation}
   f_\bullet^\text{pYB}(t)=\frac{4\nu\, e^{-t}}{1+e^t}\, \sum\limits_{p=0}^\infty \frac{\delta(t-2 \nu p)}{1+\delta_{p,0}}
     \implies  S_\bullet^\text{pYB}(\theta)=  \text{exp}\left\{
     -\frac{i \theta\nu}{\pi}-\sum\limits_{p=1}^{\infty} -\frac{2 i e^{-2 \nu  p} \sin \left(\frac{2 \theta  \nu  p}{\pi }\right)}{p \left(e^{2 \nu  p}+1\right)}
     \right\}\,,
\end{equation}
where the last terms in the exponential are negligible for physical kinematics $\theta\in\mathbb R$. Similar reasoning leads to the kernel $f_A(t)$ for antisymmetric amplitude
\begin{equation}
    f_A^\text{pYB}(t)=2-\frac{4\nu\,}{1+e^t}\, \sum\limits_{p=0}^\infty \frac{\delta(t-2\nu p)}{1+\delta_{p,0}}\implies  S_\bullet^\text{pYB}(\theta)=  -\text{exp}\left\{\frac{i \theta\nu}{\pi}-\sum\limits_{p=1}^{\infty} \frac{2 i \sin \left(\frac{2 \theta  \nu  p}{\pi }\right)}{p \left(e^{2 \nu  p}+1\right)}\right\}\,.
\end{equation}
For the minimum form factors we find
\begin{align}
    F_{\text{min},\bullet}(\theta)&= \text{exp}\left\{
    -\frac{(\pi +i \theta )^2 \nu }{4 \pi ^2} +\sum\limits_{p=1}^{\infty} \frac{2 e^{-2 \nu  p} \text{csch}(2 \nu  p) \sin ^2\left(\frac{(-\theta +i \pi ) \nu  p}{\pi }\right)}{p \left(e^{2 \nu  p}+1\right)}
    \right\} \,, \label{eq:FminpYB0} \\
    F_{\text{min},A}(\theta)&= i\sinh\(\frac{\theta}{2}\)\text{exp}\left\{
\frac{(\pi +i \theta )^2 \nu }{4 \pi ^2} - \sum\limits_{p=1}^{\infty} \frac{2 \text{csch}(2 \nu  p) \sin ^2\left(\frac{(-\theta +i \pi ) \nu  p}{\pi }\right)}{p \left(e^{2 \nu  p}+1\right)}
     \right\} \,.
\end{align}

\begin{table*}[t!]
\centering
\begin{tabular}{ |m{12em}|c|c| } 
 \hline
  & $c_2$ & $k_2$ \\ 
 \hline
 $+$free    & $N$ & $\infty$ \\ 
 $-$free & $N/2$ & 1 \\
 $+$const & $N$ & 1 \\ 
 $-$const & $N/2$ & $\infty$ \\
 $+$NLSM& 5.2124 ($N$=7)& 15.6380 ($N$=7)\\ 
 $-$NLSM & 3.9137 ($N$=7)& 1.1380 ($N$=7)\\
 $+$pYB & $\infty$ & 2.1043 ($N$=7)\\ 
 $-$pYB & $\infty$ & 1.6090 ($N$=7)\\
  $(N=2)$ $+$pYB/$-$NLSM & 1.5649  & 1.4889\\
   $(N=2)$ $-$pYB/$+$NLSM & 0.9875 & 1.6852\\
 \hline
\end{tabular}
\caption{Selected values of $c_2$ and $k_2$ for models with exact amplitudes. Numerical values are approximate since the integrations are performed numerically with a cutoff.}
\label{table:c2k2values}
\end{table*}

\subsection{\texorpdfstring{Computing $c_2$ and $k_2$}{Computing c2 and k2}}
Having the exact S-matrix and the corresponding function $f(t)$ for a given theory, we can compute the two-particle form factors and their contribution the central charges coming from the sum rules \eqref{eq:csumrule} and \eqref{eq:ksumrule}. As explained below \eqref{eq:FminINT}, there is a polynomial freedom in constructing the full two-particle form factor  $F_2(\theta)=Q(\theta)/D(\theta) \times F_\text{min}(\theta)$. The polynomial can be fixed by requiring that: 1) the form factor is analytic except for the s-channel two-particle cut, and 2) the sum rule integral converges. For most of the amplitudes treated in this section
it is enough to consider the redundancy $\left[4m^2\sinh^2(\theta/2)\right]^{\pm1}$ or equivalently $(s-4)^{\pm1}$, where the only potential dangerous points happen at threshold and the maximum degree of the polynomial in $s$ is set to 1 for convergence of the sum rule (see section~\ref{sec:analytic_structure_S}). We can also trace back this polynomial form to the $2\pi n$ redundancy in the argument of the S-matrix. 
For the $c$ sum rule of periodic Yang-Baxter however, it is not possible to use this freedom to get a convergent integral, since the minimum form factor \eqref{eq:FminpYB0} grows as $F_{\text{min},\bullet}(\theta)\sim e^{\theta^2}$. Except for simple cases like free theories and $N=3$ NLSM showed above, one needs to compute the integrals for the form factor \eqref{eq:FminINT} and sum rules numerically. This is what we did to find the endpoints of the minimum $c$ and $k$ sections in figures~\ref{fig:csections} and \ref{fig:ksections}. The values we find for the two particle contribution to the central charge are given in the table~\ref{table:c2k2values}. 

\begin{figure}[t!h!]
\begin{subfigure}{0.49\textwidth}
\includegraphics[width=1\textwidth]{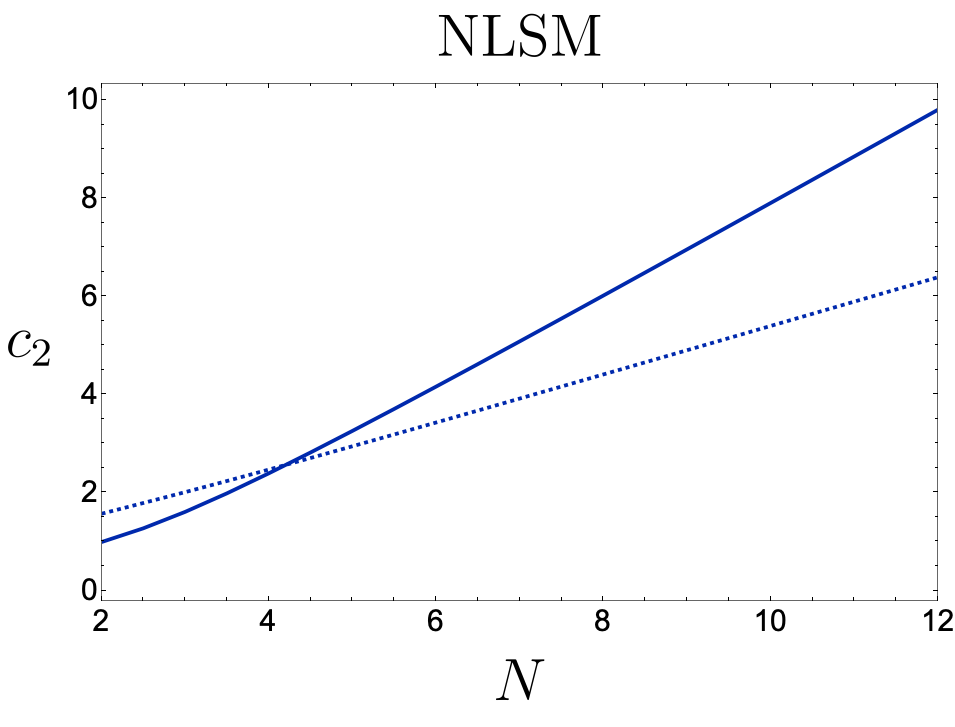}
\caption{}
\end{subfigure}
\vspace{0.2cm}
\begin{subfigure}{0.49\textwidth}
\includegraphics[width=1\textwidth]{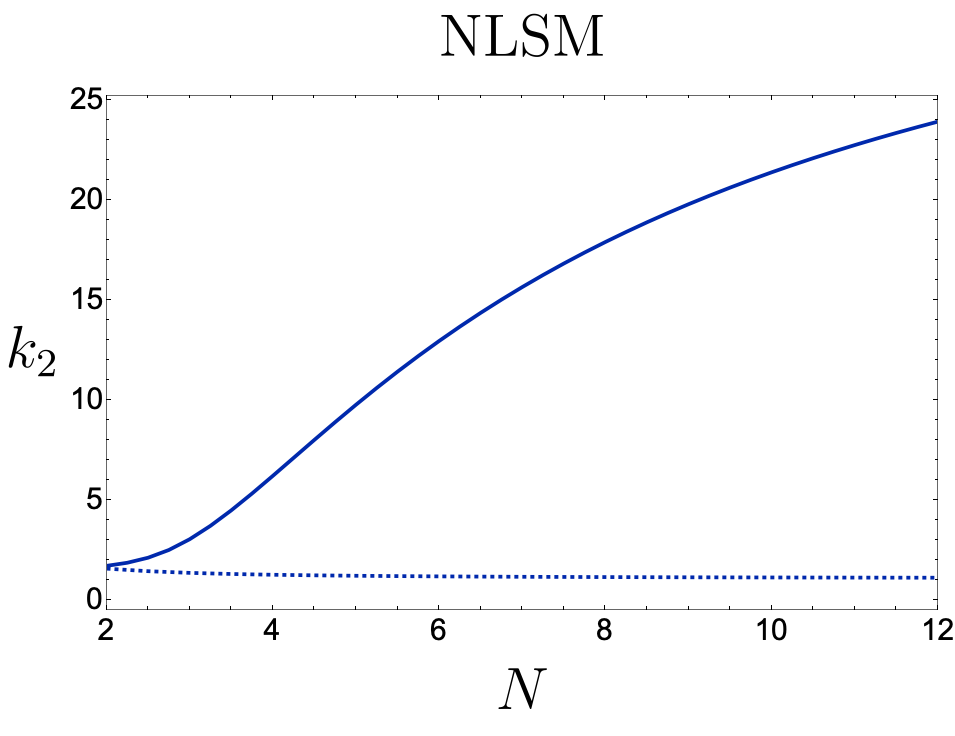}
\caption{}
\end{subfigure}
\vspace{0.2cm}
\centering
\begin{subfigure}{0.49\textwidth}
\includegraphics[width=1\textwidth]{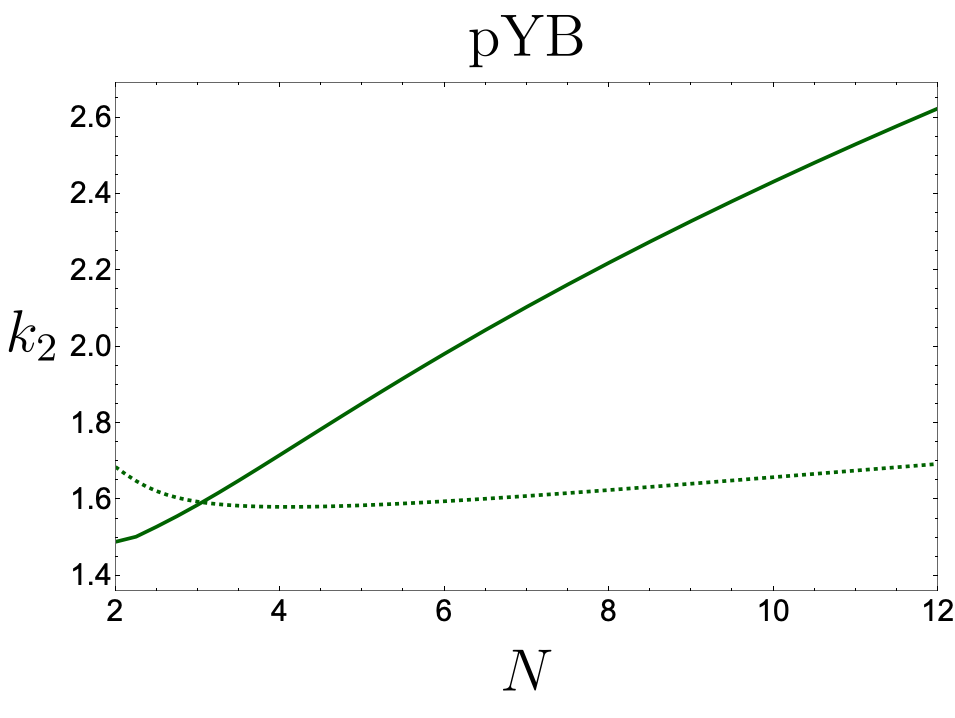}
\caption{}
\end{subfigure}
\caption{Contribution of the two-particle form factors to the central charges as a function of the number of flavors $N$. (a) Central charge $c_2$ for +NLSM (solid) and -NLSM (dashed).  (b) Central charge $k_2$ for +NLSM (solid) and -NLSM (dashed).  (c) Central charge $k_2$ for +pYB (solid) and -pYB (dashed). In the limit $N\rightarrow2$ the amplitudes $\pm$NLSM and $\mp$pYB coincide, so that the corresponding values $k_2(N=2)$ agree.
}
\label{fig:c2NLSM}
\end{figure}

In figures~\ref{fig:c2NLSM} we show the central charges for $\pm$NLSM and $\pm$pYB for different values of $N$.

\begin{figure}[t!]
\begin{subfigure}{0.49\textwidth}
\includegraphics[width=1\textwidth]{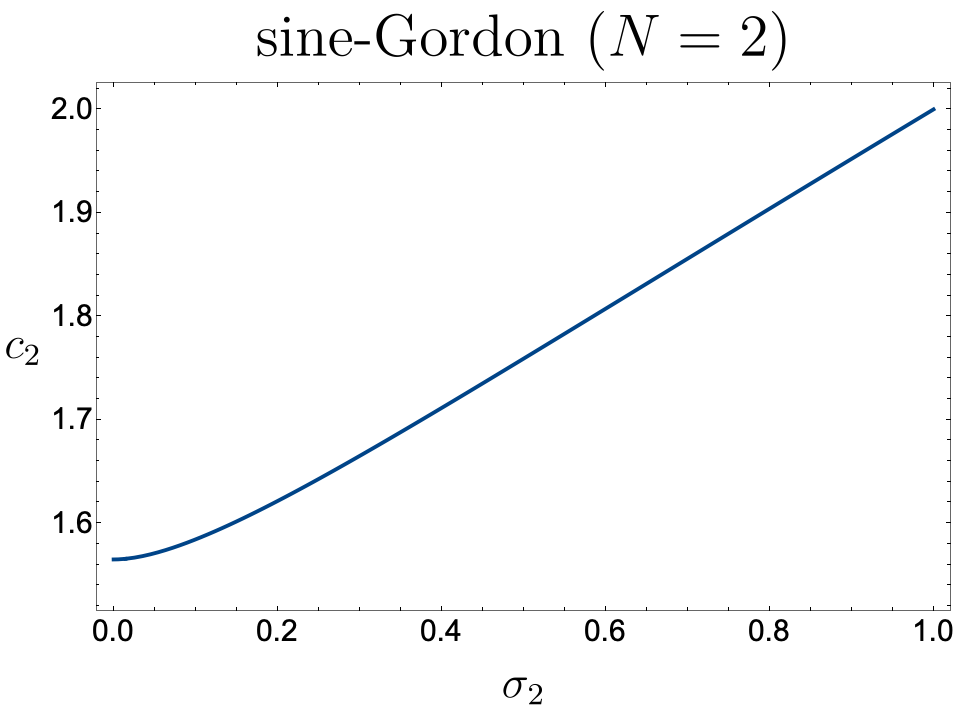}
\caption{}
\end{subfigure}
\begin{subfigure}{0.49\textwidth}
\includegraphics[width=1\textwidth]{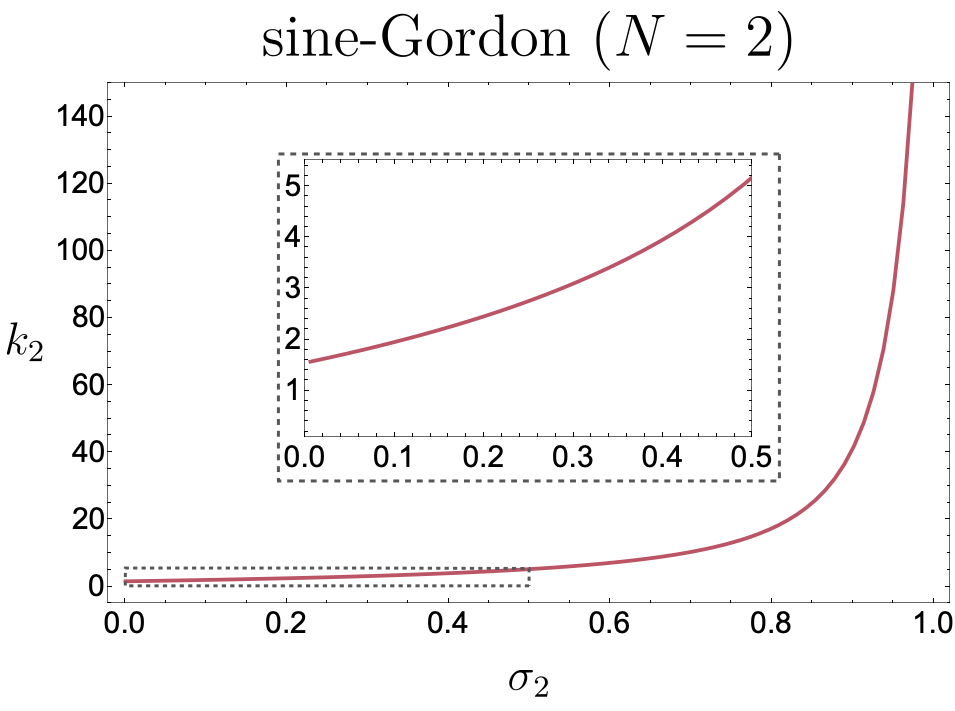}
\caption{}
\end{subfigure}
\caption{Central charges $c_2$ and $k_2$ for the sine-Gordon model with $\gamma\geq8\pi$. The horizontal axis is the position $\sigma_2$ along the $N=2$ monolith as we vary the coupling $\gamma$ from $8\pi$ ($\sigma_2=1$) to infinity ($\sigma_2=0$).
}
\label{fig:c2k2sG}
\end{figure}

As a last example we show how to get the two-particle contributions to $c$ and $k$ for the sine-Gordon model in the regime where the spectrum contains only kinks $\gamma\geq 8\pi$. As shown in appendix~E of \cite{Cordova:2019lot}, the kinks/antikinks amplitudes cover the boundary of the $N=2$ monolith, more precisely the two-dimensional section spanned by $\sigma_i(s=2m^2)$. Using the kernels \cite{Karowski:1978vz}
\begin{align}
    f(t)^\text{sG}_\bullet&= -\frac{2 \left(\sinh \left(\frac{t}{2}\right)-\sinh \left(\left(\frac{3}{2}-\frac{\gamma }{8 \pi }\right) t\right)\right)}{\sinh \left(\frac{t}{2}-\frac{\gamma  t}{8 \pi }\right)-\sinh \left(\frac{\gamma  t}{8 \pi }+\frac{t}{2}\right)}\,,\\
    f(t)^\text{sG}_A&= \frac{2 \left(-2 \sinh \left(\frac{1}{8} \left(4-\frac{\gamma }{\pi }\right) t\right)+\sinh \left(\frac{(\gamma +4 \pi ) t}{8 \pi }\right)+\sinh \left(\frac{3 t}{2}\right)\right)}{\sinh \left(\frac{t}{2}-\frac{\gamma\,  t}{8 \pi }\right)-\sinh \left(\frac{\gamma\,  t}{8 \pi }+\frac{t}{2}\right)}\,.
\end{align}
we can get the central charges for the boundary in the upper quadrant of the monolith $\sigma_{1,2}>0$ as shown in figure~\ref{fig:c2k2sG}. We see the interpolation from the $N=2$ limits of +pYB and -NLSM corresponding to $\sigma_2=0$ and $\gamma\rightarrow\infty$ and free bosons with $\sigma_2=1$ and  $\gamma=8\pi$ (c.f. table~\ref{table:c2k2values}). The rest of the sections at the boundary can be computed from the mappings described in \cite{Cordova:2019lot}.

\newpage
\section{Notation and useful formulae}

\label{sec:conventions}

\subsection{Conventions and derivation of inner products}\label{app:conventions}

Our conventions for particle states follow from \cite{Karateev:2019ymz}. In particular one particle states are defined as unitary irreducible representations of the 2 dimensional Poincar\'e group $ISO(1,1)$ with discrete eigenvalues of $P^2$, and are normalized as 
\be
{}_a \langle p_1 | p_2 \rangle_b = 2 p_1^0 \delta_{ab} 2 \pi \delta( \bm{p}_1-\bm{p}_2),
\label{eq:states_irrep}
\ee
where $a=1,...,N$ is a color index in the vector representation of O(N) and $p = (p^0, \bm{p})$. Two particle states are decomposed into singlet, antisymmetric and traceless symmetric irreducible representations of O(N) as 

\be 
\ket{p_1 p_2}_{ab} = \frac{ \delta_{ab}}{\sqrt{N}} \ket{p_1,p_2}^\bullet + \ket{p_1,p_2}^A_{[a,b]} + \ket{p_1,p_2}^S_{(a,b)}
\ee
where 
\be
\ket{p_1,p_2}^\bullet &\equiv \frac{1}{\sqrt{N}} \ket{p_1 p_2}_{a,a}, \\
\ket{p_1,p_2}^A_{[a,b]} &\equiv \frac{1}{2} (\ket{p_1,p_2}_{a,b}- \ket{p_1,p_2}_{b,a}),\\ 
\ket{p_1,p_2}^S_{(a,b)} &\equiv \frac{1}{2} (\ket{p_1,p_2}_{a,b}+ \ket{p_1,p_2}_{b,a})-\frac{1}{\sqrt{N}} \ket{p_1 p_2}_{a,b}.
\ee 

The two particle form factor of the trace of the stress energy tensor $\Theta \equiv T^\mu_\mu$ is 
\be F_\Theta (s) \equiv {}_{out} \langle 0 | \Theta(0) | p_1,p_2\rangle_{\bullet}^{in}
\ee 
and its spectral density is 
\be \rho_\Theta(s) \equiv \int d^2x e^{-ip \cdot x} \langle 0 | \Theta(x) \Theta(0) | 0 \rangle. \ee
The 2 particles form factors of conserved currents, being in the adjoint, only take non vanishing values with states in the antisymmetric representation and are defined as 
\begin{equation}
    F^\mu_{ab,cd}(p_1,p_2) \equiv _{out} \bra{0} J^\mu_{[ab]}(0) \ket{p_1,p_2}^{A}_{[cd],in} = i T^A_{ab,cd}(p_1^\mu-p_2^\mu)F_J(s),
\end{equation}

where the second equality follows from Lorentz and O(N) symmetry and conservation of the currents\footnote{The second Lorentz vector that could appear would be $p_1^\mu + p_2^\mu$, but the conservation equation $\partial_\mu J^\mu =0$ implies $(p_1+p_2)_\mu F^\mu_{ab,cd}=0$, which follows from differentiating $_{out} \bra{0} J^\mu_{[ab]}(x) \ket{p_1,p_2}^{A}_{[cd],in}= e^{i (p_1+p_2)\cdot x} F^\mu_{ab,cd}$.}, and we used $T^A_{ab,cd} \equiv \frac{1}{2}( \delta_{ac} \delta_{bd} - \delta_{ad} \delta_{bc} )$.
Similarly the spectral density of those currents reads 
\begin{equation}
     2\pi \rho^{\mu \nu}_{J;ab,cd} \equiv \int d^2 x e^{-ipx} \bra{0} J^\mu_{[ab]}(x)J^\nu_{[cd]}(0) \ket{0}  = T^A_{ab,cd} 2\pi \rho_J(s) (p^\mu p^\nu-\eta^{\mu \nu}p^2).
\end{equation}

 We want to compute it in terms of the form factors. Inserting a complete set of states between the currents, focusing on the two particles contribution sector, translating the operators and changing variables with 
 \be 
 \frac{1}{2} \frac{d \bm{p}}{2p^0 (2\pi)}\frac{d \bm{q}}{ 2q^0 (2\pi)} =  \frac{1}{\mathcal{N}_2} \frac{ d^2 (p+q)}{(2\pi)^2}
 \ee 
 we get for $s>4m^2$ 

 \be 
  T^A_{ab,cd} 2\pi \rho_J(s) (p^\mu p^\nu-\eta^{\mu \nu}p^2) = \frac{1}{\mathcal{N}_2} T^A_{ab,mn}T^A_{cd,mn}(p_1^\mu-p_2^\mu)(p_1^\nu-p_2^\nu) |F_J(s)|^2 \theta(s-4m^2) +...
 \ee 
 where $\theta$ is the Heaviside step function and the dots indicate contributions from higher particles form factors. We will denote $\rho_J^{(2)}(s)$ for the two particles contribution only.
 Using  $T^A_{ab,mn}T^A_{cd,mn} = T^A_{ab,cd}$, contracting both sides with $q^\nu \equiv (p_1^\nu-p_2^\nu)$ and collecting the components of $q^\mu$ we get
 \be 
 -2\pi \rho_J^{(2)} p^2 = \frac{1}{\mathcal{N}_2} q^2 |F_J|^2 \theta(s-4m^2).
 \ee
  We can finally use $p^2 = -s$ and $q^2 = s-4m^2$ to get 
 \be
  2\pi \rho_J^{(2)} = \frac{1}{\mathcal{N}_2} \frac{s-4m^2}{s} |F_J(s)|^2 \theta(s-4m^2)
\ee 
This is just the equivalent of Watson's equation that can be derived by a similar procedure for the stress tensor form factor and reads
\be 2\pi \rho_\Theta^{(2)} = \frac{1}{ \mathcal{N}_2} |F_\Theta|^2.\ee

We now have everything to derive the inner products for the components of the matrices $B_a$ in \eqref{eq:uniB}. The upper 2x2 blocks and the entries of the $B_\bullet$ matrix were already computed in section 3.1 in \cite{Karateev:2019ymz}. Here we focus on the remaining components of the $B_A$ matrix.

The entry (3,3) is 
\begin{equation}
\begin{split}
     _{[ab]}\braket{\psi_3}{\psi_3}_{[cd]} &= \int d^2x d^2y e^{-ipx+ip'y} \frac{q_\mu q'_\nu}{q^4} \bra{0} J^{\mu}_{[ab]}(x)J^{\nu}_{[cd]}(y) \ket{0} \\&  = (2\pi) \delta^{(2)}(p-p') T^A_{ab,cd} \frac{q_\mu q_\nu}{q^4} (2\pi)\rho^{\mu \nu}_J\\
     &=  (2\pi)^2 \delta^{(2)}(p-p') T^A_{ab,cd} \frac{(- p^2)}{q^2}(2\pi)\rho_J(s).
     \end{split}
\end{equation}
Note that 
\begin{equation}
    s=-(p_1+p_2)^2, \qquad q^2 =(p_1-p_2)^2 =p^2 -4 p_1\cdot p_2 = s-4m^2.
\end{equation}
We therefore have 
\begin{equation}
     _{[ab]}\braket{\psi_3}{\psi_3}_{[cd]} = (2\pi)^2 \delta^{(2)}(p-p') T^A_{ab,cd} \frac{s}{s-4m^2}(2\pi)\rho_J(s)
\end{equation}
The off diagonal components (2,3) are 
\begin{equation}
\begin{split}
    _{[ab]}\bra{\psi_2}\ket{\psi_3}_{[cd]} &= \int d^2x e^{ipx} \frac{\omega}{q^2} \ {}^{out}_{[ab]}\bra{p'_1 p'_2} q_\nu J^\nu_{[cd]}(x) \ket{0} = \int d^2 x \frac{\omega}{q^2} e^{ix(p-p')} q_\nu F^\nu_{ab,cd}(p_1',p_2')\\
    &= (2\pi)^2 \delta^{2}(p-p') i T^A_{ab,cd}  \omega F_J(s).
\end{split}
\end{equation}
Similarly we have for the entry (1,3)\footnote{Note that the order of the particles is important for the antisymmetric channel, and since $ (\ket{p_1,p_2})^\dagger = \bra{p_2,p_1}$ we get an extra minus sign when permuting to get the same tensor structure.}
\begin{equation}
\begin{split}
    _{[ab]}\bra{\psi_1}\ket{\psi_3}_{[cd]} &= \int d^2x e^{ipx} \frac{\omega}{q^2} \ {}^{in}_{[ab]}\bra{p'_1 p'_2} q_\nu J^\nu_{[cd]}(x) \ket{0} = \int d^2 x \frac{\omega}{q^2} e^{ix(p-p')} q_\nu (F^\nu_{ba,cd})^*(p_1',p_2')\\
    &= (2\pi)^2 \delta^{2}(p-p') i T^A_{ab,cd}  \omega F_J^*(s).
\end{split}
\end{equation}

\subsection{Sum rules}\label{app:sumrules}

Using complex coordinates the conservation of the currents becomes
\be 
\partial_\mu J^\mu =0 \implies \bar{\partial} J + \partial \bar{J}=0, \ee
where $\partial \equiv \partial_z$, $\bar{\partial} \equiv \partial_{\bar{z}}$, $J \equiv J_z$ and $\bar{J} \equiv J_{\bar{z}}$. By rotational invariance we can then write\footnote{The factor of $\pi^2$ extracted from $G(\bar{z}z)$ will make the resulting sum rule agree with the free fermion result (and subsequently all other analytical results).}
\be \langle J(z,\bar{z}) J(0) \rangle = \frac{F(z \bar{z})}{z^2}, \qquad \langle \bar{J}(z,\bar{z}) J(0) \rangle = \pi^2 \frac{G(z \bar{z})}{z \bar{z}}. \ee
At a conformal fixed point these functions become 
\be F_{CFT} = k_{CFT}, \qquad G_{CFT}=0. \ee
The conservation equation implies 
\be \langle \bar{\partial}J(z,\bar{z}) J(0) \rangle + \langle \partial \bar{J}(z,\bar{z}) J(0) \rangle = 0 \implies F' + \pi^2 G' = \pi^2 \frac{G}{\bar{z} z}. \ee
Integrating both sides we obtain a sum rule for $k$
\be k_{UV}- k_{IR} = \pi^2 \int_0^\infty dr^2 \frac{G(r^2)}{r^2} = \pi \int d^2 z\, \langle \bar{J}(z,\bar{z}) J(0) \rangle. \ee
We can relate the RHS to the spectral density $\rho_J$ by considering the (Euclidean) K{\"a}ll{\'e}n-Lehmann spectral decomposition of a spin 1 field \cite{Karateev:2020axc} 
\be \langle J_\mu(x) J_\nu(0) \rangle =    \int_0^\infty ds\, \rho_J(s)\left(  \delta_{\mu \nu} s - \partial_\mu \partial_\nu \right) \Delta_E(x;s), \ee
where the Euclidean scalar propagator is 
\be \Delta_E(x;s) \equiv \int \frac{d^2 p}{(2\pi)^2} \frac{e^{ip\cdot x}}{p^2+s}. \ee
Evaluating the trace in complex coordinates we get 
\be 4  \langle \bar{J}(z,\bar{z}) J(0) \rangle =   \int_0^\infty ds\, \rho_J(s) \left(2s- \Box \right) \Delta_E(x;s). \ee 
The sum-rule therefore becomes\footnote{Here we used that the spectral density has non vanishing support only from the two-particle threshold $s \geq 4m^2$. In principle there is also a contribution from the stable particle at $s=m^2$, but it is proportional to the one particle form factor $|F_1|^2$ that here vanishes due to O(N) symmetry. This is also true for the $c$ sum rule.}
\be k_{UV}- k_{IR} = \frac{\pi}{4} \int_0^\infty ds\  \rho_J(s)  \int d^2p \frac{2s+p^2}{p^2+s} \int \frac{d^2x}{(2\pi)^2}e^{ip\cdot x} = \frac{\pi}{2} \int_{4m^2}^\infty ds\, \rho_J(s). \label{eq:sumrulek}\ee 
As shown in more details in \cite{Correia:2022dyp,Karateev:2020axc} we can do the same for the stress tensor where conservation reads 
\begin{equation}
    \bar{\partial} T + \frac{\pi}{2} \partial \Theta = 0,
\end{equation}
and rotational invariance gives 
\begin{equation}
        \langle T(z, \bar{z}) T(0,0) \rangle = \frac{F(z \bar{z})}{z^4}, \qquad
         \langle T(z, \bar{z}) \Theta(0,0) \rangle = \frac{G(z \bar{z})}{z^3 \bar{z}}, \qquad
          \langle \Theta(z, \bar{z}) \Theta(0,0) \rangle = \frac{H(z \bar{z})}{z^2 \bar{z}^2},
          \label{invrot}
\end{equation}
Using the UV behaviour 
\begin{equation}
    F_{UV} = \frac{c}{2}, \qquad G_{UV}=H_{UV}=0,
    \label{F_UV}
\end{equation}
similar steps lead to 
\begin{equation}
    c = 12\pi  \int_{4m^2}^\infty ds \frac{\rho_\Theta(s)}{s^2} ,
    \label{c_sum_rule}
\end{equation}

\subsection{Normalization of form factors}\label{app:norm}
The currents are related to conserved charges by 
\begin{equation}
    Q_{[ab]} \equiv \int d \bm{x}\, J^0_{[ab]}(\bm{x},t).
\end{equation}
Those conserved charges generate O(N) transformations on the Hilbert space and therefore act on one particle states as 
\begin{equation}
    Q_{[ab]}\ket{p}_c = -i \delta_{bc} \ket{p}_a +i \delta_{ac} \ket{p}_b.
\end{equation}
Different normalizations for this symmetry will lead to different normalizations for the form factor. In particular let us consider the matrix element 
\begin{equation}
     \mathcal{M} = \sum_{a,b}\int d \bm{x} \tensor[_b]{\bra{p_1} J^0_{[ab]}(\bm{x},0) \ket{p_2}}{_a}.
\end{equation}
On the one hand $\mathcal{M}$ can be evaluated by using the conserved charge and its action on one particle states as 
\begin{align}
    \mathcal{M} &=\ {}_b\langle{p_1} |Q_{[ab]}| {p_2}\rangle_a = \sum_{a,b} (-i \delta_{ab} \ \ {}_b\langle{p_1} | {p_2} \rangle_a+ i \delta_{aa}\ {}_b\langle{p_1} | {p_2} \rangle_b) \\ &= i (N^2-N) (2\pi) (2 p_1^0) \delta(\bm{p}_1-\bm{p}_2).
\end{align}
On the other hand we can translate the current with $J(x) = e^{-iPx}J(0) e^{iPx}$ and use the analytic continuation $p_2\to -p_2$ to get 
\begin{equation}
    \mathcal{M} = \int d \bm{x} e^{i \bm{x} (\bm{p_1}-\bm{p_2})} F^0_{ab,ab}(p_1,-p_2) = (2\pi) \delta(\bm{p}_1-\bm{p}_2) i \frac{N(N-1)}{2} (p_1^0 +p_2^0) F_J(4m^2-s).
\end{equation}
Comparing the two results we obtain 
\begin{equation}
    \delta(\bm{p}_1-\bm{p}_2) \left( \frac{1}{2} F_J(4m^2-s) (p_1^0+p_2^0) (N^2-N) - 2p_1^0(N^2-N) \right) = 0 
\end{equation}
Evaluating this constraint in the center of mass frame, we therefore deduce
\begin{equation}
    F_J(s=0) = 2. \label{eq:normFA}
\end{equation}
The case of the stress tensor is similar and derived by starting from 
\begin{equation}
    P^\mu \ket{ p} = \int dx T^{0\mu}(x) \ket{p} = p^\mu \ket{p}.
\end{equation}
We then consider the matrix element 
\be \mathcal{M}^\mu = \sum_{a,b} \int d \bm{x} \ {}_b \langle p_1 | T^{0 \mu} (\bm{x},0) | p_2 \rangle_a
\ee 
We then have on one hand, using the charge $P^\mu$, 
\be \mathcal{M}^\mu = N \delta( \bm{p}_1-\bm{p}_2) 2p_1^0 p_1^\mu, \ee
and on the other hand, using the definition of form factors,
\be \mathcal{M}^\mu =  2\pi \sqrt{N} \delta( \bm{p}_1-\bm{p}_2) F_{T^{\mu 0}}(p_1,-p_2). \ee 
Then using that by Lorentz invariance 
\begin{equation}
 F_{T^{\mu \nu}}(s) = B(s) q^\mu q^\nu, \qquad F_\Theta(s) = (s-4m^2)B(s),
  \label{fnorm1}
\end{equation}
where we  used $q^\mu=p_1^\mu-p_2^\mu$, we get 
\begin{equation}
    \left( \sqrt{N} B(s-4m^2) (p_1^\mu +p_2^\mu) - N p_2^\mu  \right) \delta(\bm{p}_1-\bm{p}_2) = 0,
\end{equation}
and therefore $B(0) = \sqrt{N}/2$ leading to 
\begin{equation}
    F_\Theta(0) = - 2 \sqrt{N} m^2. \label{eq:normF0}
\end{equation}

\section{Dual setup and optimization problems}
\label{app:dualproblems}
\subsection{General setup}

Let us now discuss the dual implementation. We begin by analyticity and crossing. Defining the dual function $W_a$ to be crossing antisymmetric, $W_a(4-s)C_{ab}=-W_b(s)$, with a cut for $s \geq 4$ and a pole at $s = 2$ (and the crossing symmetrics of those singularities), and sufficiently fast decay at infinity, we get the dispersion relation
\begin{equation}
0 = \oint \frac{1}{2\pi i} W_a S_a = -S_a(2) \mathop{\mathrm{Res}}_{s = 2} W_a(s) + \frac{2}{\pi} \int_4^\infty ds \operatorname{Im} (W_a S_a ).
\end{equation}
A similar argument for the form factor $F \equiv F_{\Theta / J}$ gives
\begin{equation}
0 = \oint \frac{1}{2\pi i} W_F F = -F_\Theta(0) \mathop{\mathrm{Res}}_{s = 0} W_F(s) + \frac{1}{\pi} \int_4^\infty ds \operatorname{Im} ( W_F F ),
\end{equation}
where $W_F$ is analytic except for a pole at $s=0$ and a branch point at $s=4$, with a cut along the positive real axis.
Note that the residues on the poles are fixed by blowing up $0 = \oint W_a ds = -2\pi i \mathrm{Res}_{s=2} W_a + 2i(1+C^T) \int_4^\infty \operatorname{Im} W_a ds$, and $0 = \oint W_F ds = -2\pi i \mathrm{Res}_{s=0} W_F + 2i \int_4^\infty \operatorname{Im} W_F ds$. 

Now we discuss unitarity for which the constraint $B_a \succeq 0$ is written with Lagrange multipliers packaged in matrices $\lambda_a$ \cite{Guerrieri:2020kcs}
\begin{equation}
    \int_4^\infty ds\  \operatorname{Tr} B_a(s) \Lambda_a(s),
\end{equation}
where $\Lambda_a$ are hermitian and negative semidefinite parametrized as 
\begin{equation}
    \Lambda_\bullet = \begin{pmatrix} \lambda_{1,\bullet} & \lambda_{4,\bullet} &  \lambda_{6,\bullet}  \\  \lambda_{4,\bullet}^*  &  \lambda_{2,\bullet}  &  \lambda_{5,\bullet}  \\  \lambda_{6,\bullet}^*  &  \lambda_{5,\bullet}^*  &  \lambda_{3,\bullet}  \end{pmatrix}, \qquad \Lambda_A = \begin{pmatrix} \lambda_{1,A} & \lambda_{4,A} &  \lambda_{6,A}  \\  \lambda_{4,A}^*  &  \lambda_{2,A}  &  \lambda_{5,A}  \\  \lambda_{6,A}^*  &  \lambda_{5,A}^*  &  \lambda_{3,A}  \end{pmatrix}, \qquad \Lambda_\textbf{S} = \begin{pmatrix} \lambda_{1,S} & \mu_S \\ \mu_S^* & \lambda_{2,S}  \end{pmatrix}.
\end{equation}
Some components can be eliminated by noticing that at the optimum we need to satisfy $\operatorname{Tr} \Lambda B =0 \implies \Lambda B=0$. Solving those equations for the 3x3 matrices gives
\begin{equation}
\begin{gathered}
    \lambda_{5,\bullet} = \lambda_{6,\bullet}^*, \qquad \lambda_{1,\bullet} = \lambda_{2,\bullet} \equiv \lambda_\bullet,  \\
    \lambda_{5,A} = -\lambda_{6,A}^*, \qquad \lambda_{1_A} = \lambda_{2,A} \equiv \lambda_A, \\ 
  2\lambda_{1,a} |\lambda_{6,a}|^2 - \lambda_{3,a} \lambda_{1,a}^2 - 2 \operatorname{Re}(\lambda_{4,a}^* \lambda_{6,a}^2)+ \lambda_{3,a}|\lambda_{4,a}|^2 = 0, \qquad a= \bullet,A.
    \label{nonlinearconstraint3x3}
\end{gathered}
\end{equation}
For the 2x2 matrices the equations fix 
\begin{equation}
    \lambda_S \equiv \lambda_{1,S}=\lambda_{2,S} =-|\mu_S|.
    \label{nonlinearconstraint2x2}
\end{equation}
The unitarity constrain can then be expanded 
\begin{equation}
\begin{gathered}
    \operatorname{Tr} \Lambda_\bullet B_\bullet = 2 \lambda_\bullet + 2 \operatorname{Re} \lambda_{4,\bullet} S_\bullet + 4 \omega \operatorname{Re} \lambda_{6,\bullet} F_\Theta + 2 \pi \rho_\Theta \lambda_{3, \bullet}, \\ 
    \operatorname{Tr} \Lambda_A B_A = 2 \lambda_A + 2 \operatorname{Re} \lambda_{4,A} S_A + 4 \omega \operatorname{Im} \lambda_{6,A} F_J + 2 \pi \frac{s}{s-4} \rho_J \lambda_{3,A}, \\
    \operatorname{Tr} \Lambda_S B_S = 2 \lambda_S + 2 \operatorname{Re} \mu_S S_S.
\end{gathered}
\end{equation}
The remaining equations coming from $\Lambda B=0$ give the primal observables as function of the optimal dual functions as 
\begin{equation}
\begin{gathered}
    S_\bullet = \frac{\lambda_{6,\bullet}^* \lambda_\bullet - \lambda_{6,\bullet}\lambda_{4,\bullet}^*}{\lambda_{6,\bullet} \lambda_\bullet- \lambda_{6,\bullet}^*\lambda_{4,\bullet}}, \qquad \omega F_\Theta = \frac{|\lambda_{4,\bullet}|^2-\lambda_\bullet^2}{\lambda_{6,\bullet} \lambda_\bullet - \lambda_{6,\bullet}^*\lambda_{4,\bullet}}, \qquad 2\pi \rho_\Theta = |\omega F_\Theta|^2, \\
    S_A = -\frac{\lambda_{6,A}^* \lambda_A + \lambda_{6,A}\lambda_{4,A}^*}{\lambda_{6,A} \lambda_A+ \lambda_{6,A}^*\lambda_{4,A}}, \qquad i \omega (s-4) F_J = \frac{\lambda_{A}^2-|\lambda_{4,A}|^2}{{\lambda_{6,A}} \lambda_A + \lambda_{6,A}^*\lambda_{4,A}}, \qquad 2\pi \rho_J = \frac{s-4}{s}|\omega F_J|^2, \\  
     S_\textbf{S} = -\frac{\mu_\textbf{S}^*}{\lambda_\textbf{S}} .
\end{gathered}
\end{equation}

\subsection{\texorpdfstring{Minimization of the central charge $c$ on the monolith}{Minimization of the central charge c on the monolith}}
To minimize the central charge $c$ on top of the monolith we start with the constrained problem
\begin{equation}
\begin{split}
    \mathcal{L} &= 12 \pi \int_4^\infty \frac{\rho_\Theta(s)}{s^2}  - \pi F_\Theta(0) \mathop{\mathrm{Res}}_{s = 0} W_{F_\Theta}(s) -\frac{\pi}{2}S_a(2) \mathop{\mathrm{Res}}_{s = 2} W_a(s)\\ &+ \int_4^\infty ds \left( \operatorname{Im}(W_aS_a + W_{F_\Theta} F_\Theta) + \operatorname{Tr}\Lambda_a B_a \right).
\end{split}
\end{equation}
The equations of motion coming from functional variations with respect to the primal functions give 
\begin{gather}
    \lambda_{4,\bullet} = \frac{i}{2} W_\bullet, \qquad \lambda_{6,\bullet} = \frac{i}{4 \omega} W_{F_\Theta}, \qquad \lambda_{3,\bullet} = - \frac{6}{s^2}, \\
    \mu_A = \frac{i}{2} W_A, \qquad \mu_S = \frac{i}{2} W_S.
\end{gather}

In terms of the coordinates $(\sigma_1, \sigma_2)$ on the monolith (at $s=2$), the Lagrangian  becomes
\begin{equation}
\begin{split}
    \mathcal{L} &= 2\pi \sqrt{N} \mathop{\mathrm{Res}}_{s = 0} W_{F_\Theta} - \frac{\pi}{2}((N+1)\sigma_1+\sigma_2)\mathop{\mathrm{Res}}_{s = 2} W_\bullet -\frac{\pi}{2} (\sigma_1+\sigma_2)\mathop{\mathrm{Res}}_{s = 2} W_S\\& - \frac{\pi}{2} (\sigma_2-\sigma_1)\mathop{\mathrm{Res}}_{s = 2} W_A + \int_4^\infty (2\lambda_\bullet +2\lambda_A +2\lambda_{S}).
\end{split}
\end{equation}
This leads us to the dual problem ready to be implemented in SDPB : 

\underline{Dual Problem (Minimization of the central charge $c$ on top of the monolith)} 
\begin{equation}
\underset{\{\lambda_a, W_a, W_{F_\Theta}\}}{\text{Maximize}}\mathcal{L}
\end{equation}
Constrained by  
\begin{equation}
    \begin{pmatrix}
    \lambda_\bullet& \frac{i}{2} W_\bullet &  \frac{i}{4 \omega} W_{F_\Theta} \\ -\frac{i}{2} W_\bullet^*  & \lambda_\bullet & - \frac{i}{4 \omega} W_{F_\Theta}^* \\ -\frac{i}{4 \omega} W_{F_\Theta}^* &  \frac{i}{4 \omega} W_{F_\Theta} & -\frac{6}{s^2}
    \end{pmatrix} \preccurlyeq 0, \qquad \begin{pmatrix}
    \lambda_A & \frac{i}{2} W_A  \\ -\frac{i}{2} W_A^*  & \lambda_A 
    \end{pmatrix} \preccurlyeq 0,\qquad \begin{pmatrix}
    \lambda_S & \frac{i}{2} W_S  \\ -\frac{i}{2} W_S^*  & \lambda_S
    \end{pmatrix} \preccurlyeq 0.
\end{equation}

\subsection{\texorpdfstring{Minimization of the current central charge $k$ on the monolith}{Minimization of the current central charge k on the monolith}}
Analogously to the $c$ case we start with
\begin{equation}
\begin{split}
    \mathcal{L} &= \frac{\pi}{2} \int_4^\infty \rho_J(s)  - \pi F_J(0) \mathop{\mathrm{Res}}_{s = 0} W_{F_J}(s) -\frac{\pi}{2}S_a(2) \mathop{\mathrm{Res}}_{s = 2} W_a(s)\\ &+ \int_4^\infty ds \left( \operatorname{Im}(W_aS_a + W_{F_J} F_J) + \operatorname{Tr}\Lambda_a B_a \right),
\end{split}
\end{equation}
The equations of motion of the primal functions yield
\begin{gather}
    \lambda_{4,A} = \frac{i}{2} W_A, \qquad \lambda_{6,A} = -\frac{1}{4 \omega} W_{F_J}, \qquad \lambda_{3,A} = - \frac{s-4}{4s}, \\
    \mu_\bullet = \frac{i}{2} W_\bullet, \qquad \mu_S = \frac{i}{2} W_S.
\end{gather}
The Lagrangian  becomes
\begin{equation}
\begin{split}
    \mathcal{L} &= -2\pi  \mathop{\mathrm{Res}}_{s = 0} W_{F_J} - \frac{\pi}{2}((N+1)\sigma_1^*+\sigma_2^*)\mathop{\mathrm{Res}}_{s = 2} W_\bullet -\frac{\pi}{2} (\sigma_1^*+\sigma_2^*)\mathop{\mathrm{Res}}_{s = 2} W_S\\& - \frac{\pi}{2} (\sigma_2^*-\sigma_1^*)\mathop{\mathrm{Res}}_{s = 2} W_A + \int_4^\infty (2\lambda_\bullet +2\lambda_A +2\lambda_{S}),
\end{split}
\end{equation}
and the dual problem is

\underline{Dual Problem (Minimization of $k$ on top of the monolith)} 
\begin{equation}
\underset{\{\lambda_a, W_a, W_{F_J}\}}{\text{Maximize}}\mathcal{L}
\end{equation}
Constrained by  
\begin{equation}
\begin{pmatrix}
    \lambda_\bullet & \frac{i}{2} W_\bullet  \\ -\frac{i}{2} W_\bullet^*  & \lambda_\bullet 
    \end{pmatrix} \preccurlyeq 0,
    \qquad
    \begin{pmatrix}
    \lambda_A& \frac{i}{2} W_A &  -\frac{1}{4 \omega} W_{F_J} \\ -\frac{i}{2} W_A^*  & \lambda_A &  \frac{1}{4 \omega} W_{F_J}^* \\ -\frac{1}{4 \omega} W_{F_J}^* &  \frac{1}{4 \omega} W_{F_J} & - \frac{s-4}{4s}
    \end{pmatrix} \preccurlyeq 0, 
    \qquad 
    \begin{pmatrix}
    \lambda_S & \frac{i}{2} W_S  \\ -\frac{i}{2} W_S^*  & \lambda_S
    \end{pmatrix} \preccurlyeq 0.
\end{equation}

\subsection{\texorpdfstring{Minimization of $k$ for fixed $c$}{Minimization of k for fixed c}}

We start with
\begin{equation}
\begin{split}
    \mathcal{L} &= \frac{\pi}{2} \int_4^\infty \rho_J(s)   - \pi F_J(0) \mathop{\mathrm{Res}}_{s = 0} W_{F_J}(s) - \pi F_\Theta(0) \mathop{\mathrm{Res}}_{s = 0} W_{F_\Theta}(s) \\ &+ \int_4^\infty ds \left( \operatorname{Im}(W_aS_a + W_{F_J} F_J + W_{F_\Theta} F_\Theta) + \operatorname{Tr}\Lambda_a B_a \right) \\
     & + C_\rho (c_{UV}- 12 \pi \int_4^\infty \frac{\rho_\Theta(s)}{s^2}),
\end{split}
\end{equation}
The equations of motion of the primal functions are
\begin{gather}
\lambda_{4,\bullet} = \frac{i}{2} W_\bullet, \qquad \lambda_{6,\bullet} = \frac{i}{4 \omega} W_{F_\Theta}, \qquad \lambda_{3,\bullet} = - C_\rho\frac{6}{s^2}, \\
    \lambda_{4,A} = \frac{i}{2} W_A, \qquad \lambda_{6,A} = -\frac{1}{4 \omega} W_{F_J}, \qquad \lambda_{3,A} = - \frac{s-4}{4s}, \\
     \mu_S = \frac{i}{2} W_S.
\end{gather}
The Lagrangian  becomes
\begin{equation}
\begin{split}
    \mathcal{L} &=  2\pi \sqrt{N} \mathop{\mathrm{Res}}_{s = 0} W_{F_\Theta} -2\pi  \mathop{\mathrm{Res}}_{s = 0} W_{F_J} + C_\rho c_{UV}\\
     & - \frac{\pi}{2}((N+1)\sigma_1^*+\sigma_2^*)\mathop{\mathrm{Res}}_{s = 2} W_\bullet -\frac{\pi}{2} (\sigma_1^*+\sigma_2^*)\mathop{\mathrm{Res}}_{s = 2} W_S\\& - \frac{\pi}{2} (\sigma_2^*-\sigma_1^*)\mathop{\mathrm{Res}}_{s = 2} W_A + \int_4^\infty (2\lambda_\bullet +2\lambda_A +2\lambda_{S}),
\end{split}
\end{equation}
and the dual problem is

\underline{Dual Problem (Minimization of $k$ with fixed $c$)} 
\begin{equation}
\underset{\{\lambda_a, W_a, W_F, C_\rho\}}{\text{Maximize}}\mathcal{L}
\end{equation}
Constrained by  
\begin{gather}
\begin{pmatrix}
   \lambda_\bullet& \frac{i}{2} W_\bullet &  \frac{i}{4 \omega} W_{F_\Theta} \\ -\frac{i}{2} W_\bullet^*  & \lambda_\bullet & - \frac{i}{4 \omega} W_{F_\Theta}^* \\ -\frac{i}{4 \omega} W_{F_\Theta}^* &  \frac{i}{4 \omega} W_{F_\Theta} & -\frac{6 C_\rho}{s^2}
    \end{pmatrix} \preccurlyeq 0,
    \qquad
    \begin{pmatrix}
    \lambda_A& \frac{i}{2} W_A &  -\frac{1}{4 \omega} W_{F_J} \\ -\frac{i}{2} W_A^*  & \lambda_A &  \frac{1}{4 \omega} W_{F_J}^* \\ -\frac{1}{4 \omega} W_{F_J}^* &  \frac{1}{4 \omega} W_{F_J} & - \frac{s-4}{4s}
    \end{pmatrix} \preccurlyeq 0, \\
    \begin{pmatrix}
    \lambda_S & \frac{i}{2} W_S  \\ -\frac{i}{2} W_S^*  & \lambda_S
    \end{pmatrix} \preccurlyeq 0.
\end{gather}

\section{Numerical implementation}
\label{sec:numerics}
To implement this problem we need ans{\"a}tze for the variables. We use the disc variable 
\be
\rho(s,s_0) \equiv \frac{ \sqrt{4-s_0}- \sqrt{4-s}}{ \sqrt{4-s_0}+ \sqrt{4-s}}.
\ee 
\subsection{Primal}\label{app:primal}
We follow the same conventions as \cite{Karateev:2019ymz}, were we make the following ans{\"a}tze for the S-matrix, the two-particle form factor F$_{\Theta}$ and the spectral density $\rho_{\Theta}$:
\begin{align}\label{eq:anstaz_NumPrimal}
S_a(s) &= S_a(2) + \sum_{n=1}^{N_{max}} (a_a^{(n)} \rho(s,2)^n + C_{ab}a_b^{(n)} \rho(4-s,2)^n) \nonumber \\ 
F_{\Theta }(s) &= F^{Norm}_{\Theta } + \sum_{n=1}^{N_{max}} b^{(n)} \rho(s,0)^n \nonumber\\
\rho_{\Theta }(s) &=  \sum_{n=0}^{1}c^{(n)} \frac{ \rho(s)^n+ \rho(s)^{n,*}}{2} + \sum_{n=1}^{N_{max}} d^{(n)} \frac{ \rho(s)^n- \rho(s)^{n,*}}{2 i}\,.
\end{align}
The condition \eqref{eq:SmatrixCrossing} fixes $S_a(2)$ in terms of two unknown values $\sigma_{1,2}$  that characterize the position in the monolith that we want to target. For obtaining the global minimal value of $c$ we can left them as an unfixed parameter.The values $F^{Norm}_{\Theta}$ can be fixed by normalization to be $ -2\sqrt{N}$. 

The optimization problem then can then be cast using \eqref{eq:anstaz_NumPrimal} as an input in \eqref{eq:csumrule} as
\begin{equation}\label{eq:cmin_NumPrimal}
    c_{\text{min}}=6 \pi c^{(0)}  - \frac{3}{2} \pi^2 d^{(1)}
\end{equation}
In the case of the anti-symmetric current we need to modify \eqref{eq:anstaz_NumPrimal} to account for a different normalization for the two-particle form factor $F^{Norm}_J=2$ and re-scale the spectral density by
\begin{equation}
    \rho_{J}=\frac{\rho_\Theta(s)}{s^2}\,,
\end{equation}
to obtain a finite constrain for the sum-rule. Again to obtain the minimal value of $\eqref{eq:ksumrule}$ we can write the minimization requirement as
\begin{equation}\label{eq:Kmin_NumPrimal}
    k_{\text{min}}=\frac{\pi}{4} c^{(0)}  - \frac{\pi^2}{12}  d^{(1)}
\end{equation}
We can now also fix a value of $c$, using \eqref{eq:cmin_NumPrimal} to fix one of the two coefficients in the stress-enrgy spectral density and again obtain the minimal value of $k(c)$. 

\subsection{Dual}\label{app:dual}
 The crossing constraints on the dual functions are $W_a(4-s)C_{ab} = - W_b(s)$.
If we call $g_a$ the residue at $s=2$, those constraints give
\begin{equation}
    g_a = \frac{1}{2} (g_a + C_{ba}g_b) \Leftrightarrow g = \frac{1+C^T}{2}g,
    \end{equation}
    in particular $g$ is an eigenvector of $C^T$ with eigenvalue 1. A convenient basis of such eigenvectors is 

\begin{equation}
v_1 = \begin{pmatrix} -\frac{1}{N-1}\\1\\0 \end{pmatrix}, v_2 = \begin{pmatrix} \frac{1}{N-1}\\0\\1 \end{pmatrix} \implies g  = \begin{pmatrix} \frac{g_S-g_A}{N-1}\\g_A\\g_S \end{pmatrix}.
\end{equation}
Therefore we propose the ans{\"a}tze \footnote{The large $s$ behaviour compatible with the non linear constraints in \eqref{nonlinearconstraint3x3} and \eqref{nonlinearconstraint2x2}.}
\begin{equation}
    W_a(s) = g_a\frac{2}{(s-2) \sqrt{s} \sqrt{4-s}} +  \frac{1}{(s-2) \sqrt{s} \sqrt{4-s}}\sum_{n=1}^{N_{max}}\left( a_a^{(n)} \rho^n(s,2) + C_{ba} a_b^{(n)} \rho^n(4-s,2)\right),
    \end{equation}
\begin{equation}
    W_{F_{\Theta,J}}(s) =  \frac{1}{s^2 \sqrt{4-s} }\sum_{n=1}^{N_{max}}b^{(n)}_{\bullet,A} \rho^n(s,0),
    \end{equation}

\begin{equation}
    \lambda_a(s) = \frac{1}{s \sqrt{s} \sqrt{s-4}} \left( \sum_{n=0}^{1}c^{(n)}_a ( \rho(s)^n+ \rho(s)^{n,*}) + \sum_{n=1}^{N_{max}} d^{(n)}_a i ( \rho(s,0)^n- \rho(s,0)^{n,*}) \right),
    \label{lambda1}
\end{equation}

Therefore the optimization problems over the infinitely many parameters characterizing the functions is truncated to a finite dimensional optimization problem over the variables $g_a, a_a^{(n)}, b_a^{(n)}, c_a^{(n)}$ and $d_a^{(n)}$.

\section{Central charges for free boson and free fermion}
\label{sec:analytics}

\subsection{Free Majorana fermion}\label{app:freefermion}
 This theory can be described by the Lagrangian
 \be 
 \mathcal{L} = 2im \psi_+^a \psi_-^a + \psi_+^a\partial_+ \psi_+^a + (\partial_- \psi_-^a)\psi_-^a,
 \ee
 where $\partial_\pm = \partial_0 \pm \partial_1$ and $a=1,...,N$ is the O(N) index in the vector representation There is a O(N) symmetry under the simultaneous rotation of the two components $\psi_\pm$. The components of the Noether current are 
 \be 
 J^0_{[ab]} = \frac{1}{2}T_{[ab]}^{mn}( \psi_+^m \psi_+^n + \psi_-^n \psi_-^m), \qquad  J^1_{[ab]} = \frac{1}{2} T_{[ab]}^{mn}( \psi_+^m \psi_+^n + \psi_-^m \psi_-^n),
 \ee 
 where $T_{[ab]}^{mn} = ( \delta_{am} \delta_{bn} - \delta_{an} \delta_{bm}) = 2 T^A_{ab,mn}$  are the generators of the adjoint representation.
We compute the form factor by expanding the Majorana fields in creation and annihilation operators. The expansion comes from plane wave solutions of the Dirac equation\footnote{We work in a basis of $\gamma$ matrices \be \gamma^0 = \begin{pmatrix} 0 & 1 \\ 1 & 0 \end{pmatrix}, \qquad \gamma^1 = \begin{pmatrix} 0 & 1 \\ -1 & 0 \end{pmatrix}. \ee } and reads 
 \be 
 \psi^a_\pm(x)= \int \frac{d \bm{p}}{2p^0 (2\pi)} \left(a_p^a \sqrt{p_-} e^{ip \cdot x}  \pm a_p^{\dagger,a} \sqrt{p_+} e^{-ip \cdot x} \right),
 \ee
 where $p_\pm = p^0 \pm \bm{p}$. For the spatial component of the form factor we obtain 
 \be F^1_{ab,cd} = \frac{1}{4}[T_{[ab]}]^{mn} \int \frac{d \bm{p}}{2p^0 (2\pi)}\frac{d \bm{q}}{ 2q^0 (2\pi)} ( \sqrt{p_-q_-} + \sqrt{p_+q_+}) \bra{0}a_p^m a_q^n ( a_{p_1}^{\dagger,c}a_{p_2}^{\dagger,d}-a_{p_1}^{\dagger,d}a_{p_2}^{\dagger,c}) \ket{0}.
 \ee 
The matrix element is computed by using the anti-commutator relations
\be
\bra{0}a_p^m a_q^n  a_{p_1}^{\dagger,c}a_{p_2}^{\dagger,d} \ket{0} =\{a_p^m,a_{p_2}^{\dagger,d}\}\{a_q^n,a_{p_1}^{\dagger,c}\}- \{a_p^m,a_{p_1}^{\dagger,c}\}\{a_q^n,a_{p_2}^{\dagger,d}\}  ,
\ee 
and 
\be 
\{a_p^a,a_q^{\dagger,n}\} = i 2\pi 2p^0 \delta(\bm{p}-\bm{q}) \delta^{ab}.
\ee 
It yields 
\be F^1_{ab,cd} = - \frac{1}{2}[T_{[ab]}]^{mn} (\delta^{nc}\delta^{md}-\delta^{mc}\delta^{nd}) \left( \sqrt{p_1^+ p_2^+} + \sqrt{p_1^-p_2^-}\right).
 \ee 
To proceed we evaluate the last parenthesis in the center of mass frame (COM). We get 
 \be F^1_{ab,cd} = -4m T^A_{ab,cd}. \ee 
 This is to compare with the generic expression  
 \be 
  F^1_{ab,cd} = i T^{A}_{ab,cd} (\bm{p}_1-\bm{p}_2) F_J(s) = 2i T^{A}_{ab,cd} \bm{p} F_J(s),
  \ee 
  where the second equality holds in the COM. The expressions match if 
  \be F_J(s) = 2i \frac{m}{\bm{p}} = i \frac{4m}{\sqrt{s-4m^2}}. \ee 
  As there isn't any non-vanishing $n$ particles form factor for $n>2$, it is then straightforward to obtain the spectral density as 
   \be
   \label{fermion_kSD}
  2\pi \rho_J = \frac{1}{\mathcal{N}_2} \frac{s-4}{s} |F_J(s)|^2\theta(s-4m^2) = \frac{8m^2}{ s \sqrt{s} \sqrt{s-4m^2}}\theta(s-4m^2),
\ee 
and the corresponding value for $k$ 
\be 
 k =  \frac{\pi}{2} \int_{4m^2}^\infty ds\,\rho_J(s) = 1.
 \ee 
 For the trace of the stress energy tensor we get 
 \begin{equation}
\label{tracefermion}
    \Theta(x) = T^0_0 + T^1_1 = im\psi^a_+\psi^a_-,
\end{equation}
which following identical steps as above yields 
\begin{equation}
    F_\Theta(p_1,p_2) = \sqrt{N} m\left( \sqrt{p_{1-}p_{2+}} - \sqrt{p_{2-}p_{1+}} \right) = im \sqrt{N}\sqrt{s-4m^2},
\end{equation}
where the last equality comes fromes evaluating the expression in the COM frame. Again there is no $n>2$ form factors, so this gives the spectral density 
\begin{equation}
\label{fermionSD}
\rho_\Theta(s) = \frac{ |F_\Theta |^2}{4\pi \sqrt{s}\sqrt{s-4m^2}}\theta(s-4m^2) = \frac{ N m^2\sqrt{s-4m^2}}{4\pi \sqrt{s}}\theta(s-4m^2).
\end{equation}
We therefore deduce the central charge $c$
\be c = \int_{4m^2}^\infty ds \frac{\rho_\Theta}{s^2} = \frac{N}{2}. \ee 
Let us comment about the large $s$ behaviour. For a generic operator of conformal dimension $\Delta$ we expect the high energy regime to be dictated by the CFT, for which we obtain  \cite{Karateev:2019ymz} 
\be  \lim_{s \to \infty} \rho_\mathcal{O}(s) = \text{const} \times  s^{\Delta-d/2}, \qquad \lim_{s \to \infty} F_\mathcal{O}(s) \lesssim s^{1+ \frac{\Delta-d}{2}}.
\label{constraintCFT}
\ee 
The spectral densities \eqref{fermion_kSD} and \eqref{fermionSD} seem to be in contradiction with this result. It is because our operators $J^\mu$ and $\Theta$ are not generic operators. For $J$, it is a spin 1 conserved current for which the "const" in the above equation, being proportional to $\Delta_J-1$, vanishes \cite{Karateev:2020axc}, and indeed our result $\rho_J$ vanishes when $s \to \infty$.\footnote{As explained in \cite{Karateev:2020axc}, in general spectral densities of vector operators also have a spin 0 component, which for us was set to zero due to the conservation of the currents. This is compatible with equation (4.10) in \cite{Karateev:2020axc} where "const" is proportional to $\Delta_J-d+1$ which vanishes for conserved currents in $d=2$.} As for $\Theta$, since it is the trace of the stress energy tensor, its value is just zero in the CFT and the result \eqref{constraintCFT} does not apply.

\subsection{Free boson}\label{app:freeboson}
The free boson theory is described by the Lagrangian 
\be 
\mathcal{L} = -\frac{1}{2} \partial_\mu \phi^a \partial^\mu \phi^a - \frac{1}{2} \phi^a \phi^a.
\ee 
The Noether currents are
\be
J^\mu_{[ab]} = [T_{[ab]}]^{mn} \partial^\mu \phi^m \phi^n.
\ee
The mode expansion of the bosonic field reads 
\be
\phi^a(x) = \int \frac{d \bm{p}}{2p^0 (2\pi)} \left( a_p^a e^{ip \cdot x} + a_p^{\dagger,a} e^{-ip \cdot x} \right).
\ee
Plugging it in the definition of the form factor we get
\begin{equation}
    F^\mu_{ab,cd} = [T_{[ab]}]^{mn} \frac{i}{4} \int \frac{d \bm{p}}{2p^0 (2\pi)}\frac{d \bm{q}}{ (2\pi)}  \bra{0}a_p^m a_q^n ( a_{p_1}^{\dagger,c}a_{p_2}^{\dagger,d}-a_{p_1}^{\dagger,d}a_{p_2}^{\dagger,c}) \ket{0}
\end{equation}
The matrix element can be evaluated by using the commutator relations 
\be
\bra{0}a_p^m a_q^n  a_{p_1}^{\dagger,c}a_{p_2}^{\dagger,d} \ket{0} = [a_p^m,a_{p_1}^{\dagger,c}][a_q^n,a_{p_2}^{\dagger,d}] + [a_p^m,a_{p_2}^{\dagger,d}][a_q^n,a_{p_1}^{\dagger,c}],
\ee 
and 
\be 
[a_p^a,a_q^{\dagger,n}] = 2\pi 2p^0 \delta(\bm{p}-\bm{q}) \delta^{ab}.
\ee
We get
\be 
 F^\mu_{ab,cd} = [T_{[ab]}]^{mn} \frac{i}{2} ( \delta^{nd}\delta^{mc} - \delta^{nc}\delta^{md}) (p_1^\mu-p_2^\mu) = 2i T_{A}^{ab,cd} (p_1^\mu-p_2^\mu).
 \ee
Comparing to the general decomposition for the form factor
\be 
 F^\mu_{ab,cd} = i T^A_{ab,cd} (p_1^\mu-p_2^\mu) F_J(s),
 \ee
 we identify
 \be 
 F_J(s) = 2.
 \ee
We deduce the spectral density 
\be
\rho_J(s) = \frac{ \sqrt{s-4}}{\pi s \sqrt{s}} \theta(s-4m^2).
 \ee 
 We get $k$ by using the sum rule yielding
 \be 
 k= \frac{\pi}{2} \int_{4m^2}^\infty ds\, \rho_J = \infty,
 \ee 
which can be understood by noticing that the free boson $\phi$ is not a primary and the current two point function has a $\operatorname{log}(z)/z^2$ leading singularity.
Moving to the stress energy tensor we get 
\be 
\Theta(x) = -m^2 \phi^a \phi^a, \ee
leading to the form factor and spectral density 
\begin{equation}
\label{bosonFF}
    F_\Theta(p_1,p_2) = -2 \sqrt{N}m^2, \qquad \rho_\Theta(s) = \frac{N m^4}{\pi \sqrt{s} \sqrt{s-4m^2}}\theta(s-4m^2).
\end{equation}
The central charge $c$ is therefore 
\be
c = 12 \pi \int_{4m^2}^\infty ds \frac{\rho_\Theta}{s^2} = N. 
\ee

\end{appendices}

\bibliographystyle{JHEP}
\bibliography{ONFFrefs}

\providecommand{\href}[2]{#2}\begingroup\raggedright\begin{thebibliography}{10}

\bibitem{Paulos:2016but}
M.F.~Paulos, J.~Penedones, J.~Toledo, B.C.~van Rees and P.~Vieira, \emph{{The
  S-matrix bootstrap II: two dimensional amplitudes}},
  \href{https://doi.org/10.1007/JHEP11(2017)143}{\emph{JHEP} {\bfseries 11}
  (2017) 143} [\href{https://arxiv.org/abs/1607.06110}{{\ttfamily
  1607.06110}}].

\bibitem{Doroud:2018szp}
N.~Doroud and J.~Elias~Mir\'o, \emph{{S-matrix bootstrap for resonances}},
  \href{https://doi.org/10.1007/JHEP09(2018)052}{\emph{JHEP} {\bfseries 09}
  (2018) 052} [\href{https://arxiv.org/abs/1804.04376}{{\ttfamily
  1804.04376}}].

\bibitem{Paulos:2017fhb}
M.F.~Paulos, J.~Penedones, J.~Toledo, B.C.~van Rees and P.~Vieira, \emph{{The
  S-matrix bootstrap. Part III: higher dimensional amplitudes}},
  \href{https://doi.org/10.1007/JHEP12(2019)040}{\emph{JHEP} {\bfseries 12}
  (2019) 040} [\href{https://arxiv.org/abs/1708.06765}{{\ttfamily
  1708.06765}}].

\bibitem{He:2018uxa}
Y.~He, A.~Irrgang and M.~Kruczenski, \emph{{A note on the S-matrix bootstrap
  for the 2d O(N) bosonic model}},
  \href{https://doi.org/10.1007/JHEP11(2018)093}{\emph{JHEP} {\bfseries 11}
  (2018) 093} [\href{https://arxiv.org/abs/1805.02812}{{\ttfamily
  1805.02812}}].

\bibitem{Cordova:2018uop}
L.~C\'ordova and P.~Vieira, \emph{{Adding flavour to the S-matrix bootstrap}},
  \href{https://doi.org/10.1007/JHEP12(2018)063}{\emph{JHEP} {\bfseries 12}
  (2018) 063} [\href{https://arxiv.org/abs/1805.11143}{{\ttfamily
  1805.11143}}].

\bibitem{Guerrieri:2018uew}
A.L.~Guerrieri, J.~Penedones and P.~Vieira, \emph{{Bootstrapping QCD Using Pion
  Scattering Amplitudes}},
  \href{https://doi.org/10.1103/PhysRevLett.122.241604}{\emph{Phys. Rev. Lett.}
  {\bfseries 122} (2019) 241604}
  [\href{https://arxiv.org/abs/1810.12849}{{\ttfamily 1810.12849}}].

\bibitem{Homrich:2019cbt}
A.~Homrich, J.~Penedones, J.~Toledo, B.C.~van Rees and P.~Vieira, \emph{{The
  S-matrix Bootstrap IV: Multiple Amplitudes}},
  \href{https://doi.org/10.1007/JHEP11(2019)076}{\emph{JHEP} {\bfseries 11}
  (2019) 076} [\href{https://arxiv.org/abs/1905.06905}{{\ttfamily
  1905.06905}}].

\bibitem{EliasMiro:2019kyf}
J.~Elias~Mir\'o, A.L.~Guerrieri, A.~Hebbar, J.a.~Penedones and P.~Vieira,
  \emph{{Flux Tube S-matrix Bootstrap}},
  \href{https://doi.org/10.1103/PhysRevLett.123.221602}{\emph{Phys. Rev. Lett.}
  {\bfseries 123} (2019) 221602}
  [\href{https://arxiv.org/abs/1906.08098}{{\ttfamily 1906.08098}}].

\bibitem{Paulos:2018fym}
M.F.~Paulos and Z.~Zheng, \emph{{Bounding scattering of charged particles in
  $1+1$ dimensions}},
  \href{https://doi.org/10.1007/JHEP05(2020)145}{\emph{JHEP} {\bfseries 05}
  (2020) 145} [\href{https://arxiv.org/abs/1805.11429}{{\ttfamily
  1805.11429}}].

\bibitem{Karateev:2019ymz}
D.~Karateev, S.~Kuhn and J.~Penedones, \emph{{Bootstrapping Massive Quantum
  Field Theories}}, \href{https://doi.org/10.1007/JHEP07(2020)035}{\emph{JHEP}
  {\bfseries 07} (2020) 035}
  [\href{https://arxiv.org/abs/1912.08940}{{\ttfamily 1912.08940}}].

\bibitem{Bercini:2019vme}
C.~Bercini, M.~Fabri, A.~Homrich and P.~Vieira, \emph{{S-matrix bootstrap:
  Supersymmetry, $Z_2$, and $Z_4$ symmetry}},
  \href{https://doi.org/10.1103/PhysRevD.101.045022}{\emph{Phys. Rev. D}
  {\bfseries 101} (2020) 045022}
  [\href{https://arxiv.org/abs/1909.06453}{{\ttfamily 1909.06453}}].

\bibitem{Cordova:2019lot}
L.~C\'ordova, Y.~He, M.~Kruczenski and P.~Vieira, \emph{{The O(N) S-matrix
  Monolith}}, \href{https://doi.org/10.1007/JHEP04(2020)142}{\emph{JHEP}
  {\bfseries 04} (2020) 142}
  [\href{https://arxiv.org/abs/1909.06495}{{\ttfamily 1909.06495}}].

\bibitem{Kruczenski:2020ujw}
M.~Kruczenski and H.~Murali, \emph{{The R-matrix bootstrap for the 2d O(N)
  bosonic model with a boundary}},
  \href{https://doi.org/10.1007/JHEP04(2021)097}{\emph{JHEP} {\bfseries 04}
  (2021) 097} [\href{https://arxiv.org/abs/2012.15576}{{\ttfamily
  2012.15576}}].

\bibitem{Guerrieri:2020bto}
A.L.~Guerrieri, J.~Penedones and P.~Vieira, \emph{{S-matrix bootstrap for
  effective field theories: massless pions}},
  \href{https://doi.org/10.1007/JHEP06(2021)088}{\emph{JHEP} {\bfseries 06}
  (2021) 088} [\href{https://arxiv.org/abs/2011.02802}{{\ttfamily
  2011.02802}}].

\bibitem{Guerrieri:2020kcs}
A.L.~Guerrieri, A.~Homrich and P.~Vieira, \emph{{Dual S-matrix bootstrap. Part
  I. 2D theory}}, \href{https://doi.org/10.1007/JHEP11(2020)084}{\emph{JHEP}
  {\bfseries 11} (2020) 084}
  [\href{https://arxiv.org/abs/2008.02770}{{\ttfamily 2008.02770}}].

\bibitem{Correia:2020xtr}
M.~Correia, A.~Sever and A.~Zhiboedov, \emph{{An analytical toolkit for the
  S-matrix bootstrap}},
  \href{https://doi.org/10.1007/JHEP03(2021)013}{\emph{JHEP} {\bfseries 03}
  (2021) 013} [\href{https://arxiv.org/abs/2006.08221}{{\ttfamily
  2006.08221}}].

\bibitem{Hebbar:2020ukp}
A.~Hebbar, D.~Karateev and J.~Penedones, \emph{{Spinning S-matrix bootstrap in
  4d}}, \href{https://doi.org/10.1007/JHEP01(2022)060}{\emph{JHEP} {\bfseries
  01} (2022) 060} [\href{https://arxiv.org/abs/2011.11708}{{\ttfamily
  2011.11708}}].

\bibitem{Sinha:2020win}
A.~Sinha and A.~Zahed, \emph{{Crossing Symmetric Dispersion Relations in
  Quantum Field Theories}},
  \href{https://doi.org/10.1103/PhysRevLett.126.181601}{\emph{Phys. Rev. Lett.}
  {\bfseries 126} (2021) 181601}
  [\href{https://arxiv.org/abs/2012.04877}{{\ttfamily 2012.04877}}].

\bibitem{Guerrieri:2021ivu}
A.~Guerrieri, J.~Penedones and P.~Vieira, \emph{{Where Is String Theory in the
  Space of Scattering Amplitudes?}},
  \href{https://doi.org/10.1103/PhysRevLett.127.081601}{\emph{Phys. Rev. Lett.}
  {\bfseries 127} (2021) 081601}
  [\href{https://arxiv.org/abs/2102.02847}{{\ttfamily 2102.02847}}].

\bibitem{Correia:2021etg}
M.~Correia, A.~Sever and A.~Zhiboedov, \emph{{Probing multi-particle unitarity
  with the Landau equations}},
  \href{https://doi.org/10.21468/SciPostPhys.13.3.062}{\emph{SciPost Phys.}
  {\bfseries 13} (2022) 062}
  [\href{https://arxiv.org/abs/2111.12100}{{\ttfamily 2111.12100}}].

\bibitem{Tourkine:2021fqh}
P.~Tourkine and A.~Zhiboedov, \emph{{Scattering from production in 2d}},
  \href{https://doi.org/10.1007/JHEP07(2021)228}{\emph{JHEP} {\bfseries 07}
  (2021) 228} [\href{https://arxiv.org/abs/2101.05211}{{\ttfamily
  2101.05211}}].

\bibitem{Karateev:2022jdb}
D.~Karateev, J.~Marucha, J.a.~Penedones and B.~Sahoo, \emph{{Bootstrapping the
  $a$-anomaly in $4d$ QFTs}},
  \href{https://arxiv.org/abs/2204.01786}{{\ttfamily 2204.01786}}.

\bibitem{EliasMiro:2021nul}
J.~Elias~Mir\'o and A.~Guerrieri, \emph{{Dual EFT bootstrap: QCD flux tubes}},
  \href{https://doi.org/10.1007/JHEP10(2021)126}{\emph{JHEP} {\bfseries 10}
  (2021) 126} [\href{https://arxiv.org/abs/2106.07957}{{\ttfamily
  2106.07957}}].

\bibitem{He:2021eqn}
Y.~He and M.~Kruczenski, \emph{{S-matrix bootstrap in 3+1 dimensions:
  regularization and dual convex problem}},
  \href{https://doi.org/10.1007/JHEP08(2021)125}{\emph{JHEP} {\bfseries 08}
  (2021) 125} [\href{https://arxiv.org/abs/2103.11484}{{\ttfamily
  2103.11484}}].

\bibitem{Guerrieri:2021tak}
A.~Guerrieri and A.~Sever, \emph{{Rigorous Bounds on the Analytic S Matrix}},
  \href{https://doi.org/10.1103/PhysRevLett.127.251601}{\emph{Phys. Rev. Lett.}
  {\bfseries 127} (2021) 251601}
  [\href{https://arxiv.org/abs/2106.10257}{{\ttfamily 2106.10257}}].

\bibitem{Chowdhury:2021ynh}
S.D.~Chowdhury, K.~Ghosh, P.~Haldar, P.~Raman and A.~Sinha, \emph{{Crossing
  Symmetric Spinning S-matrix Bootstrap: EFT bounds}},
  \href{https://doi.org/10.21468/SciPostPhys.13.3.051}{\emph{SciPost Phys.}
  {\bfseries 13} (2022) 051}
  [\href{https://arxiv.org/abs/2112.11755}{{\ttfamily 2112.11755}}].

\bibitem{Chen:2022nym}
H.~Chen, A.L.~Fitzpatrick and D.~Karateev, \emph{{Nonperturbative Bounds on
  Scattering of Massive Scalar Particles in $d \geq 2$}},
  \href{https://arxiv.org/abs/2207.12448}{{\ttfamily 2207.12448}}.

\bibitem{Miro:2022cbk}
J.E.~Miro, A.~Guerrieri and M.A.~Gumus, \emph{{Bridging Positivity and S-matrix
  Bootstrap Bounds}},  \href{https://arxiv.org/abs/2210.01502}{{\ttfamily
  2210.01502}}.

\bibitem{Guerrieri:2022sod}
A.~Guerrieri, H.~Murali, J.~Penedones and P.~Vieira, \emph{{Where is M-theory
  in the space of scattering amplitudes?}},
  \href{https://arxiv.org/abs/2212.00151}{{\ttfamily 2212.00151}}.

\bibitem{Haring:2022sdp}
K.~H\"aring, A.~Hebbar, D.~Karateev, M.~Meineri and J.a.~Penedones,
  \emph{{Bounds on photon scattering}},
  \href{https://arxiv.org/abs/2211.05795}{{\ttfamily 2211.05795}}.

\bibitem{Correia:2022dyp}
M.~Correia, J.~Penedones and A.~Vuignier, \emph{{Injecting the UV into the
  Bootstrap: Ising Field Theory}},
  \href{https://arxiv.org/abs/2212.03917}{{\ttfamily 2212.03917}}.

\bibitem{Marucha:2023vrn}
J.K.~Marucha, \emph{{Bootstrapping the $a$-anomaly in $4d$ QFTs: Episode II}},
  \href{https://arxiv.org/abs/2307.02305}{{\ttfamily 2307.02305}}.

\bibitem{He:2023lyy}
Y.~He and M.~Kruczenski, \emph{{Bootstrapping gauge theories}},
  \href{https://arxiv.org/abs/2309.12402}{{\ttfamily 2309.12402}}.

\bibitem{Acanfora:2023axz}
F.~Acanfora, A.~Guerrieri, K.~H\"aring and D.~Karateev, \emph{{Bounds on
  scattering of neutral Goldstones}},
  \href{https://arxiv.org/abs/2310.06027}{{\ttfamily 2310.06027}}.

\bibitem{Chen:2021pgx}
H.~Chen, A.L.~Fitzpatrick and D.~Karateev, \emph{{Bootstrapping 2d
  \ensuremath{\phi}$^{4}$ theory with Hamiltonian truncation data}},
  \href{https://doi.org/10.1007/JHEP02(2022)146}{\emph{JHEP} {\bfseries 02}
  (2022) 146} [\href{https://arxiv.org/abs/2107.10286}{{\ttfamily
  2107.10286}}].

\bibitem{Smirnov:2016lqw}
F.A.~Smirnov and A.B.~Zamolodchikov, \emph{{On space of integrable quantum
  field theories}},
  \href{https://doi.org/10.1016/j.nuclphysb.2016.12.014}{\emph{Nucl. Phys. B}
  {\bfseries 915} (2017) 363}
  [\href{https://arxiv.org/abs/1608.05499}{{\ttfamily 1608.05499}}].

\bibitem{Camilo:2021gro}
G.~Camilo, T.~Fleury, M.~Lencs\'es, S.~Negro and A.~Zamolodchikov, \emph{{On
  factorizable S-matrices, generalized TTbar, and the Hagedorn transition}},
  \href{https://doi.org/10.1007/JHEP10(2021)062}{\emph{JHEP} {\bfseries 10}
  (2021) 062} [\href{https://arxiv.org/abs/2106.11999}{{\ttfamily
  2106.11999}}].

\bibitem{Hannesdottir:2022bmo}
H.S.~Hannesdottir and S.~Mizera, \emph{{What is the i\ensuremath{\varepsilon}
  for the S-matrix?}}, SpringerBriefs in Physics, Springer (1, 2023),
  \href{https://doi.org/10.1007/978-3-031-18258-7}{10.1007/978-3-031-18258-7},
  [\href{https://arxiv.org/abs/2204.02988}{{\ttfamily 2204.02988}}].

\bibitem{Correia:2022dcu}
M.~Correia, \emph{{Nonperturbative Anomalous Thresholds}},
  \href{https://arxiv.org/abs/2212.06157}{{\ttfamily 2212.06157}}.

\bibitem{Caron-Huot:2023ikn}
S.~Caron-Huot, M.~Giroux, H.S.~Hannesdottir and S.~Mizera, \emph{{Crossing
  beyond scattering amplitudes}},
  \href{https://arxiv.org/abs/2310.12199}{{\ttfamily 2310.12199}}.

\bibitem{Karateev:2020axc}
D.~Karateev, \emph{{Two-point functions and bootstrap applications in quantum
  field theories}}, \href{https://doi.org/10.1007/JHEP02(2022)186}{\emph{JHEP}
  {\bfseries 02} (2022) 186}
  [\href{https://arxiv.org/abs/2012.08538}{{\ttfamily 2012.08538}}].

\bibitem{Zamolodchikov:1978xm}
A.B.~Zamolodchikov and A.B.~Zamolodchikov, \emph{{Factorized s Matrices in
  Two-Dimensions as the Exact Solutions of Certain Relativistic Quantum Field
  Models}}, \href{https://doi.org/10.1016/0003-4916(79)90391-9}{\emph{Annals
  Phys.} {\bfseries 120} (1979) 253}.

\bibitem{Zamolodchikov:1986gt}
A.B.~Zamolodchikov, \emph{{Irreversibility of the Flux of the Renormalization
  Group in a 2D Field Theory}}, {\emph{JETP Lett.} {\bfseries 43} (1986) 730}.

\bibitem{Vilasis-Cardona:1994oke}
X.~Vilasis-Cardona, \emph{{Renormalization group flows and conserved vector
  currents}}, \href{https://doi.org/10.1016/0550-3213(94)00451-J}{\emph{Nucl.
  Phys. B} {\bfseries 435} (1995) 735}
  [\href{https://arxiv.org/abs/hep-th/9404150}{{\ttfamily hep-th/9404150}}].

\bibitem{Simmons-Duffin:2015qma}
D.~Simmons-Duffin, \emph{{A Semidefinite Program Solver for the Conformal
  Bootstrap}}, \href{https://doi.org/10.1007/JHEP06(2015)174}{\emph{JHEP}
  {\bfseries 06} (2015) 174}
  [\href{https://arxiv.org/abs/1502.02033}{{\ttfamily 1502.02033}}].

\bibitem{Landry:2019qug}
W.~Landry and D.~Simmons-Duffin, \emph{{Scaling the semidefinite program solver
  SDPB}},  \href{https://arxiv.org/abs/1909.09745}{{\ttfamily 1909.09745}}.

\bibitem{Watson:1954uc}
K.M.~Watson, \emph{{Some general relations between the photoproduction and
  scattering of pi mesons}},
  \href{https://doi.org/10.1103/PhysRev.95.228}{\emph{Phys. Rev.} {\bfseries
  95} (1954) 228}.

\bibitem{Cardy:1986ie}
J.L.~Cardy, \emph{{Operator Content of Two-Dimensional Conformally Invariant
  Theories}}, \href{https://doi.org/10.1016/0550-3213(86)90552-3}{\emph{Nucl.
  Phys. B} {\bfseries 270} (1986) 186}.

\bibitem{Smirnov:1992vz}
F.A.~Smirnov, \emph{{Form-factors in completely integrable models of quantum
  field theory}}, vol.~14 (1992).

\bibitem{Omnes:1958hv}
R.~Omnes, \emph{{On the Solution of certain singular integral equations of
  quantum field theory}}, \href{https://doi.org/10.1007/BF02747746}{\emph{Nuovo
  Cim.} {\bfseries 8} (1958) 316}.

\bibitem{PhysRev.137.B720}
N.N.~Khuri and T.~Kinoshita, \emph{Real part of the scattering amplitude and
  the behavior of the total cross section at high energies},
  \href{https://doi.org/10.1103/PhysRev.137.B720}{\emph{Phys. Rev.} {\bfseries
  137} (1965) B720}.

\bibitem{Bronzan:1974jh}
J.B.~Bronzan, G.L.~Kane and U.P.~Sukhatme, \emph{{Obtaining Real Parts of
  Scattering Amplitudes Directly from Cross-Section Data Using Derivative
  Analyticity Relations}},
  \href{https://doi.org/10.1016/0370-2693(74)90432-8}{\emph{Phys. Lett. B}
  {\bfseries 49} (1974) 272}.

\bibitem{1987CzJPh..37..297F}
J.~{Fischer} and P.~{Kol{\'a}{\v{r}}}, \emph{{Differential forms of the
  dispersion integral}},
  \href{https://doi.org/10.1007/BF01597257}{\emph{Czechoslovak Journal of
  Physics} {\bfseries 37} (1987) 297}.

\bibitem{Chester:2019wfx}
S.M.~Chester, \emph{{Weizmann Lectures on the Numerical Conformal Bootstrap}},
  \href{https://arxiv.org/abs/1907.05147}{{\ttfamily 1907.05147}}.

\bibitem{Appadu:2017bnv}
C.~Appadu, T.J.~Hollowood, D.~Price and D.C.~Thompson, \emph{{Yang Baxter and
  Anisotropic Sigma and Lambda Models, Cyclic RG and Exact S-Matrices}},
  \href{https://doi.org/10.1007/JHEP09(2017)035}{\emph{JHEP} {\bfseries 09}
  (2017) 035} [\href{https://arxiv.org/abs/1706.05322}{{\ttfamily
  1706.05322}}].

\bibitem{DELFINO1996327}
G.~Delfino, P.~Simonetti and J.~Cardy, \emph{Asymptotic factorisation of form
  factors in two-dimensional quantum field theory},
  \href{https://doi.org/https://doi.org/10.1016/0370-2693(96)01035-0}{\emph{Physics
  Letters B} {\bfseries 387} (1996) 327}.

\bibitem{Caron-Huot:2023tpw}
S.~Caron-Huot, A.~Pokraka and Z.~Zahraee, \emph{{Two-point sum-rules in
  three-dimensional Yang-Mills theory}},
  \href{https://arxiv.org/abs/2309.04472}{{\ttfamily 2309.04472}}.

\bibitem{Gorbenko:2020xya}
V.~Gorbenko and B.~Zan, \emph{{Two-dimensional O(n) models and logarithmic
  CFTs}}, \href{https://doi.org/10.1007/JHEP10(2020)099}{\emph{JHEP} {\bfseries
  10} (2020) 099} [\href{https://arxiv.org/abs/2005.07708}{{\ttfamily
  2005.07708}}].

\bibitem{Jacobsen:2023vdf}
J.L.~Jacobsen, R.~Nivesvivat and H.~Saleur, \emph{{On currents in the $O(n)$
  loop model}},  \href{https://arxiv.org/abs/2310.11064}{{\ttfamily
  2310.11064}}.

\bibitem{Zamolodchikov:1990dg}
A.B.~Zamolodchikov, \emph{{Exact S matrix associated with selfavoiding polymer
  problem in two-dimensions}},
  \href{https://doi.org/10.1142/S0217732391001950}{\emph{Mod. Phys. Lett. A}
  {\bfseries 6} (1991) 1807}.

\bibitem{Gorbenko:2018dtm}
V.~Gorbenko, S.~Rychkov and B.~Zan, \emph{{Walking, Weak first-order
  transitions, and Complex CFTs II. Two-dimensional Potts model at $Q>4$}},
  \href{https://doi.org/10.21468/SciPostPhys.5.5.050}{\emph{SciPost Phys.}
  {\bfseries 5} (2018) 050} [\href{https://arxiv.org/abs/1808.04380}{{\ttfamily
  1808.04380}}].

\bibitem{Gorbenko:2018ncu}
V.~Gorbenko, S.~Rychkov and B.~Zan, \emph{{Walking, Weak first-order
  transitions, and Complex CFTs}},
  \href{https://doi.org/10.1007/JHEP10(2018)108}{\emph{JHEP} {\bfseries 10}
  (2018) 108} [\href{https://arxiv.org/abs/1807.11512}{{\ttfamily
  1807.11512}}].

\bibitem{Cavaglia:2016oda}
A.~Cavagli\`a, S.~Negro, I.M.~Sz\'ecs\'enyi and R.~Tateo, \emph{{$T
  \bar{T}$-deformed 2D Quantum Field Theories}},
  \href{https://doi.org/10.1007/JHEP10(2016)112}{\emph{JHEP} {\bfseries 10}
  (2016) 112} [\href{https://arxiv.org/abs/1608.05534}{{\ttfamily
  1608.05534}}].

\bibitem{Cordova:2021fnr}
L.~C\'ordova, S.~Negro and F.I.~Schaposnik~Massolo, \emph{{Thermodynamic Bethe
  Ansatz past turning points: the (elliptic) sinh-Gordon model}},
  \href{https://doi.org/10.1007/JHEP01(2022)035}{\emph{JHEP} {\bfseries 01}
  (2022) 035} [\href{https://arxiv.org/abs/2110.14666}{{\ttfamily
  2110.14666}}].

\bibitem{Mussardo:2020rxh}
G.~Mussardo, \emph{{Statistical Field Theory}}, Oxford Graduate Texts, Oxford
  University Press (3, 2020).

\bibitem{Karowski:1978vz}
M.~Karowski and P.~Weisz, \emph{{Exact Form-Factors in (1+1)-Dimensional Field
  Theoretic Models with Soliton Behavior}},
  \href{https://doi.org/10.1016/0550-3213(78)90362-0}{\emph{Nucl. Phys. B}
  {\bfseries 139} (1978) 455}.

\end{thebibliography}\endgroup

\end{document}